\documentclass[12pt,letterpaper,twoside,singlespace]{thesis} 

\usepackage{graphicx}
\usepackage{times}
\usepackage{psfrag}

\title{\bf Two problems in spin-dependent transport in metallic magnetic 
multilayers}
\thesisauthor{Asya Shpiro}
\degree{Doctor of Philosophy}
\department{Physics}
\thesisadviser{Peter M. Levy}     
\submitmonth{January}           
\submityear{2004}                 

\begin{document}
\ifx\href\undefined\else\hypersetup{linktocpage=true}\fi

    \pagenumbering{roman}
    \maketitle
    \copyrightpage        
    \begin{preface}{}

\Large \it
\begin{center}
To my parents
\end{center} 

\end{preface}
    \begin{preface}{Acknowledgements}

I would like to express my deepest gratitude to my thesis adviser,
Professor Peter M. Levy, for his constant support and guidance, patience
and encouragement. I appreciate an opportunity to learn from his
expertise, dedication to physics and to the people under his supervision.

Almost all my research was done in a close collaboration with Professor
Shufeng Zhang. I am grateful to him for his explanations, suggestions,
and constant interest in my work. 

I would also like to acknowledge the hospitality of the Laboratoire de 
Physique des Solides at the Universite Paris-Sud in Orsay, France, where 
part of my research was done, in particular, Professor Albert Fert and 
Charles Sommers.

I am thankful to many graduate students, current and former, at Physics
Department. Their attitude toward knowledge sharing and supporting each
other was a great help. Many of them became good friends of mine. 

I am in a great debt to my parents, Polina and Vladimir Shpiro, who
first introduced me to the world of science and the world of scientists,
who influenced my choice of career, and made me realize that physicists
are the best people on Earth. My PhD is their big achievement.

Finally, I would like to thank the people who were next to me during the
last stages of my work on the dissertation, my husband Lazar Fleysher
and our daughter Sonya, for their love and understanding.

\end{preface}
    \begin{preface}{Abstract}

Transport properties of magnetic multilayers in current perpendicular to
the plane of the layers geometry are defined by diffusive scattering in
the bulk of the layers, and diffusive and ballistic scattering across
the interfaces. Due to the short screening length in metals, layer by
layer treatment of the multilayers is possible, when the macroscopic
transport equation (Boltzmann equation or diffusion equation) is solved
in each layer, while the details of the interface scattering are taken
into account via the boundary conditions.  Embedding the ballistic and
diffusive interface scattering in the framework of the diffusive
scattering in the bulk of the layers is the unifying idea behind the
problems addressed in this work: the problem of finding the interface
resistance in the multilayered structures and the problem of
current-induced magnetization switching.

In order to find the interface resistance, the method of solving 
the semiclassical linearized Boltzmann equation in CPP geometry is 
developed, allowing one to obtain the equations for the chemical 
potential everywhere in the multilayers in the presence of specular 
and diffuse scattering at the interfaces, and diffuse scattering in the 
bulk of the layers. The variation of the chemical potential within a 
mean-free path of the interfaces leads to a breakdown of the 
resistors-in-series model which is currently used to analyze experimental 
data. While the resistance of the whole system is found by adding 
resistances due to the bulk of the layers and resistances due to the 
interfaces, the interface resistances are not independent of the 
properties of the bulk of the layers, particularly, of the ratio of the 
layer thickness to the mean-free path in this layer. 

A mechanism of the magnetization switching that is driven by
spin-polarized current is studied in noncollinear magnetic multilayers.
Even though the transfer of the spin angular momentum between current
carriers and local moments occurs near the interface of the magnetic
layers, in order to determine the magnitude of the transfer one should
calculate the spin transport properties far beyond the interfacial
regions. Due to the presence of long longitudinal spin-diffusion lengths,
the longitudinal and transverse components of the spin accumulations
become intertwined from one layer to the next, leading to a significant
amplification of the spin-torque with respect to the treatments that
concentrate on the transport at the interface only, i.e., those that 
only consider the contribution to the torque from the bare current and 
neglect that arising from spin accumulation.

\end{preface}

    \tableofcontents          
    \listoffigures            
    \listofappendices        


\begin{thesisbody}

 
\chapter{\label{chap_intro} Introduction}

\section{\label{mr_effects}Magnetoresistive effects}

During the last decade, attention has been focused on a question of
magnetically controlled electrical transport. Materials can change their
resistance in response to a magnetic field. This phenomena is called
magnetoresistance (MR). All metals have an inherent, though small, MR
owing to the Lorentz force that a magnetic field exerts on moving
electrons. Metallic alloys containing magnetic atoms can have an enhanced
MR. For example, an anisotropic magnetoresistance (AMR) measures the
change in resistance as the direction of the magnetization changes
relative to the direction of the electric current. (Fig.~\ref{anis_mr})  
If $\theta$ is the angle between the magnetization and the current
direction, (Fig.~\ref{anis_mr}b)), the resistance is the following
function of $\theta$:
\begin{equation}
\label{R_amr}
R=R_0+\Delta R_{AMR}\cos^2\theta,
\end{equation}
where $R_0$ is the resistance at $\theta=\pi/2$, and $\Delta
R_{AMR}/R_0$ is the AMR ratio. The resistivity is typically smaller if
the current direction is perpendicular to the direction of magnetization
than in the condition that those are parallel due to the scattering
anisotropy of electrons. The AMR ratio is fairly small: a few percent for
Ni$_{0.8}$Fe$_{0.2}$ alloy (permalloy) at room temperature, and somewhat
larger at lower temperatures. Nevertheless, the phenomenon of anisotropic
magnetoresistance has a significant importance for technical
applications, such as magnetic sensors. As illustrated in
Fig.~\ref{anis_mr}c), if a magnet is attached on a rotating disk, an MR
sensor can detect the number of rotations or the speed of motion from the
resistance change of the MR sensor. It has also been attempted to apply
an AMR sensor for a magnetic recording technology, but for ultra-high
density recording, a very high sensitivity of reading head is necessary,
and thus the high MR ratio is required, which can't be provided by the
AMR effect.

\begin{figure}
\centering
\includegraphics[width=\textwidth]{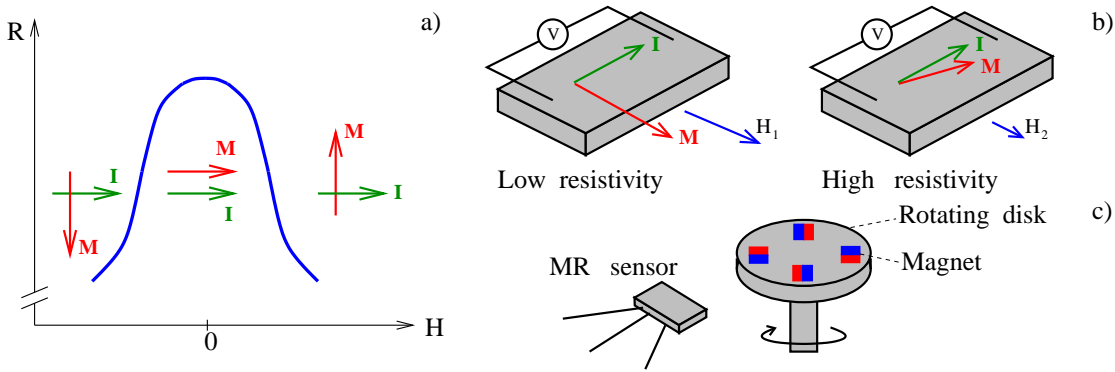}
\caption[Anisotropic magnetoresistance]{Anisotropic magnetoresistance 
(AMR): a) Resistance as a function of applied field. b) Measurement. V, 
I, M, H and $\theta$ are voltage, current, magnetization, applied field, 
and the angle between the current and the magnetization. c) Application of 
AMR sensor. Adapted from Ref.~\cite{sdtmn_ch1}}
\label{anis_mr}
\end{figure}

Magnetoresistive effects are more pronounced in magnetic layered
structures, or magnetic superlattices. They are formed artificially by
alternately depositing on a substrate several atomic layers of one
element, say, iron, followed by layers of another element, such as
chromium. A term Giant Magnetoresistance (GMR) is used to describe the
behavior of materials consisting of alternating layers of ferromagnetic
and nonmagnetic metals deposited on an insulating substrate.  The GMR
effect has been observed in 1988 in the resistivity measurements on Fe/Cr
multilayers,~\cite{baib_prl_88} as shown in Fig.~\ref{gmr_exp}.  At 4.2 K
the resistivity of the Fe/Cr multilayer was decreased by almost 50$\%$ by
applying an external field. At 300 K, the decrease of resistivity reaches
17$\%$, which is significantly larger than MR changes caused by the AMR
effect.  If $\theta$ is the angle between the magnetization directions of
the neighboring ferromagnetic layers, the resistance of the system takes
the following form:
\begin{equation}
\label{R_gmr}
R=R_P+\frac{\Delta R_{GMR}}{2}(1-\cos\theta),
\end{equation}
where $R_P$ is the resistance of the system when the magnetic moments in
the alternating layers are aligned in the same direction, or parallel
($\theta=0$), and $\Delta R_{GMR}=R_{AP}-R_P$, where $R_{AP}$ is the
resistance of the system when the magnetic moments of the neighboring
layers are oppositely aligned, or antiparallel ($\theta=\pi$).  
Resistance is the greatest for the antiparallel (AP) configuration of the
magnetizations, and smallest for the parallel (P)  
one~(Fig.~\ref{gmr_schem}). The GMR ratio is defined as
$(R_{AP}-R_{P})/R_{AP}*100\%$ or, alternatively, as
$(R_{AP}-R_{P})/R_{P}*100\%$. GMR effects has been obtained in two 
geometries (Fig.~\ref{cip_cpp}). In the first one the current is applied 
in the plane of the layer (Current In Plane, or CIP geometry), while in 
the second one the current flows perpendicular to the plane of the 
layers (Current Perpendicular to the Plane, or CPP geometry). The 
CPP-MR is larger than the CIP-MR, and it exists at much larger 
thicknesses of the samples. 
\begin{figure}
\centering
\includegraphics[width=3.5 in]{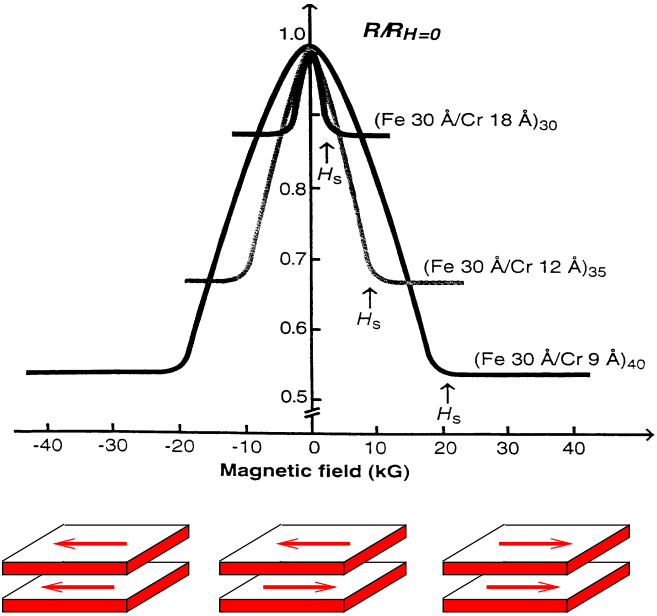}
\caption[Magnetoresistance of Fe/Cr superlattices]{Magnetoresistance of 
three Fe/Cr superlattices at 4.2 K. The 
current and the applied field are along the same [110] axis in the 
plane of the layers. Adapted from Ref.~\cite{baib_prl_88}}
\label{gmr_exp}
\end{figure}
\begin{figure}
\centering
\includegraphics[width=4.5 in]{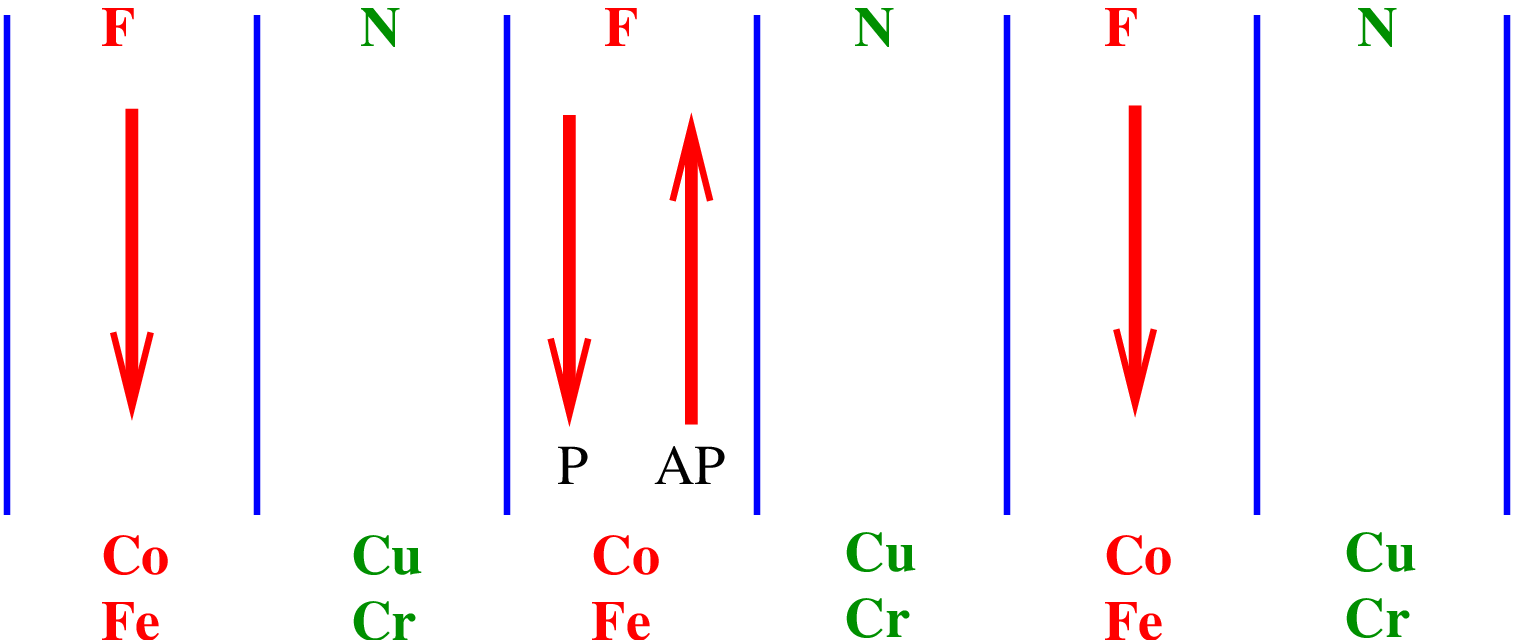}
\caption{Ferromagnetic and antiferromagnetic configuration of a
multilayered magnetic metallic structure}
\label{gmr_schem}
\end{figure}
\begin{figure}
\centering
\includegraphics[width=3.5 in]{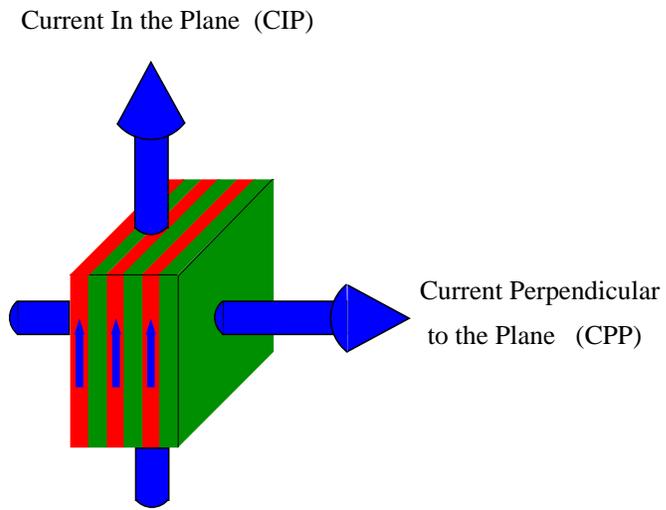}
\caption[Current In Plane of the layers and Current Perpendicular to
the Plane of the layers geometries]{Current 
In Plane of the layers and Current Perpendicular to 
the Plane of the layers geometries. Adapted from Ref~\cite{sdtmn_ch2}}
\label{cip_cpp}
\end{figure}

The mechanism of GMR is attributed to the change of the magnetic
structure induced by an external field. The role of external magnetic
field is to change the internal magnetic configuration; in cases when it
is not possible, the GMR does not appear. It is necessary to separate
magnetic regions from one another so as to be able to reorient their
magnetization, otherwise layers are too strongly coupled and ordinary
fields can not rotate magnetic moments in one layer relative to moments
in the neighboring layer. The multilayers are designed so that at zero
external field the magnetizations of the alternating magnetic layers are
aligned opposite to each other. The external magnetic field rotates the
magnetization, so that all ferromagnetic layers become aligned in the
same direction, and resistance changes. There are several ways to provide
the original antiparallel alignment of the neighboring FM layers.  The
first one involves choosing the metallic interlayer thickness ($\approx
1$ nm) such that the Ruderman-Kittel-Kasuya-Yoshida (RKKY)  coupling
between localized moments via the conduction electrons forces the
magnetic layers to be antiferromagnetically coupled.  The other is to
make the successive magnetic layers with different coercivities. Smaller
field may switch a layer with the smaller coercivity, while the layer
with a bigger one remains in the previous direction; it will rotate at
larger field, and there is a range of external fields when the alignment
is antiparallel.  Finally, a phenomena known as "exchange bias" may be
used, when an AF metal in a contact with FM metal pins the magnetization
of the later in the direction opposite to the magnetization of the last
layer of AF so as to produce zero net magnetization. The second FM layer,
separated from the pinned one by the nonmagnetic layer remains free to
rotate its magnetization under the influence of the external field.

The primary source of GMR is spin-dependent scattering of conducting
electrons, or, more precisely, change in the scattering rate as the
magnetic configuration changes by an external magnetic field.  In the
antiparallel magnetic structure, conduction electrons are much more
scattered that in the parallel magnetic structure. Spin-dependent
scattering can make a large contribution to the resistivity. In the
following section, I will discuss a phenomenon of spin-dependent
scattering and its effect on the electron transport.

Similarly to AMR sensors, GMR sensors can also be used to detect rotations
and to measure the speed of motion. They have an advantage of the full
angular dependence. Whereas for an AMR-type sensor opposite field
directions produce the same signal, for a GMR-type sensor parallel and
antiparallel magnetization alignments yield different resistivities. High
sensitivity of GMR sensors make them attractive as the sensors for
magnetic fields, particularly, as read-out heads in hard disk drives in
computers. The fact that GMR is mainly an interface effect (see below)  
allows the sensor to be made thinner, which leads to the improvement of
the spatial resolution in read-out. GMR-based hard drives are already
being used in computer industry. The magnetic random access memories
(MRAMs) based on GMR effect have also been realized, but they are not
favorable because of the relatively small resistance of the GMR memory
element compared with the current leads connecting it with the processing
unit.

Besides metallic multilayer systems, similar phenomena are found in 
granular systems where small ferromagnetic clusters are dispersed in 
non-magnetic matrices. Extremely large MR effect has been found in 
manganese perovskite oxides and is called the colossal MR (CMR) 
effect. Another class of structures where high MR ratios has been 
obtained is the tunneling junctions, where the layers of ferromagnetic 
metal are separated by a thin insulating barrier; the MR effect in these 
structures is called tunnel magnetoresistance (TMR). 

\section{Spin-dependent transport in metallic magnetic multilayers and 
the GMR}

The transport in the GMR devices studied to date is {\it diffusive}. In a
macroscopic sample, electron undergoes scattering off a large number of
impurities. One is in the regime of diffusive transport when the
concentration of impurities requires the averaging of the scattering
potential. The averaging leads to the loss of the memory of the electron
momentum direction and, hence, to a resistance. The characteristic
lengthscale over which electrons retain a memory of their momentum is
called the mean free path, $\lambda_{mfp}$. Due to the large transverse
size of the layered systems, the macroscopic transport equations, such as
the Boltzmann equation or diffusion equation, see
Sec.~\ref{chap_transp_mltlrs}, can be applied in order to describe
electron transport in multilayers, even if the thickness of the layers is
of the order or smaller than the mean free path. In the multilayers, the
presence of interfaces leads to additional scattering. At the interface
between layers of different metals, electrons experience {\it specular}
scattering due to the band mismatch. The direction of the electron
momentum changes, but it does so in a predictable and reproducible
fashion. Specular scattering at the interface between two metals is
treated in detail in Appendix~\ref{app_refl_transm}. Roughness of the
interfaces leads to an additional, {\it diffusive} scattering, when the
incoming electron unpredictably changes the direction of its momentum,
and the information about the electron momentum is lost. Due to the short
screening length in metals (typically about 1-3~A, which is at least an
order of magnitude smaller than the layer thicknesses in the structures
currently being investigated), a layer by layer treatment of the
multilayers is possible, where the macroscopic transport equation is
solved in each layer, while the details of the interface scattering are
taken into account via the boundary conditions.

In ferromagnetic metals, the transport properties of electrons are {\it
spin-dependent}. This dependency arises from the unbalance of the spin
populations at the Fermi level due to the splitting between the up and
down spin states (exchange splitting). As illustrated in
Fig.~\ref{spin_polar} for cobalt, the majority (spin parallel to the
magnetization, or spin-up) d-states are all filled, and the d-electron
states at the Fermi level contain entirely minority (spin anti-parallel
to the magnetization, or spin-down) electrons. Although there are also s
and p electrons at the Fermi level, a significant number of the carriers
are the more highly polarized d electrons, which produces a current
which is partially spin-polarized. The imbalance of spin-up and
spin-down electrons also results in the different probabilities of the
s$\rightarrow$d transition for majority and minority electrons and
consequently in different resistivities.
\begin{figure}
\centering
\includegraphics[width=2.5 in]{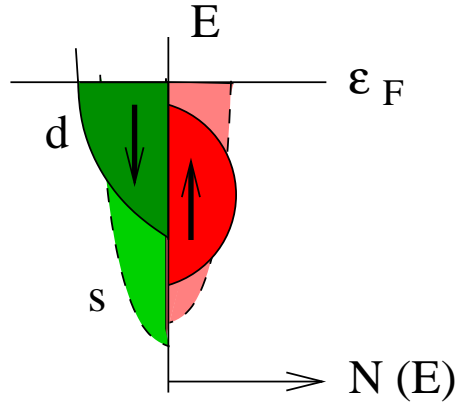}
\caption[Density of states for up and down electrons for 
cobalt]{Density of states $N(E)$ for up and down s and p electrons for 
cobalt. $\epsilon_F$ is the Fermi energy.}
\label{spin_polar}
\end{figure}

Spin-dependent scattering occurs both in the bulk of the layers and at
the interfaces. In a multilayer, electrons move in the potential which
reflects the band mismatch at the interfaces between the magnetic and
nonmagnetic layers. The exchange splitting of the up- and down d-bands
in the ferromagnetic layers results in different heights of the steps
seen by up- and down- conduction electrons (Fig.~\ref{spin_dep_scat}a)).
In the P configuration, the height of the steps is the same in all
layers but different for majority and minority spins. In the AP state,
small and large steps alternate for each spin direction. Another
contribution to the scattering is due to the presence of impurities in
the layers, and due to the roughness at the interfaces. This scattering
is also spin-dependent in ferromagnetic metals, and it is stronger for
minority electrons (Fig.~\ref{spin_dep_scat}b)), resulting in the
different resistance in P and AP states and the GMR effect. 
\begin{figure}
\centering
\includegraphics[width=4.5 in]{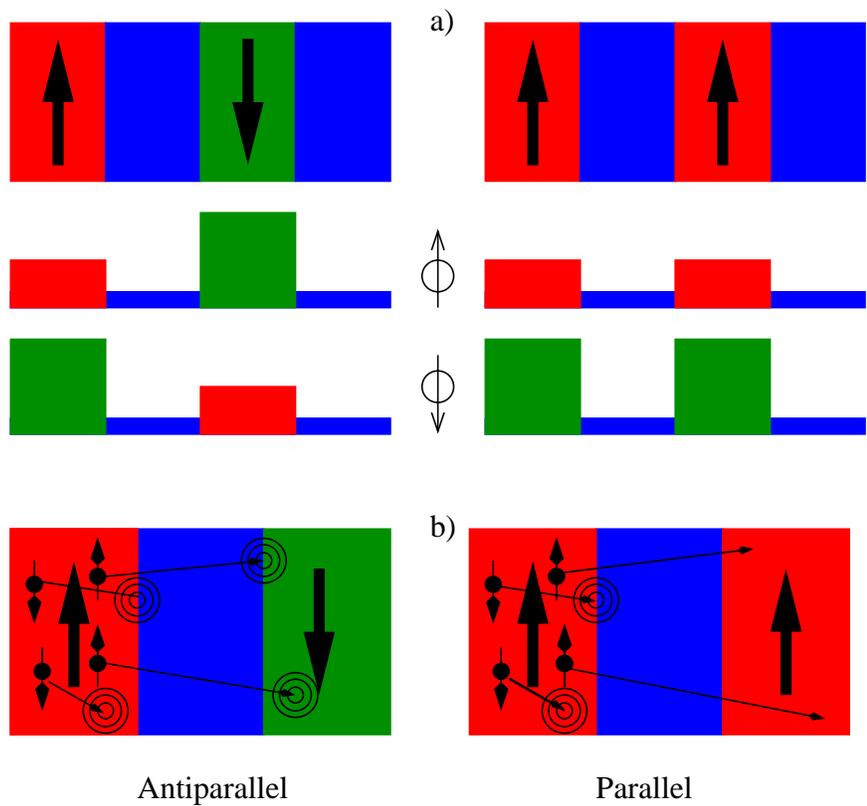}
\caption[Spin-dependent scattering]{a) Spin dependent potentials in a 
magnetic multilayer for the AP and P configurations for majority and 
minority electrons. Adapted from Ref.~\cite{sdtmn_ch2}. b) Illustration 
of spin-dependent electron scattering in GMR multilayers. Adapted from 
Ref.~\cite{sens_act}.}
\label{spin_dep_scat}
\end{figure}

A remarkable property of the electron transport in magnetic multilayers
is that at low temperatures most of the scattering encountered by the
electrons does not flip their spin, as it costs energy. Even though the
carrier may undergo many scattering events, the orientation of its spin
can be very long-lived. The characteristic time when the electron
remembers its spin is the spin-flip time $\tau_{sf}$, and the
characteristic length scale is the spin-diffusion length
$\lambda_{sdl}=\sqrt{D\tau_{sf}}$, where $D$ is the diffusion constant. 
The spin-diffusion length is usually of the order of ten times larger
then the electron mean-free path. The slow spin relaxation leads to a
formation of a steady-state non-equilibrium build-up of magnetic
moments, or spin-accumulation, extending over a length $\lambda_{sdl}$
from the interface between the layers with antiparallel magnetization
directions. This magnetization acts as a bottleneck for spin transport
across the interface, which in turn hinders the flow of charge and
results in the higher resistance in the antiparallel configuration
compared with the parallel configuration.~\cite{john_91}

The large spin-diffusion length in ferromagnetic metals means that the
current can be considered to be carried by two independent channels of
carriers, one for spin-up electrons with a resistivity $\rho_\uparrow$,
and the other made of spin-down electrons with a resistivity
$\rho_\downarrow$; this is commonly referred to as the {\it two current
model}. This model allows a simple illustration of how the GMR
effect works in CPP structures. If the thickness of the layers is larger
then the electron mean free path in the layers, a resistor network
analogy is appropriate (Fig.~\ref{two_curr}a)), when the resistivities of
each layer for each spin direction are added in series, while those for
two channels are added in parallel. Since the resistances of the majority
($R_M$) and minority ($R_m$) electrons are different, and $R_M<R_m$,
there is a "short-circuit" effect in the parallel configuration, and
resistance is smaller in P state than in the AP state, and the GMR effect
exists. In the CIP geometry, a simple resistor network analogy yields the
same resistances in both P and AP configurations (Fig.~\ref{two_curr}b)),
and the absence of the GMR effect. For the CIP-MR to exist, the
mean-free-path has to be larger than the thickness of a layer. In this
case, electrons, which travel parallel to the interfaces, sample the
scattering from different layers, and the current is sensitive to the
change of the magnetic configuration of the system. The electron
mean-free path is the characteristic length scale for spin-dependent
transport in the CIP geometry. In the CPP geometry, where the electric
current samples scattering in all layers, the GMR does not depend on
$\lambda_{mfp}$. The characteristic length scale for the transport in the
CPP geometry is the spin-diffusion length $\lambda_{sdl}$. Spin-flips
limit the distance over which the two spin currents are independent. If
the size of a sample is larger than $\lambda_{sdl}$, spin-up and
spin-down currents are mixed, and the CPP-MR is decreased. In this work,
I will only consider the current perpendicular to the plane of the layers
geometry.
\begin{figure}
\centering
\includegraphics[width=4.5 in]{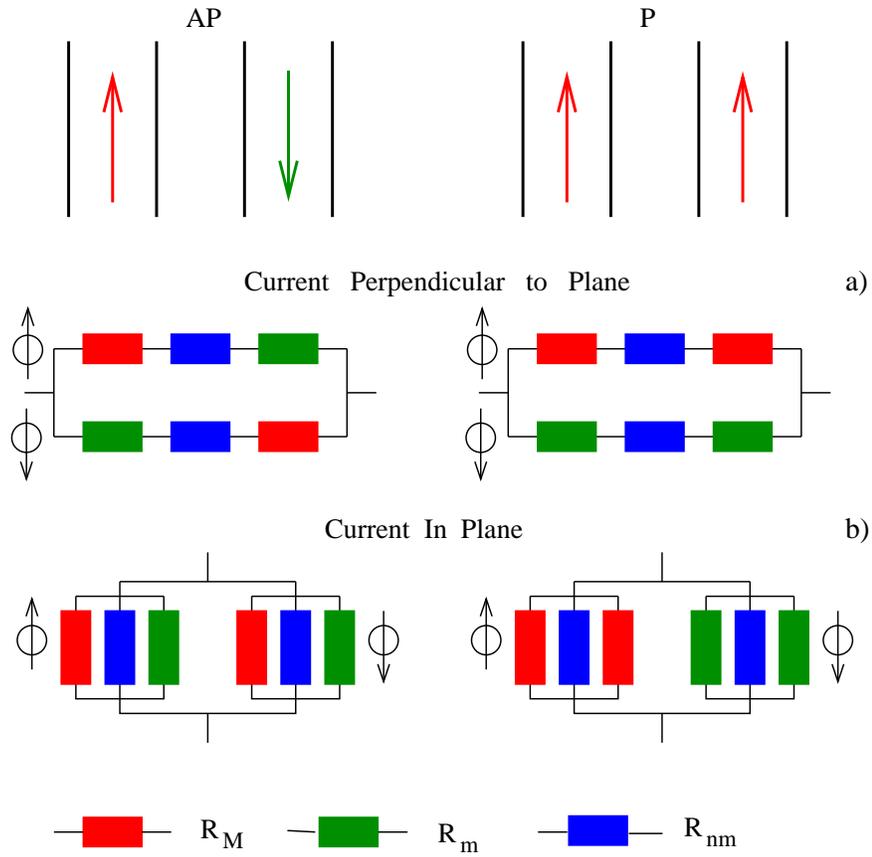}
\caption[Two current model: resistor network analogy]{Two-current model: 
resistor network analogy for the CIP and CPP resistances of the 
ferromagnetic (P) and antiferromagnetic (AP) configurations of the 
magnetic moments in a multilayer. $R_M$ (red), $R_m$ (green), and 
$R_{nm}$ (blue) stand for the resistances of the magnetic layers for 
spin parallel and antiparallel to the local magnetization, and of the 
non-magnetic layers. Adapted from Ref.~\cite{sdtmn_ch2}}
\label{two_curr}
\end{figure}

A CPP-MR model that provides a significant insight into the problem was
developed by Valet and Fert.~\cite{VF_prb} Within this model, the
Boltzmann equation, with an additional term to account for spin
accumulation, is solved at zero temperature. In the case when the mean
free path is much smaller then the spin diffusion length, the macroscopic
transport equations are obtained that relate the current and the
electrochemical potential, so that the resistance of the system can be
found. The Valet-Fert model correctly describes the main features of
magnetoresistive multilayers, i.e., increase in MR with decreasing
temperature, increase with decreasing film thickness and increase with
increasing number of layers.


\section{\label{sec_c-i-s}Current-induced switching} It has been seen
from the discussion of the GMR effect that the relative orientation of
the magnetic moments of the layers affects the electric current, causing
different resistances for different magnetic configurations. The reverse
effect, that a spin-polarized current could affect the magnetic moment
of a layer, has also been
predicted,~\cite{Sl_jmmm_96,Sl_jmmm_99,Sl_jmmm_02,Berg_prb_96,Berg_jap_01}
and experimentally demonstrated.~\cite{myers_science, tsoi_prl,
barb_kent} In the perpendicular transport geometry, the spin-polarized
currents may transfer angular momentum between the layers, resulting in
the current-driven excitations in magnetic multilayers: either reversal
of layer magnetization, or generation of spin-waves.~\cite{Bass_0310467}

Spin-polarized current exerts a torque on a ferromagnetic layer if that
layer's moment is not collinear with the direction of current
polarization. The reversal of the magnetization is due to the interaction
between the magnetization and the spin accumulation in a direction
perpendicular to the magnetization. Fig.~\ref{switch_sch} shows
schematically the five-layered structure used to measure current-induced
switching, and the directions of the torque acting on the magnetic moment
of the thin ferromagnetic layer due to the spin transfer by the current,
polarized in the direction of the magnetization of the thick
ferromagnetic layer.~\cite{myers_science}
\begin{figure}
\centering
\includegraphics[width=4.5 in]{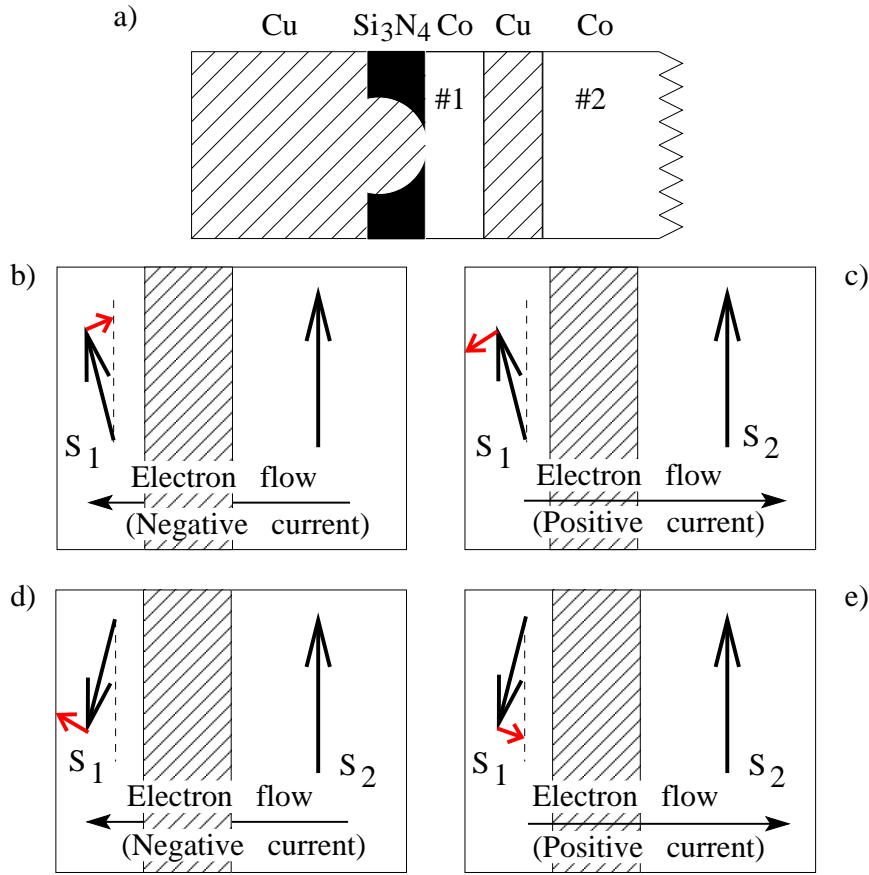} 
\caption[Current-induced magnetization switching]
{a) Schematic representation of the five-layered structure used to
measure current-induced switching. The system consists of the thin Co
layer of the variable thickness ($\#$1), separated by the 4 nm Cu layer
from the the thick (100 nm) Co layer ($\#$2). Another Cu layer contacts
the thin Co layer via an opening 5 to 10 nm in diameter in an insulating
silicon nitride membrane.
b)-e) Directions of torque (red arrows) on the magnetic 
moments in the thin FM layer due to the spin-polarized current. Adapted 
from Ref.~\cite{myers_science}} 
\label{switch_sch} 
\end{figure}

The great interest in the phenomenon of the current-driven excitations in
magnetic multilayers lies both in trying to understand the underlying
physics and in its potential for device use: magnetization reversal for
magnetic media and magnetic memories, and spin-wave generation for
production of high frequency radiation. In present magnetic devices the
moments are reversed via externally generated magnetic fields. The
reading/writing processes would be simplified by applying a polarized
current through the magnetic layer itself. Practically, though, the
magnetization reversal by spin transfer requires high current densities,
around 10$^7$~A/cm$^2$, in order to overcome magnetic damping. The
current density should be reduced by approximately an order of magnitude
so that this effect could be considered for 
applications.~\cite{fert_prepr_03}

\section{Problems to be solved}

Transport properties of magnetic multilayers in current perpendicular to
the plane of the layers geometry are defined by diffusive scattering in
the bulk of the layers, and diffusive and ballistic scattering across the
interfaces. Electrons scattered at the interfaces may be scattered again
in the bulk of the layers, and, in order to better describe transport in
the entire structure, the transport in the bulk of the layers and the
transport through the interfaces should be treated self-consistently.  
Embedding the ballistic and diffusive interface scattering in the
framework of the diffusive scattering in the bulk of the layers is the
unifying idea behind the problems that I address in my work: the problem
of finding the interface resistance and the problem of current-induced
magnetization switching.

\subsection{Interface resistance}

Chapters~\ref{chap_res_thr} and~\ref{chap_res_rslt} of this work are
devoted to the problem of finding resistance due to the interfaces in
metallic multilayered structures. The presence of the specular and
diffuse scattering at interfaces leads to a finite voltage drop
across them,~\cite{VF_jmmm,VF_prb,BF_prb,BF_jmmm} and, hence, to a
resistance. An origin of interface resistance, and the procedure of
finding a resistance of an interface between two ballistic conductors
is discussed in Chap.~\ref{chap_transp_mltlrs}. In the realistic
multilayered structure, where transport in the bulk of the layers is
diffusive, resistance of the interfaces has to be incorporated into
the resistance of the whole system. In order to do that, the
resistors-in-series model was developed, and is currently used to
analyze experimental data.~\cite{ZhL_jap_91,pratt_jmmm_93,
bass_ccmp_98,BP_jmmm_99} Within this model, resistance of an
interface is treated independently of the resistance of the bulk of
the layers, and total resistance of a multilayered system is a sum of
the bulk and interface contributions.(Fig.~\ref{res_in_ser})
\begin{figure}
\includegraphics[width=\textwidth]{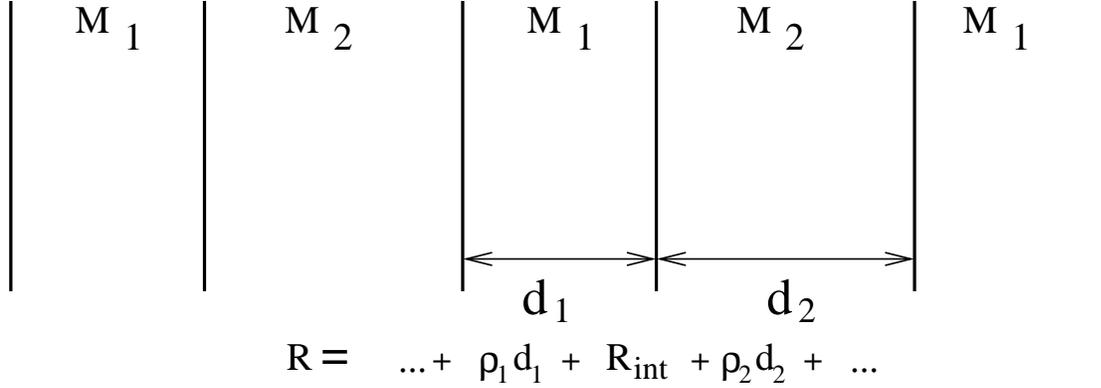}
\caption[Resistors-in-series model]{Resistors-in-series model. $\rho_1$ 
and $\rho_2$ are resistivities of the metallic layers M$_1$ and M$_2$, 
$d_1$ and $d_2$ are thicknesses of M$_1$ and M$_2$, $R_{int}$ is a 
resistance of the interface between M$_1$ and M$_2$.}
\label{res_in_ser}
\end{figure}

But the assumption that the resistance of the interface and resistance of
the bulk are independent is not correct. An electron reflected from the
interface can be scattered in the bulk and go back to the interface, not
to the reservoir, hence changing the current through the interface. In
terms of chemical potentials it means that in addition to an abrupt drop
at an interface, there is a gradual change in the chemical potential
within a mean-free-path from the
interface.~\cite{Zh_L,Butl,Penn_Stls,LkhtmLri,Kunze,land_prb_95}
Fig.~\ref{chempot_schem} shows schematically a chemical potential profile
at an interface between two semi-infinite metallic layers. In
Fig.~\ref{chempot_schem}, as well as in the all following pictures of
chemical potentials, only the potential drop due to the scattering from
the interface is shown; the drop due to the resistivity of the adjacent
layers is not shown. The resistance measured far from the interface,
$R_1$, is different from the one that would be measured in its immediate
vicinity, $R_{0}$. In the multilayers with layer thicknesses of the order
of electron mean-free path, the chemical potential doesn't reach its
asymptotic value within a layer. This leads to a breakdown of the
resistors-in-series model. While the resistance of the whole system is
still found by adding resistances due to the bulk of the layers and
resistances due to the interfaces, interface resistances are not
independent of the properties of the bulk of the layers, in particular, of
the ratio of the layer thickness to the electron mean-free path in this
layer.
\begin{figure}
\centering
\includegraphics[width=4.5in]{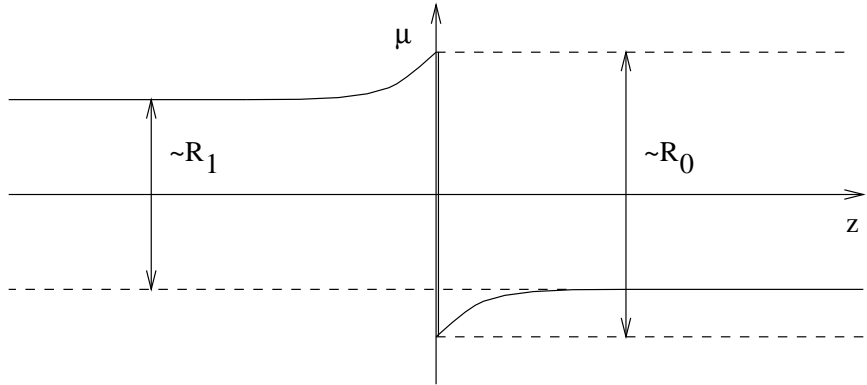}
\caption[Chemical potential profile in two semi-infinite metallic
layers]{Chemical potential profile in two semi-infinite metallic layers. 
Only the potential drop due to the scattering from the interface is
shown; the drop due to the resistivity of the adjacent layers is not
shown. $R_{0}$ is the resistance measured directly at the interface;
$R_1$ is the resistance measured far from the interface.}
\label{chempot_schem} 
\end{figure}

In order to determine the contribution of the interfaces to the resistance
of the whole multilayered structure with layer thicknesses of the order or
less the electron mean-free path, the form of chemical potential
everywhere in the system has to be considered, and there was no such study
for the systems with both specular and diffuse scattering at interfaces.
In Ref.\cite{Butl} the variation of the chemical potential was studied,
however without examining how diffuse scattering at the interface altered
its pattern, while in Ref.~\cite{Schep} the interface resistance $R_1$
(their $R_{A/B}$) coming from the specular reflection at the interface
together with diffuse scattering in the bulk of the layers is calculated
far from the interface. More recently the change of interface resistance
$R_{A/B}$ with diffuse scattering at the interface was calculated in
Ref.~\cite{Xia}, however only far from the interface. Contrary to the
works~\cite{Butl,Schep,Xia}, where ab-initio calculations of the interface
resistance due to specular reflections at interfaces are performed, simple
free-electron model for reflections at the interfaces is used in this
work. In Chap.~\ref{chap_res_thr}, a procedure allowing one to find a
chemical potential profile everywhere in the multilayered structure in CPP
geometry, including the close vicinity of the interfaces, and take into
account both specular and diffuse scattering at the interfaces will be
developed.~\cite{ShpL_prb_00} Results of the calculation of the
resistances due to the interfaces in various systems will be presented in
Chap.~\ref{chap_res_rslt}.

\subsection{Current-induced magnetization switching} 

Chapters~\ref{chap_swt_thr} and~\ref{chap_swt_rslt} of this work are
devoted to the problem of the magnetization switching that is driven by
spin-polarized current in noncollinear magnetic multilayers.  It is known
that the transfer of spin angular momenta between current carriers and
local moments occurs near the interface of magnetic layers when their
moments are non-collinear. The specular scattering of the current at
interfaces between magnetic and nonmagnetic layers that is attendant to
ballistic transmission can create spin
torque.~\cite{Sl_jmmm_96,Sl_jmmm_99,Sl_jmmm_02,Berg_prb_96,Berg_jap_01,
Wain_prb_00,Brat_prl_00,Hern_prb_00} However, to determine the magnitude
of the transfer, one should calculate the spin transport properties far
beyond the interface regions.~\cite{ZhLF_prl_02} In this work, I consider
the effect that the diffuse scattering in the bulk of the magnetic layers
and the diffuse scattering at interfaces have on the spin-torque;  the
spin transfer that occurs at interfaces is self-consistently determined by
embedding it in the globally diffusive transport
calculations.~\cite{ShpLZh_prb_03} The ballistic component of transport
can be accommodated within the formalism described below, but this
requires the knowledge of the band structure in the layers and is outside
the scope of this study.~\cite{JZhL}

A model system to calculate the spin torque is a magnetic multilayer whose
essential elements consist of a thick magnetic layer, whose primary role
is to polarize the current, a thin magnetic layer that is to be switched,
a nonmagnetic spacer layer so that there is no interlayer exchange
coupling between the thick and thin layers, and a nonmagnetic layer or
lead on back of the thin magnetic layer; see Fig.~\ref{pic_multi}. The
transport in the multilayer is considered as a diffusive process, and the
interfaces are taken into account via the boundary conditions.  In order
to discuss the angular momentum transfer between the spin-polarized
current and the magnetic background, one has to consider the exchange
interaction between the accumulation and the background, or "sd"
interaction, described by the Hamiltonian operator $H_{int}=-J{\bf
m}\cdot{\bf M}_d$.  The term $(J/\hbar){\bf m}\times{\bf M}_d$ can be
shown to exist in the equation of motion of the spin-accumulation by
considering the quantum Boltzmann equation for the distribution function
which takes the spin of an electron into account.\cite{ZhLAnt} It
describes a deterministic or ballistic precession of the accumulation due
to the "sd" interaction when the magnetization directions of the
spin-accumulation and the local moments are not parallel. The exchange
coupling parameter $J$ is the difference between the energies of the
spin-up and spin-down electrons for each particular value of the electron
momentum, $J(k)=\epsilon_\uparrow (k)-\epsilon_\downarrow (k)$ (see
Fig.~\ref{pic_exch}), averaged over the Fermi surface. The value of $J$
can be obtained from the band structure calculations. For cobalt, I use
the value of $J=$0.3~eV.~\cite{antr_unpub} For permalloy, $J$ has been
directly measured to be about 0.1~eV.~\cite{Coop}
\begin{figure}
\centering
\includegraphics[width=4.5in]{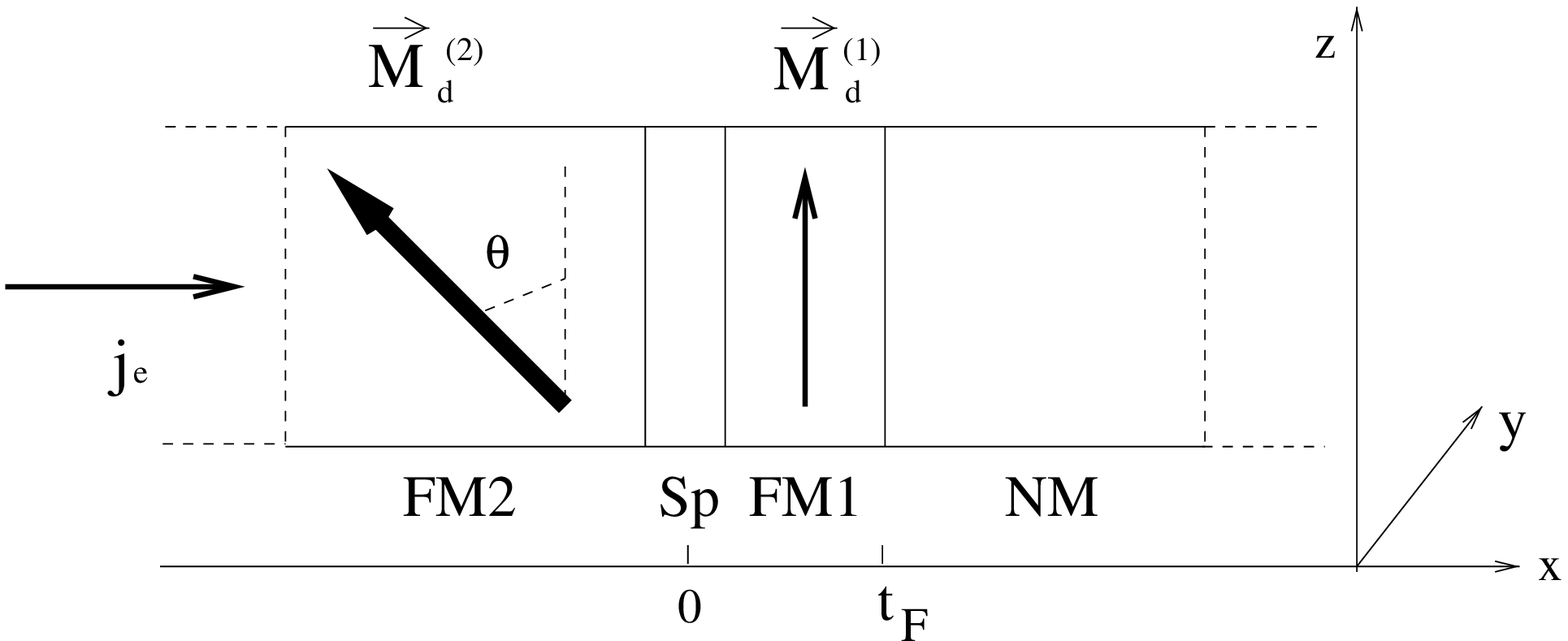}
\caption[Multilayered pillar-like
structure used for current induced reversal of a magnetic layer]
{Multilayered pillar-like structure used for current induced reversal of
a magnetic layer. FM2 is a thick ferromagnetic layer with the thickness
exceeding $\lambda^F_{sdl}$ and local magnetization ${\bf
M}_d^{(2)}=\cos\theta{\bf e}_z-\sin\theta{\bf e}_y$, Sp is a thin
nonmagnetic spacer, FM1 is a thin ferromagnetic layer with the thickness
$t_F$ and local magnetization ${\bf M}_d^{(1)}={\bf e}_z$, and NM is a
nonmagnetic back layer.}
\label{pic_multi}
\end{figure}
\begin{figure}
\centering
\includegraphics[width=3in]{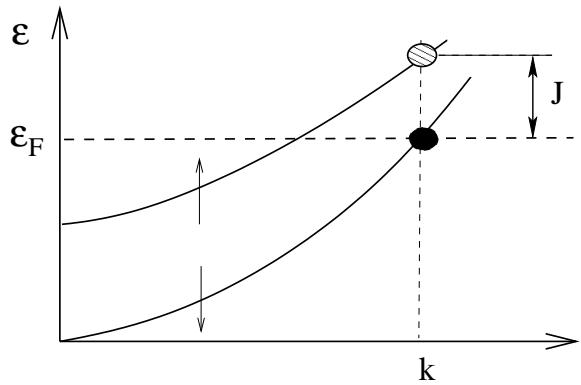}
\caption{Exchange splitting}
\label{pic_exch}
\end{figure}

The exchange interaction leads to the appearance of the component of the
spin accumulation transverse to the local magnetization direction. The
transverse spin accumulation produces two effects simultaneously: one is
to create a magnetic "effective field" acting on a local magnetization in
the layer, and the other, the "spin torque", is to increase or decrease
the angle between the magnetizations in the ferromagnetic layers.  
Contrary to the previous
treatments,~\cite{Sl_jmmm_96,Sl_jmmm_99,Sl_jmmm_02,Berg_prb_96,
Berg_jap_01} both these effects enter the equation of motion for the local
magnetization (the Landau-Lifshitz-Gilbert equation, Eq.~(\ref{LLG})) on
an equal footing.

One of the consequences of the global approach to evaluate the effect the
polarized current has on the background magnetization is that due to the
presence of the long longitudinal spin diffusion length, longitudinal and
transverse components (to the local magnetization direction) of spin
accumulations are intertwined from one layer to the next. As will be shown
in Chap.~\ref{chap_swt_rslt}, this leads to a large amplification of the
spin torque acting on the free ferromagnetic layer. The angular momentum
transferred to a thin layer {\it far exceeds} the transverse component (to
the orientation of the magnetization of the free layer) of the bare portion
of the incoming spin polarized current, i.e., the part proportional to the
electric field. This, in turn, may lead to the reduction of the critical
current necessary to switch spintronics devices (see
Sec.~\ref{sec_c-i-s}).\cite{fert_prepr_03}

The transfer of angular momentum from the polarized current has an effect
on the voltage drop across the multilayer being studied.~\cite{Sl_jmmm_96}
The experimental data on several multilayered structures has confirmed
that there are corrections to the simple $\cos^2(\theta/2)$ dependence of
the CPP resistance, where $\theta$ is the angle between the magnetization
directions in the magnetic layers.~\cite{Daug, Pratt_priv} The formalism
described in this work allows to evaluate these corrections.

In Chap.~\ref{chap_swt_thr}, I will present a spin transfer model in
which the equation of motion of the spin accumulation (essentially a
spin-diffusion equation) is solved in order to describe the effect of the
spin-polarized current on the background magnetization. Based on this
equation, a simplified system of two thick magnetic layers separated by
a non-magnetic spacer, and, in Chap.~\ref{chap_swt_rslt}, a realistic
multilayered structure depicted in Fig.~\ref{pic_multi} will be
studied.

\chapter{\label{chap_transp_mltlrs} Description of transport in 
multilayers}
In this chapter, the different formalisms for calculating electrical
transport in solids will be reviewed.  There are three most widely used
ways to describe transport - semiclassical Boltzmann equation approach,
Landauer formalism, and Kubo approach.
 
The semi-classical Boltzmann equation approach combines Newton's law
with a probabilistic description of random scattering forces. It can be
used to describe transport in the systems where the quantum interference
effects do not play a significant role, i.e., the electron mean-free
path is large compared to the lattice constant, and small relative to
the macroscopic dimensions of the system. The Boltzmann equation approach
gives conductivity in terms of classical parameters of electrons -
velocity, mean-free path, density. This formalism will be discussed in
Sec.~\ref{sec_boltz_eqn} below in detail as it is the main tool used 
to obtain results in the present work.

In the Landauer approach,~\cite{land_orig,land_repeat} current flow can
be viewed as a consequence of the injection of carriers at contacts of a
sample, and their probability of reaching the other end.  In this
approach, a current through a conductor, and conductivity, are expressed
in terms of scattering properties of the system, or in terms of the
probability that an electron can transmit through a
conductor.~\cite{land_cm,datta} The Landauer formula for the conductance
of a localized scatterer (an interface between two metallic layers, for
example) will be derived and discussed below in Sec.~\ref{land_appr}.

The Kubo formalism is a method of calculating the response of a
many-particle system to an external potential, for instance, the
electrical current in response to an electric field. The external field
is treated as a small perturbation on the equilibrium state of the
system, eliciting a linear response, whose magnitude measures the
corresponding transport coefficient, which is given in terms of the
equilibrium properties of the system, i.e., in zero field. The
electrical conductivity tensor for example, may be expressed abstractly
by the Kubo formula:\cite{ziman_princip} 
$$
\sigma_{\mu\nu}=\frac{1}{kT}\int_0^\infty <j_\mu(t)j_\nu(0)>dt. 
$$ 
The conductivity depends on the time correlation between a component of
the current operator $j_\nu(0)$ at time zero and the component $j_\mu(t)$
at some later time $t$, integrated over all time and evaluated as the
average of the expectation value of the product over the equilibrium
ensemble. The direct application of the Kubo formulas can be
cumbersome, but they provide an exact basis for various theorems
involving the transport coefficients. Fortunately, all three approaches
- Boltzmann equation, Kubo, and Landauer (for electrical conductivity) -
can be shown to be equivalent, and yield the same results for transport
coefficients in some limits.~\cite{datta}

\section{\label{sec_boltz_eqn} Boltzmann equation}
The Boltzmann approach assumes the existence of a distribution function
$f({\bf k},{\bf r})$, which is a local concentration of carriers in the
state ${\bf k}$ in the neighborhood $d^3r$ of the point ${\bf r}$ in 
space. The total number of the carriers in the small volume $d^3rd{\bf 
k}$ of the phase space is 
\begin{eqnarray}
2&\times&(\#\,\, of\,\, the\,\, allowed\,\, {\bf k}\,\, states\,\, 
per\,\, 
unit\,\, volume\,\, in\,\, the\,\, momentum\,\, space)
\nonumber
\\
&\times& d{\bf k} \times f({\bf k},{\bf r}) d^3r,
\nonumber
\end{eqnarray}
where the factor of $2$ enters because of the spin degeneracy. The number
of allowed states per unit volume $d{\bf k}$ is inverse of the volume
occupied by one state in the momentum space. As follows from the
Schr{\"{o}}dinger equation for a free electron with periodic boundary
conditions in the $L_x\times L_y\times L_z$ box, the allowed ${\bf
k}$-states are spaced within $2\pi/L_i$ in each direction, so that the 
volume in the ${\bf k}$-state per one state is $(2\pi)^3/V$. The total 
number of carriers in the phase volume $d^3rd{\bf k}$ is
\begin{equation}
\label{numb_of_carr}
dN=2f({\bf k},{\bf r})\frac{d^3r}{(2\pi)^3}d{\bf k}.
\end{equation}

The distribution function can change due to the diffusion of the
carriers, under the influence of the external fields, and due to
scattering. The Boltzmann equation states that in a steady state at any
point, and for any value of {\bf k}, the net rate of change of $f({\bf
k},{\bf r})$ is zero, i.e.
$$
\left(\frac{\partial f({\bf k},{\bf r})}{\partial t}\right)_{scatt}+
\left(\frac{\partial f({\bf k},{\bf r})}{\partial t}\right)_{field}+
\left(\frac{\partial f({\bf k},{\bf r})}{\partial t}\right)_{diff}=0.
$$
The rate of the change of the distribution function due to diffusion (the
motion of the carriers in and out of the region ${\bf r})$ can be found as
follows. Suppose that ${\bf v}_{\bf k}$ is the velocity of a carrier in
state ${\bf k}$. Then, in an interval $t$, the carriers in this state move
a distance ${\bf v}_{\bf k}t$. Since the volume occupied by particles in
phase space remains invariant (Liouville's theorem), the number of carriers
in the neighborhood of ${\bf r}$ at time $t$ is equal to the number of
carriers in the neighborhood of ${\bf r}-{\bf v}_{\bf k}t$ at time $0$:
$$
f({\bf k},{\bf r},t)=f({\bf k},{\bf r}-{\bf v}_{\bf k}t, 0).
$$
This means that the rate of the change of the distribution function due to 
diffusion is
\begin{equation}
\label{df_diff}
\left(\frac{\partial f({\bf k},{\bf r})}{\partial t}\right)_{diff}=
-{\bf v}_{\bf k}\cdot\frac{\partial f({\bf k},{\bf r})}{\partial {\bf r}}=
-{\bf v}_{\bf k}\cdot\nabla_{\bf r}f({\bf k},{\bf r}).
\end{equation}
External fields change the {\bf k}-vector of each carrier, according to 
Newton's law, at the rate
$$
\dot{\bf k}=\frac{{\bf F}}{\hbar},
$$
where ${\bf F}$ is the force acting on the carriers due to, for example, an 
electric field {\bf E}, so that ${\bf F}=e{\bf E}$. According to 
Liouville's theorem in {\bf k}-space, one can write
$$
f({\bf k},{\bf r},t)=f({\bf k}-\dot{\bf k}t,{\bf r}, 0),
$$
so that the rate of the distribution function change due to the 
external fields is
\begin{equation}
\label{df_field}
\left(\frac{\partial f({\bf k},{\bf r})}{\partial t}\right)_{field}=
-\dot{\bf k}\cdot\frac{\partial f({\bf k},{\bf r})}{\partial {\bf k}}=
-\frac{e}{\hbar}{\bf E}\cdot\nabla_{\bf k}f({\bf k},{\bf r}).
\end{equation} 
The rate of change of f({\bf k},{\bf r}) due to scattering is
\begin{equation}
\label{df_scatt}
\left(\frac{\partial f({\bf k},{\bf r})}{\partial t}\right)_{scatt}=
\int[f({\bf k}^\prime,{\bf r})(1-f({\bf k},{\bf r}))P_{{\bf k}^\prime,{\bf 
k}}
    -f({\bf k},{\bf r})(1-f({\bf k}^\prime,{\bf r}))P_{{\bf k},{\bf 
k}^\prime}]
d{\bf k}^\prime.
\end{equation}
The process of scattering from ${\bf k}$ to ${\bf k}^\prime$ decreases
$f({\bf k},{\bf r})$. The probability of this process is proportional to
the number of carriers in the initial state ${\bf k}$, $f({\bf k},{\bf
r})$, and to the number of vacancies in the final state ${\bf k}^\prime$,
$1-f({\bf k}^\prime,{\bf r})$. It is also proportional to the scattering
probability $P_{{\bf k},{\bf k}^\prime}$, which measures the rate of
transition between the states ${\bf k}$ and ${\bf k}^\prime$ if the state
${\bf k}$ is known to be occupied and the state ${\bf k}^\prime$ is known
to be empty. There is also the inverse process, the scattering from ${\bf
k}^\prime$ into ${\bf k}$, which increases $f({\bf k},{\bf r})$. It's
probability is proportional to $f({\bf k}^\prime,{\bf r})(1-f({\bf k},{\bf
r}))$. The transition rate from ${\bf k}^\prime$ into ${\bf k}$ is
$P_{{\bf k}^\prime,{\bf k}}$. The summation over all possible ${\bf 
k}^\prime$ states has to be performed.
 
Combining the equations~(\ref{df_diff}), (\ref{df_field}), 
and~(\ref{df_scatt}), one obtains the Boltzmann equation for the 
distribution function:
\begin{eqnarray}
\label{boltz_eq}
{\bf v}_{\bf k}\cdot\nabla_{\bf r}f({\bf k},{\bf r})+
\frac{e}{\hbar}{\bf E}\cdot\nabla_{\bf k}f({\bf k},{\bf r})&=&
\int[f({\bf k}^\prime,{\bf r})(1-f({\bf k},{\bf r}))P_{{\bf k}^\prime,{\bf 
k}}\\ \nonumber
    &-&f({\bf k},{\bf r})(1-f({\bf k}^\prime,{\bf r}))P_{{\bf k},{\bf 
k}^\prime}]
d{\bf k}^\prime.
\end{eqnarray}
In this form, the Boltzmann equation is a nonlinear integrodifferential
equation. In general, $P_{{\bf k},{\bf k}^\prime}$ may depend on the
distribution function $f({\bf k},{\bf r})$, and on the distribution of
the scatterers. The nonlinearity may be removed provided that the
principle of microscopic reversibility is valid, or the symmetry
$P_{{\bf k},{\bf k}^\prime}=P_{{\bf k}^\prime,{\bf k}}$ exist. This is
usually the case if the crystal and scattering potentials are real and
invariant under spatial inversion.~\cite{AshcMerm_16} If the scatterers
are sufficiently dilute and the potential describing the interaction
between a carrier and a scatterer is sufficiently weak, $P_{{\bf k},{\bf
k}^\prime}$ is independent of the distribution function $f({\bf k},{\bf
r})$. 

The Boltzmann equation may be simplified to a linear partial differential 
equation by the relaxation-time approximation. This approximation assumes 
that there exist a relaxation time $\tau$ such that an electron 
experiences a collision in an infinitesimal time interval $dt$ with 
probability $dt/\tau$. In general, the collision rate $1/\tau$ may depend 
on the position and the momentum of the electron: $\tau=\tau({\bf k},{\bf 
r})$. The additional assumptions are necessary to express the fact that 
collisions drive the electronic system toward local equilibrium. First, 
the distribution of electrons emerging from collisions at any time is 
assumed not to depend on the structure of the nonequilibrium distribution 
function $f({\bf r},{\bf k},t)$ just prior to collision. Second, if the 
electrons in a region about ${\bf r}$ have the equilibrium distribution 
function
\begin{equation}
\label{dist_FD}
f({\bf r},{\bf k},t)=f^0({\bf r},\epsilon)=
\frac{1}{e^{\frac{\epsilon({\bf k})-\mu({\bf r})}{k_BT({\bf r})}}+1},
\end{equation}
where both the chemical potential $\mu$ and the
temperature $T$ may depend on the coordinate, the collisions will not 
alter the form of the  distribution function.~\cite{AshcMerm_13} 
Assuming a uniform temperature distribution in space, the equilibrium 
distribution function takes the following form:
\begin{equation}
\label{dist_unif_T}
f({\bf r},{\bf k},t)=f^0(\epsilon)=
\frac{1}{e^{\frac{\epsilon-\epsilon_F}{k_BT}}+1},
\end{equation}
where $\epsilon_F$ is an electron Fermi energy.
 
It can be shown~\cite{AshcMerm_16} that in the relaxation-time 
approximation the collision term in the Boltzmann equation simplifies to 
$$
\left(\frac{\partial f({\bf k},{\bf r})}{\partial t}\right)_{scatt}=
-\frac{f({\bf k},{\bf r})-f^0(\epsilon)}{\tau({\bf k},{\bf r})}.
$$
This form of the scattering term reflects the fact that the role of the 
collisions is to bring the system to an equilibrium. The 
Boltzmann equation in the relaxation-time approximation takes the form
\begin{equation}
\label{boltz_rel}
{\bf v}_{\bf k}\cdot\nabla_{\bf r}f({\bf k},{\bf r})+
\frac{e}{\hbar}{\bf E}\cdot\nabla_{\bf k}f({\bf k},{\bf r})=
-\frac{f({\bf k},{\bf r})-f^0(\epsilon)}{\tau({\bf k},{\bf r})}.
\end{equation}
The relaxation-time approximation provides the same description as the 
full Boltzmann equation if it is applied to spatially homogeneous 
disturbances in an isotropic metal with isotropic elastic 
scattering.~\cite{AshcMerm_16} In this case, the relaxation time is defined as
$$
\frac{1}{\tau({\bf k})}=\int P_{{\bf k},{\bf 
k}^\prime}(1-\hat{{\bf k}}\cdot\hat{{\bf k}^\prime})d{\bf k}^\prime.
$$
This relaxation time is called the transport relaxation time.

Within the relaxation-time approximation, the Boltzmann
equation~(\ref{boltz_rel}) may be solved, and the electrical
conductivity, which is the proportionality constant between the
electrical current and the field, ${\bf  j}=\sigma{\bf  E}$, may be
found. In the homogeneous medium kept at constant temperature, the terms
proportional to the spacial gradient $\nabla$ are equal to zero, and the
distribution function $f({\bf r},{\bf k})$ turns out to be
\begin{equation}
\label{sol_df}
f({\bf k})=f^0(\epsilon({\bf k}))-\frac{\partial 
f^0(\epsilon)}{\partial\epsilon}\tau({\bf k}){\bf v}_{{\bf k}}\cdot 
e{\bf E}.
\end{equation}
This equation may be written as
$$
f({\bf k},{\bf r})=f^0(\epsilon({\bf k}))-
\frac{\partial f^0({\bf k})}{\partial\epsilon({\bf k})}
\frac{\partial\epsilon({\bf k})}{\partial{\bf k}}
\cdot\frac{e\tau({\bf k})}{\hbar}{\bf E}=f^0(\epsilon({\bf 
k}-\frac{e\tau({\bf 
k})}{\hbar}{\bf E})).
$$
Assuming that the electron relaxation time does not depend on it's
momentum, $\tau({\bf k})=\tau$, it looks as if the whole Fermi
surface had been shifted by the amount $(e\tau/\hbar){\bf E}$ in ${\bf
k}$-space (Fig~\ref{fermi_surf}). But, in fact, only the electrons close
to the surface of the Fermi sphere have moved, the ones near the bottom
of the conductivity band, deep within the Fermi sphere, are not really
affected by the field. One can say that only the electrons with energies
close to the Fermi energy, and, hence, the velocities close to the Fermi
velocity participate in transport. The conductivity of a metal depends
only on the properties of the electrons at the Fermi level, not on the
properties of all electrons in the metal.
\begin{figure}
\centering
\includegraphics[width=2in]{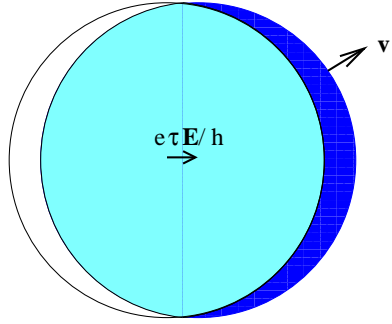}
\caption{Displacement of the Fermi surface in the momentum space under 
the influence of the electric field}
\label{fermi_surf}
\end{figure}

The electrical current density $j$ is defined as
\begin{equation}
\label{el_curr}
{\bf  j}=e\int{\bf v}f({\bf v},{\bf r})d{\bf v},
\end{equation}
where $f({\bf v},{\bf r})$ is the density of the carriers in the $({\bf 
r},{\bf v})$ space, so that the number of electrons in the volume 
$d^3rd{\bf v}$ is 
$$
dN=f({\bf v},{\bf r})d^3rd{\bf v}.
$$
Comparing this expression with the equation~(\ref{numb_of_carr}), the
following relationship between $f({\bf v},{\bf r})$ and $f({\bf k},{\bf
r})$ may be obtained:
\begin{equation}
\label{fk_fv}
f({\bf v},{\bf r})=\frac{1}{4\pi^3}
\left(\frac{m}{\hbar}\right)^3f({\bf k},{\bf r}),
\end{equation}
where $m$ is the electron mass, and the relation $m{\bf v}=\hbar{\bf k}$ 
is used. 
Considering the case where both ${\bf  j}$ and ${\bf  E}$ are in the 
$x$-direction, the relaxation time $\tau$ is independent of the momentum, 
and substituting $f({\bf k},{\bf r})$ from the equation~(\ref{sol_df}), 
one obtains the following expression for the conductivity: 
\begin{equation}
\label{conduct}
\sigma=\frac{1}{4\pi^3}\frac{e^2\tau}{3}\left(\frac{m}{\hbar}\right)^3
\int\left(-\frac{\partial f(\epsilon)}{\partial\epsilon}\right)v^2d^3v,
\end{equation}
where the fact that $\int{{\bf v}}_{\bf k}f^0(\epsilon({\bf k}))d{\bf 
v}=0$ is
used, since $f^0(\epsilon({\bf k}))$ is isotropic in ${\bf k}$. The 
factor $1/3$
appears in the expression since $v_x^2$ has been averaged over the Fermi
surface, so that $v_x^2=\frac{1}{3}v^2$. In a metal for the temperature
close to 0$^\circ$K, the function $(-\partial
f^0(\epsilon)/\partial\epsilon)$ behaves like a delta-function at the
Fermi level, $\delta(\epsilon-\epsilon_F)$, and the expression for the
conductivity~(\ref{conduct}) reduces to
\begin{equation}
\label{conduct_2}
\sigma=\frac{k_F^3}{3\pi^2}\frac{e^2\tau}{m}=\frac{ne^2\tau}{m},
\end{equation}
where $n=k_F^3/3\pi^2$ is the electron density, and $m$ is the electron
mass. Strictly speaking, the above expression for the electrical
conductivity is only valid for free electrons. The free-electron 
description of a metal will be used throughout this work, so the
equation~(\ref{conduct_2}) will be used to calculate the conductivity.

An alternative form of the scattering term in the Boltzmann equation may
be obtained as follows. Assuming the rate of transition between the
states ${\bf k}$ and ${\bf k}^\prime$ to be independent of ${\bf k}$ and
${\bf k}^\prime$, $P_{{\bf k},{\bf k}^\prime}=P_0$, the scattering term
Eq.~(\ref{df_scatt}) may be written as
$$
\left(\frac{\partial f}{\partial t}\right)_{scatt}=
-\frac{f({\bf r},{\bf k})-\bar f({\bf r})}{\tau},
$$
where
$$
\frac{1}{\tau}=\int P_0 d{\bf k},
$$
\begin{equation}
\label{f_bar}
\bar f({\bf r})=\frac{1}{4\pi}\int_{FS} f({\bf r},{\bf k})d{\bf k}.
\end{equation}
The linearized Boltzmann equation then takes the form
\begin{equation}
\label{boltzm_fbar}
ev_{\bf k}\cdot{\bf E}\frac{\partial f^0}{\partial\epsilon}
+v_{\bf k}\cdot\nabla f({\bf r},{\bf k})=
-\frac{f({\bf r},{\bf k})-\bar f({\bf r})}{\tau}.
\end{equation}
Introducing $g({\bf r},{\bf k})$ such that  
$f({\bf r},{\bf k})=f^0(\epsilon({\bf k}))-\frac{\partial 
f^0}{\partial\epsilon} g({\bf r},{\bf k})$, 
the function $\bar f({\bf r})$ may be written as
\begin{equation}
\label{fbar_gbar}
\bar f({\bf r})=f^0(\epsilon({\bf k}))-\frac{\partial
f^0}{\partial\epsilon}\frac{1}{4\pi}\int_{FS}g({\bf r},{\bf 
k})d{\bf k},
\end{equation}
and the equation for the function $g({\bf r},{\bf k})$ takes the form
\begin{equation}
\label{boltzm_g}
v_{\bf k}\cdot\nabla g({\bf r},{\bf k})
+\frac{g({\bf r},{\bf k})}{\tau}=ev_{\bf k}\cdot{\bf E}
+\frac{\mu(\bf r)}{\tau},
\end{equation}
where $\mu(\bf r)$ is defined as
\begin{equation}
\label{mu_g}
\mu({\bf r})=\frac{1}{4\pi}\int_{FS} g({\bf r},{\bf k})d{\bf k},
\end{equation}  
so that $g({\bf r},{\bf k})$ may be written as 
$$
g({\bf r},{\bf k})=\mu({\bf r})+\tilde g({\bf r},{\bf k}),
$$
where
\begin{equation}
\label{zero_int_g}
\int_{FS} \tilde g({\bf r},{\bf k})d{\bf k}=0.
\end{equation}
The function $\mu({\bf r})$ defined in Eq.~(\ref{mu_g}) is the 
non-equilibrium part of the chemical potential, since the distribution 
function $f({\bf r},{\bf k})$ may be written as
\begin{equation}
\label{full_f}
f({\bf r},{\bf k})=f^0(\epsilon({\bf k}))-\frac{\partial 
f^0}{\partial\epsilon}
\mu({\bf r},{\bf k})
-\frac{\partial f^0}{\partial\epsilon}\tilde g({\bf r},{\bf k})\approx
\frac{1}{e^{\frac{\epsilon({\bf k})-(\epsilon_F+\mu({\bf r}))}{k_BT({\bf 
r})}}+1}
-\frac{\partial f^0}{\partial\epsilon}\tilde g({\bf r},{\bf k})
\end{equation}
Equations~(\ref{boltzm_g}) and~(\ref{mu_g}) are the starting point of  
Chap.~\ref{chap_res_thr}, where the equations for the chemical 
potential profile in the multilayered metallic structures are derived.

To conclude the discussion of the Boltzmann approach to the electrical 
transport in solids, the diffusion equation on the concentration of 
carriers will be derived below. The {\it non-stationary} linearized 
Boltzmann equation takes the following form:
\begin{equation}
\label{boltz_nonstat}
\frac{\partial f(x,{\bf k},t)}{\partial t}
+v_x\frac{\partial f(x,{\bf k},t)}{\partial x}
-e\frac{\partial f^0(x,\epsilon)}{\partial\epsilon}E_xv_x=
-\frac{f(x,{\bf k},t)-{\bar f}(x)}{\tau}, 
\end{equation}
where, for simplicity, electrical field is assumed to have only one 
component, $E_x$, and spacial gradient in $x$-direction only is 
considered. Integrating Eq.~(\ref{boltz_nonstat}) over all ${\bf 
v}$-space, and using the definitions of the concentration and current 
density
\begin{equation}
\label{concentr}
n({\bf r},t)=\int f({\bf r},{\bf v},t)d{\bf v},
\end{equation}
and
\begin{equation}
\label{elcurr}
{\bf j}=e\int {\bf v}f({\bf r},{\bf v},t)d{\bf v},
\end{equation}
where $m$ is the electron mass, the continuity equation
\begin{equation}
\label{cont}
\frac{\partial n(x,t)}{\partial t}+
\frac{1}{e}\frac{\partial j_x(x,t)}{\partial x}=0
\end{equation}
may be obtained. The term proportional to $\int v_x(\partial 
{\bar f}/\partial\epsilon)d{\bf v}$ can be shown to be zero, and the term  
proportional to $\int(f-{\bar f})d{\bf v}$ is zero due to the 
condition~(\ref{zero_int_g}). Multiplying the Boltzmann 
equation~(\ref{boltz_nonstat}) by $v_x$ and integrating over all ${\bf
v}$-space again, the following equation is obtained:
\begin{eqnarray}
\label{first_mom}
\frac{1}{e}\frac{\partial j_x(x,t)}{\partial t}+
\int v_x^2\frac{\partial f(x,{\bf v},t)}{\partial x}d{\bf v}&-&
\frac{1}{4\pi^3}\left(\frac{m}{\hbar}\right)^3eE_x
\int v_x^2\frac{\partial f^0(x,\epsilon)}{\partial\epsilon}d{\bf v} 
\nonumber \\
&=&
-\frac{1}{e}\frac{j_x(x,t)}{\tau}.
\end{eqnarray}
The term proportional to $\int v_xf^0d{\bf v}$ can be shown to be zero. 
Assuming the electrical current to be independent of time, and 
writing the second term, 
$\int v_x^2({\partial f}/{\partial x})d{\bf v}$, as 
$<v_x^2>\frac{\partial}{\partial x}\int fd{\bf 
v}=\frac{1}{3}v_F^2\frac{\partial}{\partial x}\int fd{\bf v}$, the 
following expression for the current is obtained:
\begin{equation}
\label{curr_lin_resp}
j_x(x)=\sigma E_x-eD\frac{\partial n(x)}{\partial x},
\end{equation}
where $D=\frac{1}{3}v_F^2\tau$ is the diffusion constant, 
$\sigma=\frac{k_F^3}{3\pi^2}\frac{e^2\tau}{m}$ is the electrical  
conductivity. Equation~(\ref{curr_lin_resp}) states that within a linear 
response model, the current is proportional to the electrical field and 
the gradient of the carriers concentration with the negative sign. 
Substituting the equation~(\ref{curr_lin_resp}) into the continuity 
equation~(\ref{cont}), the diffusion equation is obtained:
\begin{equation}
\label{diffusion}
\frac{\partial n(x,t)}{\partial t}=D\frac{\partial^2 n(x,t)}{\partial 
x^2}.
\end{equation} 
Equations~(\ref{cont}) and~(\ref{curr_lin_resp}), generalized to the
spinor form, are the starting point of Chap.~\ref{chap_swt_rslt},
where the equations for the spin-accumulation are obtained.


\section{\label{land_appr} Landauer approach - interface resistance.} 
The conductance $G$ due to elastic scattering of an obstacle, characterized 
by transmission and reflection coefficients $T$ and $R$, is given by   
\begin{equation}
\label{land_form} 
G=\frac{e^2}{\pi\hbar}\frac{T}{R}, 
\end{equation}
where $e$ is electron charge, and $\hbar$ is Plank's constant. At zero
temperature $T$ and $R$ are evaluated at the Fermi energy. Equation
~(\ref{land_form}) applies to samples of arbitrary shape and structural
complexity, and may be used, for example, to calculate the conductance
of a planar barrier, such as an interface between two metals. Below,
equation ~(\ref{land_form}) will be derived within an approximation of a
single conductivity channel,~\cite{datta,but_prb,land_cm,imry_rmp} which 
makes a system effectively one-dimensional. Expression (\ref{land_form}) 
can be generalized for the case of many independent conducting channels, 
and for nonzero temperature,~\cite{but_prb, datta} but these 
generalizations won't be covered here.

Consider a barrier connected through ideal 1D wires to some external
source (a pair of reservoirs with different chemical potentials $\mu_1$
and $\mu_2$) which drives a current $I$ through the system. The barrier
is characterized by its transition coefficient $T$ and reflection
coefficient $R=1-T$. (Fig.~\ref{land_barr})
\begin{figure}
\includegraphics[width=\textwidth]{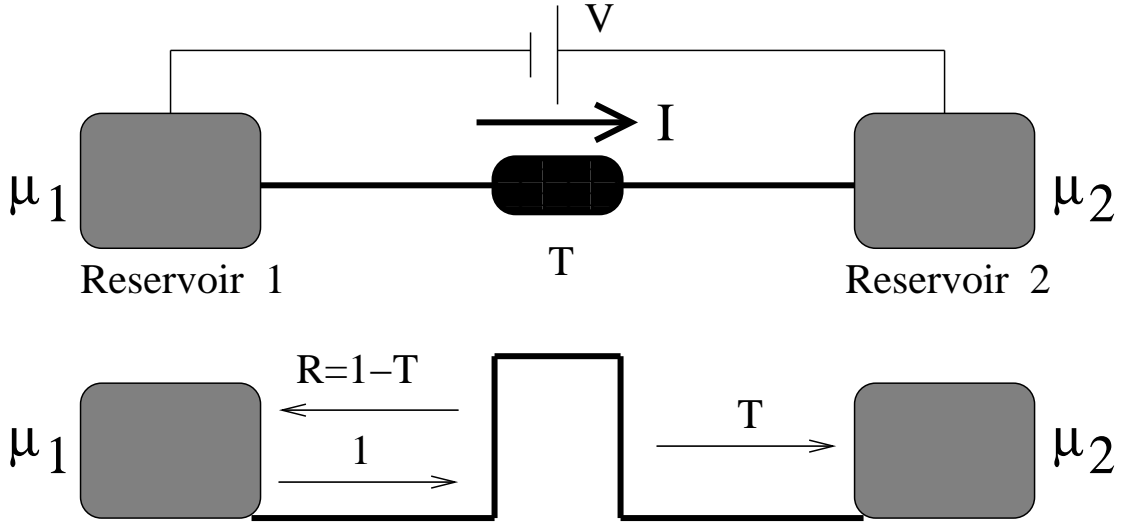}
\caption[One-dimensional barrier with the transition coefficient $T$ and
reflection coefficient $R=1-T$, connected to an external source]
{One-dimensional barrier with the transition coefficient $T$ and 
reflection coefficient $R=1-T$, connected to an external source (a 
pair of reservoirs with different chemical potentials $\mu_1$ and 
$\mu_2$.)}
\label{land_barr}
\end{figure}

Consider first two reservoirs without a barrier. Left reservoir emits
electrons with energies $E$ up to a quasi-Fermi energy (chemical potential)
$\mu_1$, Right reservoir emits electrons with energies $E$ up to the
chemical potential $\mu_2$. Energies are measured relative to equilibrium
chemical potential (common for both reservoirs). Current emitted by the
left reservoir and going to the right is
$$I^+=-evn(E-\mu_1)\approx -evn_0(E)+ev\frac{\partial n}{\partial 
E}\mu_1,$$ 
where $n(E)$ is electron density per unit length. $v$ is the velocity
component along the wire at the Fermi surface. Current emitted by the right
reservoir to the left is
$$I^-=evn(E-\mu_2)\approx evn_0(E)-ev\frac{\partial n}{\partial E}\mu_2.$$ 
Then, total current is given by the sum of $I^+$ and $I^-$ and is equal to
$$I=ev\frac{\partial n}{\partial E}(\mu_1-\mu_2).$$ 
$\partial n/\partial E$ is the density of states for two spin directions
and for carriers with positive velocity. In one-dimensional case, $\partial
n/\partial E=1/\pi\hbar v$~\cite{AshcMerm} and
$$I=\frac{e}{\pi\hbar}(\mu_1-\mu_2).$$ 
This is the total current emitted by the left reservoir due to the
difference in the quasi-Fermi levels. Carriers have a probability $T$ for
traversal of the sample and a probability $R$ of being reflected back.
Assuming that $T$ and $R$ don't depend on the electron energy in the energy
range $\mu_1 > E > \mu_2$, the net current flow may be written as
$$I=\frac{e}{\pi\hbar}T(\mu_1-\mu_2).$$ 
The difference in chemical potentials of the two reservoirs $\mu_1-\mu_2$
can be identified with the drop of electron potential $eV$ across the
source. Conductance, given by $G=I/V$ is
$$G=\frac{I}{(\mu_1-\mu_2)/e}=\frac{e^2}{\pi\hbar}T,$$ 
and resistance of the system is 
\begin{equation} 
\label{barr_cont}
G^{-1}=\frac{\pi\hbar}{e^2}\frac{1}{T}.  
\end{equation} 
Expression (\ref{barr_cont}) gives a non-zero resistance in the case of
completely transparent barrier, $T=1$. The resistance in a system
consisting only of the ideal wires and reservoirs is a finite quantity
\begin{equation}
\label{cont_res} G_C^{-1}=\frac{\pi\hbar}{e^2}=12.9 \,\,\, k\Omega.
\end{equation} 
This quantity is called the contact resistance. It occurs between a large
reservoir with the infinite number of states that an electron can have,
and a narrow wire with only one conductivity channel in the same way as
traffic jam occurs when 4-lane highway transforms into 2-lane street. 
If a larger number $M$ of the conductivity channels is allowed to exist 
in a wire (when the wire is three-dimensional but still very thin, so 
that the perpendicular motion of an electron is quantized), the contact 
resistance can be shown to have the form~\cite{datta}    
$$G_c^{-1}=\frac{\pi\hbar}{e^2}\frac{1}{M}.$$ 
The wider the wire, the larger the number of transverse modes $M$ can
propagate in it and the smaller the contact resistance is, and it is 
negligible for thick contacts. 

The resistance of the barrier alone can be calculated by subtracting the
contact resistance (equation~(\ref{cont_res})) from the total resistance
of the "barrier-contacts-reservoirs" system
(equation~(\ref{barr_cont})), and is equal to
\begin{equation}
\label{barr_res}
G^{-1}=\frac{\pi\hbar}{e^2}\frac{R}{T};
\end{equation}
this is equivalent to the expression~(\ref{land_barr}) for the 
conductivity when one uses $1-T=R$.  

Resistance is usually associated with energy loss. In the considered model,
scattering both at the barrier and in the wires is perfectly elastic
($R+T=1$), so that there seems to be no energy loss in this system.
Nevertheless, resistance exists. The resistance of the barrier can't be
considered without connecting the barrier to the reservoirs. Electrons
reflected from the barrier go back to the reservoir. Instead of reissuing
the electron that entered the reservoir, a new one comes out, so that
information about the momentum of the electron is lost in a large
reservoir, the outcoming electron does not have the same direction of the
momentum as the incoming one had. This loss of information is the origin of
the resistance of the barrier.~\cite{Don_Sond} The existence of resistance
in the Landauer approach requires the presence of the reservoirs, but its
magnitude depends only on the elastic events at the barrier.

\chapter{\label{chap_res_thr} Interface resistance in multilayers - 
theory} 

The Landauer formula~(\ref{barr_cont}) is derived for an obstacle connected
to the source via ideal scattering-free wires.  The purpose of this work,
though, is to find a resistance due to interfaces in a realistic
multilayered structure consisting of several diffusive metallic layers and
interfaces between them. As discussed in Chap.~\ref{chap_intro},
scattering at the interfaces is affected by the presence of the diffuse
scattering in the bulk of the layers, and contributions from the bulk of
the layers and the interfaces to the total resistance of the structure
can't simply be added. In this section, the procedure allowing one to find
a chemical potential profile and resistance due to the interfaces in a
multilayered metallic structure in CPP geometry that takes into account an
interaction of the scattering at the interfaces and scattering in the bulk
of that layers will be developed. Transport in the bulk of the layers will
be described by semiclassical Boltzmann equation, and interfaces will be
taken into account via boundary conditions. Both specular and diffuse
scattering at the interfaces will be considered. While Landauer's result
for the interface resistance won't be explicitly used in this work, only
scattering properties of the interfaces, their reflection and transmission
coefficients will be used, which is a signature of the Landauer approach.
 
First, a general solution of the semiclassical linearized Boltzmann
equation in CPP geometry is presented. Next, boundary conditions at an
interface between two layers in the presence of both specular and
diffuse scattering at the interface will be discussed.  Then, a system
of integral equations describing a chemical potential profile in two
semi-infinite metallic layers will be derived.~\cite{ShpL_prb_00}
Integral equations on the chemical potentials in the systems consisting
of three and five layers are presented in Appendices~\ref{app_three_lrs}
and~\ref{app_five_lrs}. This chapter is completed with determining the
special form of diffuse scattering such that resistances measured at the
interface and far from it are the same for a system consisting of two
semi-infinite layers of identical metals and an interface between them.
The forms of diffuse scattering for which a prediction can be made as to
whether the potential drop at the interface is bigger or smaller than
that measured far from it will be indicated.

\section{General solution of the Boltzmann equation}

In the relaxation-time approximation the linearized Boltzmann equation
in CPP geometry takes the form (see Eq.~(\ref{boltzm_g}) and $\mu (z)$ 
denoting the non-equilibrium part of the chemical 
potential):
\begin{equation}
\frac{\partial g(\theta,z)}{\partial z}+\frac{g(\theta,z)}{v_{z}\tau }%
=eE_{z}+\frac{\mu (z)}{v_{z}\tau },
\label{eq_on_g}
\end{equation}
where $e$ is an electron charge, $\tau$ is an electron relaxation rate,
$v_z$ and $E_z$ are an electron velocity and an electric field in the
direction $z$ perpendicular to the plane of the layers, and $\theta$ is
the angle that the electron momentum ${\bf k}$ makes with $z$ axis which
takes values from $0$ to $\pi$. Chemical potential $\mu (z)$ is the
average of $g(\theta,z)$ in momentum space (see
Eq.~(\ref{mu_g})):
\begin{equation}
\mu (z)=\frac{1}{2}\int_{0}^{\pi }g(\theta, z)\sin \theta d\theta . 
\label{def_of_chem_pot}
\end{equation}

Equation~(\ref{eq_on_g}) is an ordinary first-order differential equation. 
Since the chemical potential $\mu(z)$ enters the right hand side of this 
equation, its solution $g(\theta,z)$ in the region between $z_1$ and 
$z_2$ of $z$ axis will depend on $\mu(z)$ via an 
integral:
\begin{equation}
\label{sol_g}
g(\theta,z)=\int_{z_1}^{z_2} dz^\prime k(\theta,z,z^\prime) \mu(z^\prime) 
            + y(\theta, z),
\end{equation}
where $ k(\theta,z,z^\prime)$ and $y(\theta, z)$ are functions defined by
the geometry of the problem and the boundary conditions. By using the
definition of the chemical potential~(\ref{def_of_chem_pot}), an equation
for $\mu(z)$ can be obtained by integrating equation~(\ref{sol_g}) over
$\theta$:

\begin{equation}
\label{sol_mu}
\mu(z)=\int_{z_1}^{z_2} dz^\prime K(z,z^\prime) \mu(z^\prime) + Y(z),
\end{equation}
where $K(z,z^\prime)=
\frac{1}{2}\int_0^\pi k(\theta,z,z^\prime)\sin\theta d\theta$,
$Y(z)=\frac{1}{2}\int_0^\pi y(\theta, z)\sin\theta d\theta$.\cite{Laik_Lur}

In the structure consisting of several layers (N), boundary conditions
connect distribution function in a layer ($L_i$) with the distribution
functions in the adjacent layers ($L_j$).  Equation~(\ref{sol_mu}) can then
be generalized to describe the chemical potential everywhere in the system:
\begin{equation}
\label{sol_mu_many}
\mu_i(z)=\sum_{j=1}^N\int_{L_j}dz^\prime K_{ij}(z,z^\prime) 
\mu_j(z^\prime) + Y_i(z).
\end{equation}

The integral equation~(\ref{sol_mu_many}) is a system of Fredholm
equations of the second kind, which can be numerically solved to obtain
the chemical potential profile everywhere in the layered
structure.~\cite{ShpL_prb_00,ShpLZh_mrs_01} The functions
$K_{ij}(z,z^\prime)$ and $Y_i(z)$ are defined by the boundary conditions
at the interfaces. Numerical procedure of solving the Fredholm equation
of the second kind is presented in the Appendix~\ref{app_numerics}.
  
\section{Boundary conditions at the interface between two layers}

At the interface between two layers, an incoming electron is scattered via
reflection and transmission. There is also diffuse scattering in all
directions. In the present work, a simple model for the diffuse scattering
will be used, where a single parameter $S$ represents an amount of
electrons which are {\it not} scattered diffusely, so that the specularly
reflected and transmitted electron fluxes are scaled by $S$.  $1-S$ is the
amount of electrons scattered diffusely. Figure~\ref{scat_bound}a)  
represents schematically how an electron is scattered at an interface.
\begin{figure}
\includegraphics[width=\textwidth]{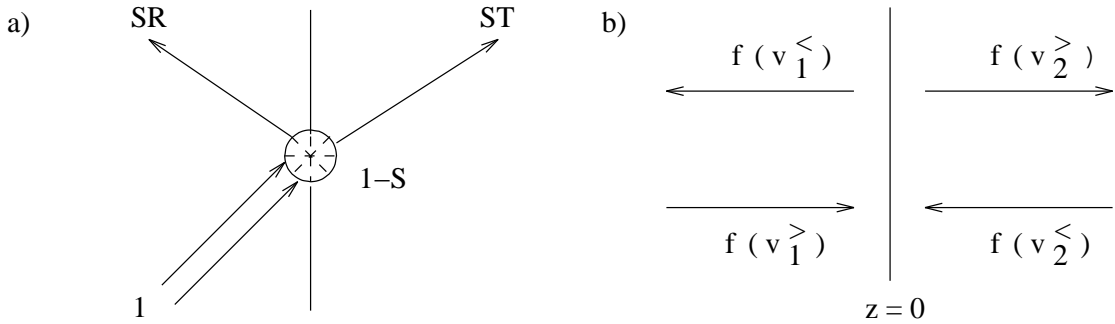}
\caption[Scattering at an interface]{Scattering at an interface: a) 
schematic representation of electron scattering, b) direction of 
electron fluxes at an interface.}
\label{scat_bound}
\end{figure}

The boundary conditions for the distribution function $f$ at the
interface $z=0$ between two semi-infinite layers in the presence of
diffuse and specular scattering take the following
form:~\cite{Zh_L,Hood_Fal}
\begin{equation}
\label{bound_cond_f}
\left\{
\begin{array}{ccl}
f(v_{2}^{>},0)&=&S(v_{1},v_{2})R_{21}(v_{1},v_{2})f(v_{2}^{<},0)
+S(v_{1},v_{2})T_{12}(v_{1},v_{2})f(v_{1}^{>},0)\\
& & +[1-S(v_{1},v_{2})]F(f),
\\ 
\\
f(v_{1}^{<},0)&=&S(v_{1},v_{2})R_{12}(v_{1},v_{2})f(v_{1}^{>},0)
+S(v_{1},v_{2})T_{21}(v_{1},v_{2})f(v_{2}^{<},0)\\
& & +[1-S(v_{1},v_{2})]F(f),
\end{array}
\right.
\end{equation}
where 
\begin{equation}
F(f)=\frac{1}{\Omega }\int_{FS} d{\bf k}
[|v_{2z}^{<}|f(v_{2}^{<},0)+|v_{1z}^{>}|f(v_{1}^{>},0)],
\label{diff_scat_term_f}
\end{equation}
$$
\Omega =\int_{FS} d{\bf k}\delta (\epsilon -\epsilon _{F})|v_{z}|,
$$
$\epsilon _{F}$ is the electron Fermi energy, $v_{1}$ $(v_{2})$ is the
electron velocity in the layer 1(2), and $v^{>}$ $(v^{<})$ refer to
$v_{z}>0$ $(v_{z}<0)$.  $R_{ij}(v_{1},v_{2})$ ($T_{ij}(v_{1},v_{2})$) is
the reflection (transmission) coefficient at an interface for an
electron travelling from the layer $i$ to the layer $j$,and
$1-S(v_{1},v_{2})$ describes the diffuse scattering. As shown at the
Fig.~\ref{scat_bound}b), an outgoing particle flux $f(v_{2}^{>})$
($f(v_{1}^{<})$) comes from the specularly reflected particle flux
(first term), the specularly transmitted particle flux (second term),
and the particle flux from all directions diffusely scattered in the
direction of $v_{2}^{>}$ ($v_{1}^{<}$) (last term). 

Similar boundary conditions, but without the last term, were
investigated by Hood and Falicov for currents in plane of the layers
geometry.~\cite{Hood_Fal} In CPP geometry, current conservation requires
the presence of the term responsible for the diffuse scattering at an
interface in the boundary conditions. The proof of the fact that the
boundary conditions~(\ref{bound_cond_f})-(\ref{diff_scat_term_f}) indeed
conserve the current across an interface is presented in
Appendix~\ref{app_curr_cons}.

At the interface between two metals, reflection and transmission
coefficients $R_{ij}$ and $T_{ij}$ have to be found from quantum mechanical
considerations, as reflection and transmission coefficients for the quantum
mechanical flux. In general, $R_{ij}$, $T_{ij}$ and $S$ depend on both
variables $v_{1}$ and $v_{2}$. In the quasi-one-dimensional layered
structures, where the component of electron velocity parallel to an
interface in conserved, only $z$ component of velocity enters reflection
and transmission coefficients. Furthermore, in the case of specular
scattering, angle of reflection and angle of transmission are uniquely
defined by the angle of incidence, so that $R_{ij}$ and $T_{ij}$ are the
functions of the cosine of an angle that an electron velocity makes with an
interface only, $R_{ij}(\cos\theta_i)$ and $T_{ij}(\cos\theta_i)$, where
$\theta_i$ is the angle of incidence of an electron at the interface in
$i$-th layer. Specular scattering at an interface between two metals is
considered in detail in Appendix~\ref{app_refl_transm}. In order to
simplify the discussion of the effects of the diffuse scattering at an
interface on the electron transport in multilayers, diffuse scattering
parameter $S$ is assumed to be constant, independent of the electron
direction, with one exception, in Sec.~\ref{sec_spec_form_of_S},
where the special forms of the diffuse scattering coefficient for which a
shape of the chemical potential profile can be predicted analytically are
considered. With these simplifications, the boundary conditions on the 
function $g$ take the following form:
\begin{equation}
\label{bound_cond}
\left\{
\begin{array}{c}
g(v_{2}^{>},0)=SR_{21}(\theta_2)g(v_{2}^{<},0)+ST_{12}(\theta_1)g(v_{1}^{>},0)
+(1-S)F(g),
\\
\\
g(v_{1}^{<},0)=SR_{12}(\theta_1)g(v_{1}^{>},0)
+ST_{21}(\theta_2)g(v_{2}^{<},0)+(1-S)F(g),
\end{array}
\right.
\end{equation}
where $F(g)$ can be shown (Appendix~\ref{duff_scat_term}) to have the 
form:
\begin{eqnarray}
\label{diff_scat_term_g}
F(g)&=&\frac{2v^2_{F1}}{v^2_{F2}+v^2_{F1}}\int_0^{\pi/2} 
g(v_{1}^{>},0)\cos\theta\sin\theta d\theta \\
& + & \frac{2v^2_{F2}}{v^2_{F2}+v^2_{F1}}\int_0^{\pi/2}
g(v_{2}^{<},0)\cos\theta\sin\theta d\theta, \nonumber
\end{eqnarray}
where $v_{F1}$ and $v_{F2}$ are the electron Fermi velocities in the 
first and the second layer correspondingly.

\section{\label{eq_mu_mult}Equations for chemical potential profile in a 
multilayer}
\begin{figure}
\centering
\includegraphics[width=4in]{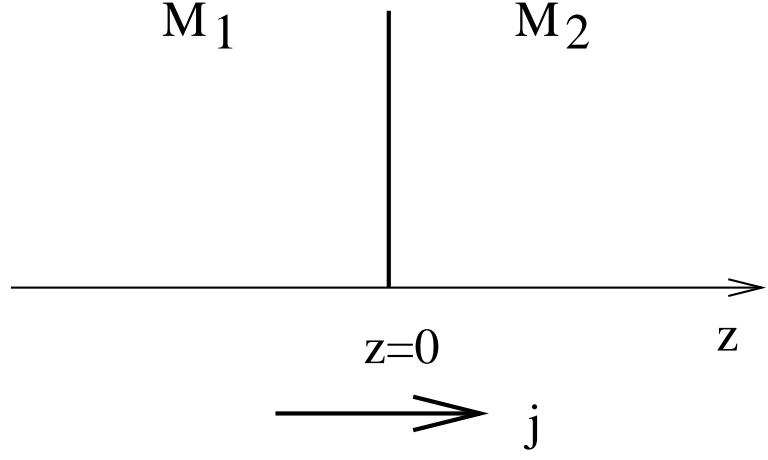}
\caption[System consisting of two layers]{System consisting of two 
layers.}
\label{two_layers}
\end{figure}

In this section, solution of the Boltzmann equation in the system
consisting of two semi-infinite metallic layers and interface between them
(Fig.~\ref{two_layers}) is presented, and equations on the chemical
potential profile are derived.  For simplicity, the same electron
relaxation time $\tau$ is assumed in both layers. Solution of the Boltzmann
equation~(\ref{eq_on_g}) for corrections to the distribution function of
the electrons moving to the right $g(v_1^>,z)$ and to the left $g(v_1^<,z)$
in the first layer, and electrons moving to the right $g(v_2^>,z)$ and to
the left $g(v_2^<,z)$ in the second layer take the following form:
\begin{equation}
\label{g_2lay}
\left\{
\begin{array}{ccl}
g(v_1^>,z)&=&\int^z_{-\infty}  
dz^\prime\exp{\left(-\frac{z-z^\prime}{|v_{z1}|\tau}\right)}
\left(eE_{z1}+\frac{\mu_1(z^\prime)}{|v_{z1}|\tau}\right)\\ \\
g(v_1^<,z)&=&A\exp{\left(\frac{z}{|v_{z1}|\tau}\right)}
+\int^z_0dz^\prime\exp{\left(\frac{z-z^\prime}{|v_{z1}|\tau}\right)}
\left(eE_{z1}-\frac{\mu_1(z^\prime)}{|v_{z1}|\tau}\right)\\ \\
g(v_2^>,z)&=&B\exp{\left(-\frac{z}{|v_{z2}|\tau}\right)}
+\int^z_0dz^\prime\exp{\left(-\frac{z-z^\prime}{|v_{z2}|\tau}\right)}
\left(eE_{z2}+\frac{\mu_2(z^\prime)}{|v_{z2}|\tau}\right)\\ \\        
g(v_2^<,z)&=&-\int_z^{\infty}
dz^\prime\exp{\left(\frac{z-z^\prime}{|v_{z2}|\tau}\right)}
\left(eE_{z2}-\frac{\mu_2(z^\prime)}{|v_{z2}|\tau}\right),
\end{array}
\right.
\end{equation}
where $\mu_1(z)$ ($\mu_2(z)$) is the chemical potential in the left (right)
layer, $E_{z1}$ ($E_{z2}$) is an electric field in the left (right) layer,
and $v_{z1}$ ($v_{z2}$) is a component of an electron velocity in the
direction $z$ in the left (right) layer. Since only the $z$ component of an
electric field is present in CPP geometry, the subscript $z$ will be
omitted from $E_{z1}$ and $E_{z2}$ everywhere below.  $A$, $B$ are
constants of integration, independent of $z$. They are defined by the
boundary conditions~(\ref{bound_cond}), and equal to
\begin{eqnarray}  
\label{const_A}
A&=&SR_{12}(\theta_1)eE_{1}l_1|\cos\theta_1|-
              ST_{21}(\theta_2)eE_{2}l_2|\cos\theta_2| \nonumber \\
&+&SR_{12}(\theta_1)\int^0_{-\infty}
\exp{\left(\frac{\xi^\prime}{\cos\theta_1}\right)}
\frac{\mu_1(\xi^\prime)}{\cos\theta_1}d\xi^\prime\\
&+&ST_{21}(\theta_2)\int_0^{\infty}
\exp{\left(\frac{-\xi^\prime}{\cos\theta_1}\right)}
\frac{\mu_2(\xi^\prime)}{\cos\theta_2}d\xi^\prime + (1-S)F(g)   \nonumber  
\end{eqnarray} 
\begin{eqnarray}  
\label{const_B}
B&=&-SR_{21}(\theta_2)eE_{2}l_2|\cos\theta_2|+
              ST_{12}(\theta_1)eE_{1}l_1|\cos\theta_1| \nonumber \\
&+&SR_{21}(\theta_2)\int_0^{\infty}
\exp{\left(\frac{-\xi^\prime}{\cos\theta_2}\right)}
\frac{\mu_2(\xi^\prime)}{\cos\theta_2}d\xi^\prime\\   
&+&ST_{12}(\theta_1)\int^0_{-\infty}
\exp{\left(\frac{\xi^\prime}{\cos\theta_2}\right)}
\frac{\mu_1(\xi^\prime)}{\cos\theta_1}d\xi^\prime + (1-S)F(g), \nonumber
\end{eqnarray}
where $l_i=v_{Fi}\tau$ is the electron mean free path in $i$-th layer,
$\theta_i$ is an angle that an electron velocity makes with $z$ axis in
$i$-th layer, and a dimensionless variable $\xi=z/l_i$, where $i=1$ or
$2$ depending on in which layer the chemical potential is considered, is
introduced. Since the boundary conditions connect the distribution
functions at both sides of the interface, both angles $\theta_1$ and
$\theta_2$ enter the expressions for $A$ and $B$. From the definition of
the chemical potential~(\ref{def_of_chem_pot}) it follows (see
Appendix~\ref{ang_av}) that
\begin{equation}
\label{def_mu}
\left\{
\begin{array}{ccl}
2\mu_1(z)&=&\int^{\pi/2}_0(g(v_1^>,z)+g(v_1^<,z))\sin\theta_1d\theta_1
\\ 
\\
2\mu_2(z)&=&\int^{\pi/2}_0(g(v_2^>,z)+g(v_2^<,z))\sin\theta_2d\theta_2,
\end{array}
\right.
\end{equation}
where the averaging is performed over the electron momentum direction
$\theta_1$ and $\theta_2$ in the first or the second layer correspondingly.
Hence, the constants of integration $A$ has to be expressed in terms of the
angle $\theta_1$ only, $A(\theta_1)$, and $B$ in terms of the angle
$\theta_2$ only, $B(\theta_2)$. It can be shown (see
Appendix~\ref{app_refl_transm}) that angles $\theta_1$ and $\theta_2$ are
related; $\theta_1$ can be expressed as a function of $\theta_2$,
$\theta_1(\theta_2)$, and vice versa, $\theta_2(\theta_1)$. It follows from
this relation that $T_{12}(\theta_1)=T_{21}(\theta_2)$ and
$R_{12}(\theta_1)=R_{21}(\theta_2)$. Note that when calculating the
function $F(g)$, change of angles is not required, which reflect the fact
that in the process of diffuse scattering angles of incidence, reflection
and transmission are not related, so a variable $\theta$ will enter the
expression for $F(g)$ without the subscript. The electrical fields $E_{1}$
and $E_{2}$ in the layers are related due to the current conservation, so
that $\sigma_i E_i=const$ throughout the system, where $\sigma_i$ is the
conductivity of the layer $i$. Since electron relaxation time $\tau$ is
assumed to be the same in all layers, and the conductivity is inversely
proportional to $v^3_F$,~\cite{AshcMerm} electrical fields $E_{1}$ and
$E_{2}$ are related via $v^2_{F1}E_{1}l_1=v^2_{F2}E_{2}l_2$.

The following equations on the chemical potentials in the left and right
layers are obtained from equations~(\ref{g_2lay})-(\ref{def_mu}):
\begin{eqnarray}
\label{eq_mu_1} 
2\mu_1(\xi)&=&eE_{1}l_1\int_0^{\pi/2}
\left[(1+SR_{12}(\theta_1))\cos\theta_1-
\frac{v^2_{F1}}{v^2_{F2}}ST_{12}(\theta_1)\cos\theta_2(\theta_1)\right]
 \nonumber \\
&\times&\exp{\left(\frac{\xi}{\cos\theta_1}\right)}\sin\theta_1d\theta_1
 \nonumber \\
&+&\int^0_{-\infty}\mu_1(\xi^\prime)d\xi^\prime  \nonumber \\
&\times&
\int_0^{\pi/2}\left[\exp{\left(\frac{-|\xi-\xi^\prime|}{\cos\theta_1}\right)}
+SR_{12}(\theta_1)\exp{\left(\frac{\xi+\xi^\prime}{\cos\theta_1}\right)}\right]
\tan\theta_1d\theta_1  \nonumber \\
&+&\int_0^{\infty}\mu_2(\xi^\prime)d\xi^\prime
\int_0^{\pi/2}ST_{12}(\theta_1)
\exp{\left(\frac{-\xi^\prime}{\cos\theta_2(\theta_1)}\right)}
\exp{\left(\frac{\xi}{\cos\theta_1}\right)}
\frac{\sin\theta_1}{\cos\theta_2(\theta_1)}d\theta_1  \nonumber \\
&+&\frac{2(1-S)}{1+v^2_{F2}/v^2_{F1}}
\int^0_{-\infty}\mu_1(\xi^\prime)d\xi^\prime  \nonumber \\
&\times&\int_0^{\pi/2}\exp{\left(\frac{\xi^\prime}{\cos\theta}\right)}
\sin\theta d\theta
\int_0^{\pi/2}\exp{\left(\frac{\xi}{\cos\theta}\right)}\sin\theta d\theta  
\nonumber \\
&+&\frac{2(1-S)}{1+v^2_{F1}/v^2_{F2}}
\int_0^{\infty}\mu_2(\xi^\prime)d\xi^\prime \nonumber \\
&\times&\int_0^{\pi/2}\exp{\left(\frac{-\xi^\prime}{\cos\theta}\right)}
\sin\theta d\theta
\int_0^{\pi/2}\exp{\left(\frac{\xi}{\cos\theta}\right)}
\sin\theta d\theta
\end{eqnarray}
 
\begin{eqnarray} 
\label{eq_mu_2}
2\mu_2(\xi)&=&-eE_{2}l_2\int_0^{\pi/2}
\left[(1+SR_{21}(\theta_2))\cos\theta_2-
\frac{v^2_{F2}}{v^2_{F1}}ST_{21}(\theta_2)\cos\theta_1(\theta_2)\right]
 \nonumber \\
&\times&\exp{\left(\frac{-\xi}{\cos\theta_2}\right)}\sin\theta_2d\theta_2
 \nonumber \\
&+&\int_0^{\infty}\mu_2(\xi^\prime)d\xi^\prime  \nonumber \\
&\times&
\int_0^{\pi/2}\left[\exp{\left(\frac{-|\xi-\xi^\prime|}{\cos\theta_2}\right)}
+SR_{21}(\theta_2)\exp{\left(\frac{-\xi-\xi^\prime}{\cos\theta_2}\right)}\right]
\tan\theta_2d\theta_2  \nonumber \\
&+&\int^0_{-\infty}\mu_1(\xi^\prime)d\xi^\prime
\int_0^{\pi/2}ST_{21}(\theta_1)
\exp{\left(\frac{\xi^\prime}{\cos\theta_1(\theta_2)}\right)}
\exp{\left(\frac{-\xi}{\cos\theta_2}\right)}
\frac{\sin\theta_2}{\cos\theta_1(\theta_2)}d\theta_2  \nonumber \\
&+&\frac{2(1-S)}{1+v^2_{F2}/v^2_{F1}}
\int^0_{-\infty}\mu_1(\xi^\prime)d\xi^\prime  \nonumber \\
&\times&\int_0^{\pi/2}\exp{\left(\frac{\xi^\prime}{\cos\theta}\right)}
\sin\theta d\theta
\int_0^{\pi/2}\exp{\left(-\frac{\xi}{\cos\theta}\right)}\sin\theta d\theta  
\nonumber \\
&+&\frac{2(1-S)}{1+v^2_{F1}/v^2_{F2}}
\int_0^{\infty}\mu_2(\xi^\prime)d\xi^\prime \nonumber \\
&\times&\int_0^{\pi/2}\exp{\left(\frac{-\xi^\prime}{\cos\theta}\right)}
\sin\theta d\theta
\int_0^{\pi/2}\exp{\left(\frac{-\xi}{\cos\theta}\right)}
\sin\theta d\theta
\end{eqnarray}

These equations describe the chemical potential profile everywhere in
the system consisting of two semi-infinite metallic layers in the
presence of both specular and diffuse scattering at the interface
between them.  In a similar fashion, equations describing the chemical
potential profiles in three- and five-layered systems can be obtained.
Because of the tedious calculations required for the derivation, only
the final equations are presented in Appendices~\ref{app_three_lrs}
and~\ref{app_five_lrs}. Even in the simplest structure consisting of two
layers of identical metals divided by an interface, the form of the
chemical potential can't be found analytically, though is this case a
certain prediction can be made. Results of the numerical solution of the
equations on the chemical potential in various systems are presented in
Chap.~\ref{chap_res_rslt}. 
 
\section{\label{sec_spec_form_of_S}Special forms of diffuse scattering}
In the simple case of identical metals at both sides of an interface,
special forms of diffuse scattering $S(\cos\theta)$ can be found, for
which the chemical potential remains constant outside the interfacial
region, so that the resistance measured at the interface is the same as
the resistance measured far from it. In the identical layers,
$v_{F1}=v_{F2}$, $\theta_1=\theta_2$, and chemical potential is
antisymmetrical around $z=0$: 
\begin{equation} \label{anti_symm}
\mu_2(\xi)=-\mu_1(-\xi)
\end{equation}
Then, equation~\ref{eq_mu_1} for the chemical potential in the left 
layer, $\mu_1(\xi)$ for example, takes the form 
\begin{eqnarray} \label{mu_symm}
2\mu_1(\xi)&=&eE_1l_1\int_0^{\pi/2}
\left(1-S(\theta_1)+2S(\theta_1)R_{12}(\theta_1)\right)
\exp{\left(\frac{\xi}{\cos\theta_1}\right)}\sin\theta_1\cos\theta_1d\theta
\nonumber \\ 
&+&\int^0_{-\infty}\mu_1(\xi^\prime)d\xi^\prime \\
&\times&\int_0^{\pi/2}
\left(\exp{\left(\frac{-|\xi-\xi^\prime|}{\cos\theta_1}\right)}
+S(\theta_1)\left(2R_{12}(\theta_1)-1\right)
\exp{\left(\frac{\xi+\xi^\prime}{\cos\theta_1}\right)}\right)
\tan\theta_1d\theta_1. \nonumber 
\end{eqnarray} 
The derivative of the chemical potential $\mu_1(\xi)$ takes the form 
\begin{eqnarray} 
\label{mu_deriv}
2\mu_1^\prime(\xi)&=&eE_1l_1\int_{0}^{\pi/2}
(1-S(\theta_1)+2S(\theta_1)R_{12}(\theta_1))
\exp{\left(\frac{\xi}{\cos\theta_1}\right)}\sin\theta_1d\theta_1
\nonumber \\ &+&\int^0_{-\infty}\mu_1(\xi^\prime)d\xi^\prime \\ \nonumber
&\times&\int_0^{\pi/2}S(\theta_1)\left(2R_{12}(\theta_1)-1\right)
\exp{\left(\frac{\xi+\xi^\prime}{\cos\theta_1}\right)}
\frac{\tan\theta_1}{\cos\theta_1}d\theta_1 \\
&-&\int^\xi_{-\infty}\mu_1(\xi^\prime)d\xi^\prime \int_0^{\pi/2}
\exp{\left(\frac{-\xi+\xi^\prime}{\cos\theta_1}\right)}
\frac{\tan\theta_1}{\cos\theta_1}d\theta_1 \nonumber \\
&+&\int^0_\xi\mu_1(\xi^\prime)d\xi^\prime
\int_0^{\pi/2}\exp{\left(\frac{\xi-\xi^\prime}{\cos\theta_1}\right)}
\frac{\tan\theta_1}{\cos\theta_1}d\theta_1. \nonumber 
\end{eqnarray}
The condition $\mu_1^\prime(\xi)=0$, so that the chemical potential in the 
left layer $\mu_1(\xi)$ is constant, can be satisfied when the 
reflection coefficient $R_{12}(\theta)$ and diffuse scattering parameter 
$S(\theta)$ are related via
\begin{equation}
\label{S_special}
S^\star(\theta)=\frac{\cos\theta-\beta}{(\cos\theta 
+\beta)(1-2R(\theta))},
\end{equation}
where $\beta$ is a constant. Then the chemical potential in the left
layer is equal to $\mu_1(z)=\beta eE_1l_1=const$, and in the right layer
$\mu_2(z)=-\beta eE_2l_2=const$, where electrical fields $E_1$ and $E_2$,
and electron mean-free paths $l_1$ and $l_2$ are the same in both layers. 
The sheet resistance across the interface $AR_0$ and measured far from it
$AR_1$ (Fig.~\ref{chempot_schem}) are
$$
AR_0=AR_1=\frac{\mu_1(z)-\mu_2{z}}{ej}=\beta\frac{6\pi^2\hbar^3}{e^2m^2v_F^2},
$$
where $m$ is an electron mass, $v_F$ is an electron Fermi velocity, same 
in both layers.

The special form of diffuse scattering (Eq.~(\ref{S_special})) which makes
$\mu(z)$ constant for $z\neq 0$ can be found for any reflection
coefficient. In particular, if an interface between metals is modeled as
a sheet delta function potential, so that the reflection coefficient
takes the form $R(\theta )=\alpha/(\alpha
+{\cos}^{2}\theta)$,~\cite{baym} it exists for {\it any} $\beta$ from $1$
to $\alpha$. These special forms of $S$ occupy region C in
Fig.~\ref{spec_form_S}. If the function $S(\theta)$ stays above this
region for all $\theta$ (region A in Fig.~\ref{spec_form_S}) then the
resistance across the interface $R_{0}$ is larger than resistance $R_s$
measured far from it. When $S(\theta)$ stays below this region for all
$\theta$ (region B in Fig.~\ref{spec_form_S}) then $R_{0}$ is smaller
then $R_1$. For $S(\theta)$ occupying both regions C and A or (and) B,
no prediction for the chemical potential shape can be made.
\begin{figure}
\centering
\includegraphics[width=4.5in]{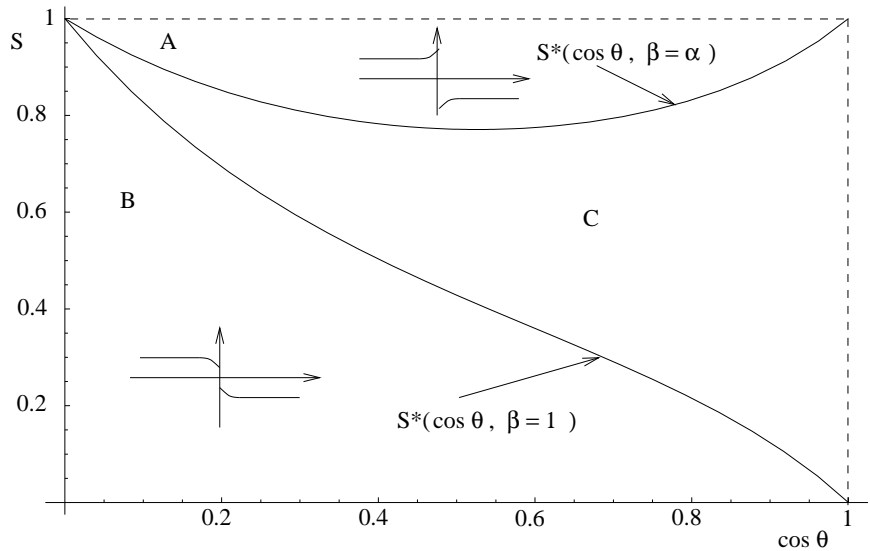}
\caption[Diffuse scattering $S(\theta)$]{Diffuse scattering $S(\theta)$. 
Region A: potential drop at the interface is bigger then potential drop
far from the interface. Region B: potential drop at the interface is
smaller then potential drop far from the interface.  Region C: region of
special forms of diffuse scattering.}
\label{spec_form_S}
\end{figure}

\chapter{\label{chap_res_rslt} Interface resistance in multilayers - 
results}
In this chapter, the numerical results for the chemical potential profile
and resistance due to the interfaces in the multilayered systems are
presented. 
 
The systems consisting of two, three, or five metallic layers, i.e., with
1, 2, or 4 interfaces, in which I consider electrons move across a 
step-like potential. Energies are referenced to the Fermi level so that
$\epsilon_F=0$, and the effective mass of the electron is assumed to be
independent of the material and spin-orientation. In the $i$-th layer,
electrons move in a constant potential $V_i=-mv^2_{iF}/2$, where $v_{iF}$
is the electron Fermi velocity in the $i$-th layer. The same electron
relaxation time $\tau$ is assumed in all layers, and an electron mean-free
path $l_i=v_{Fi}\tau$ is introduced. Current conservation requires
electric fields $E_{i}$ in the layers to be related by
$v^2_{Fi}E_{i}l_i=const$ (see Sec.~\ref{eq_mu_mult}). The reflection
coefficient at the interface between two metals takes the following form 
(see Appendix~\ref{app_refl_transm}):
\begin{equation}  
\label{R_ij}
R_{ij}(\theta_i)=\left|
\frac{1-\sqrt{1-\frac{1}{\cos^2\theta_i}+
\frac{v^2_{Fj}}{v^2_{Fi}}\frac{1}{\cos^2\theta_i}}}
{1+\sqrt{1-\frac{1}{\cos^2\theta_i}+
\frac{v^2_{Fj}}{v^2_{Fi}}\frac{1}{\cos^2\theta_i}}}\right|^2.
\end{equation}
Besides the specular scattering at the interfaces, diffuse 
scattering is present, such that the diffuse scattering parameter $S$ 
(see Chap.~\ref{chap_res_thr}) is assumed to be independent of the 
angle of incidence, and the same for all interfaces.

Equations~(\ref{eq_mu_1})-(\ref{eq_mu_2}), (\ref{eq_mu3_1})-(\ref{eq_B3})
and~(\ref{eq_mu5_1})-(\ref{eq_C5}) yield the chemical potentials in 
each layer of the multilayered system, normalized to the voltage drop 
within a mean-free path in this layer $\mu_i(z)/eE_il_i$. Resistance due to 
the interfaces is proportional to the drop of the chemical potential: 
\begin{equation}
AR=\frac{\Delta\mu}{ej},
\end{equation}
where $AR$ is the sheet resistance due to an interface, $e$ is the 
electron charge and $j$ is the electric current. In terms of the chemical 
potential drop normalized to the voltage drop within a mean-free path in 
the layer $\Delta\mu/eEl$, the interface resistance may be written 
as
\begin{equation}
\label{real_res}
AR=\frac{\Delta\mu}{e\sigma E}=\frac{\Delta\mu}{eEl}\frac{l}{\sigma}=
\frac{\Delta\mu}{eEl} \frac{6\pi^2\hbar}{e^2k^2_F}.
\end{equation}
The factor $6\pi^2\hbar/e^2k^2_F$, different for each material and 
each spin-orientation, has to be taken into account when deducing the 
resistance due to the interfaces from the chemical potential drops obtained 
by solving the equations for the chemical potential profiles.

$R_N$, where $N$ is the number of interfaces in the system, $N=1$, $2$, 
$4$, denotes the resistance due to the interfaces, measured far from 
the interfaces:
$$R_N=[\mu(-\infty)-\mu(+\infty)]/ej.$$
For the two-layered systems, resistance $R_0$ measured directly at the
interface at $z=0$ (Fig.~\ref{chempot_schem}) is introduced:
$$R_0=[\mu(0-)-\mu(0+)]/ej.$$

In this chapter, the systems consisting of two semi-infinite metallic
layers are considered. The influence of the diffuse scattering at the
interface on the chemical potential profile, and the influence of the
diffuse scattering in the bulk of the layers on the interface resistance
are investigated. The breakdown of the resistors-in-series model is
anticipated. Next, the prediction about the resistors-in-series model
breakdown is tested in the systems consisting of three and five layers. The
interface resistance dependence on the layers thicknesses relative to the
electron mean-free path is obtained for different parameters $S$ of diffuse
scattering at the interfaces. Finally, realistic multilayers consisting of
two, three and five layers of ferromagnetic and non-magnetic metals (Co-Cu
and Fe-Cr systems) are considered. Resistances due to the interfaces $R_N$
are obtained as a function of the layer thicknesses and the amount of the
diffuse scattering at the interfaces. I investigate the deviation of the
resistance $R_N$ in the three- and five-layered systems with the layer
thicknesses less than the electron mean-free path in the layers from the
resistance $NR_1$ that such systems would have if the interfaces were far
enough from each other so that the scattering from the neighboring
interfaces would not interfere.
 
The numerical procedure of solving the integral equations for the chemical
potential profile~(\ref{eq_mu_1})-(\ref{eq_mu_2}),
(\ref{eq_mu3_1})-(\ref{eq_B3}) and~(\ref{eq_mu5_1})-(\ref{eq_C5}) is
discussed in the appendix~\ref{app_numerics}.

\section{Two layers}
\subsection{Same metals}
Equations~(\ref{anti_symm}) and~(\ref{mu_symm}) are solved
numerically~\cite{ShpL_prb_00} to obtain the chemical potential profile
in the system consisting of two identical metallic layers separated by a
delta function potential, so that the reflection coefficient takes the
form $R(\theta )=\alpha/(\alpha +\cos^2\theta)$.~\cite{baym}
Fig.~\ref{iden_met}a) shows how the chemical potential profile changes
with the amount of diffuse scattering when $S$ is a constant between $0$
(only diffuse scattering is present at an interface) and $1$ (only
specular scattering is present). Interface resistance grows as the
amount of diffuse scattering decreases,~\cite{Zh_L} and the chemical
potential varies within a mean-free path of the boundary. The resistance
$R_1$ measured far from the interface can be larger or smaller then that
measured at the interface, $R_0$, (see Fig.~\ref{chempot_schem}),
depending on the value of $S$. Fig.~\ref{iden_met}b) shows the
difference between $R_0$ and $R_1$ relative to $R_1$. This difference
can be as large as 18-19\%. When there is no diffuse scattering at the
interface, $S=1$, the difference between the resistance $R_{1}$ and
$R_{0}$, agrees with the results obtained by Penn and
Stiles~\cite{Penn_Stls}. 
\begin{figure}
\includegraphics[width=\textwidth]{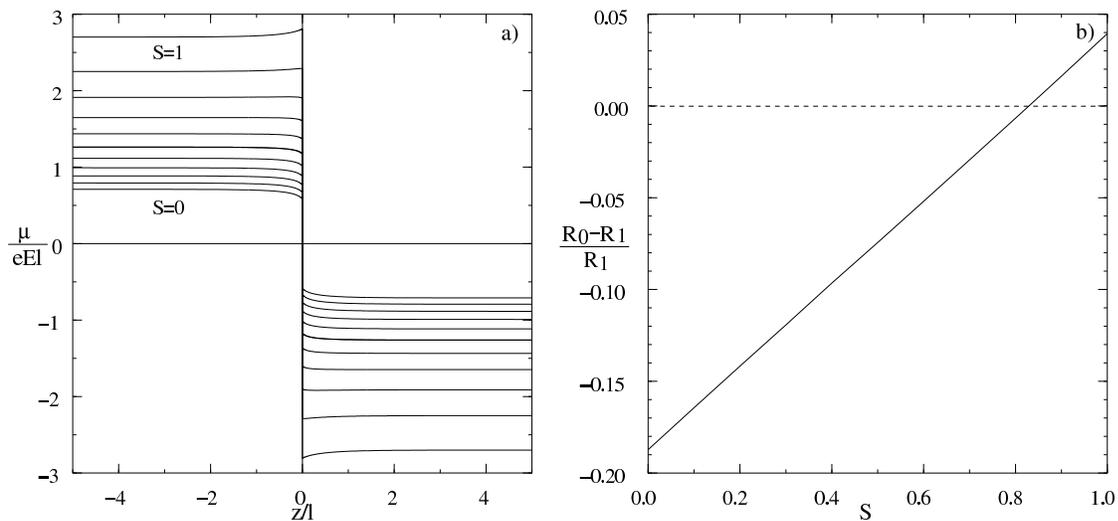}
\caption[Chemical potential and interface resistance in two layers of 
identical metals]{Chemical potential and interface resistance in two 
layers of identical metals: a) chemical potential profile for 
$S=0$, $0.1$, $\dots$, $1$, normalized with respect to potential drop 
$eEl$ within a mean-free path $l$ of the metal, b) the difference between 
$R_0$ and $R_1$ relative to $R_1$.}
\label{iden_met} 
\end{figure}

\subsection{Different metals} Next, a system consisting of two different
metallic layers and an interface between them is
considered.~\cite{ShpL_prb_00} Figure~\ref{two_lrs_pot} shows the
step-like potential experienced by electrons in such a system. For $z<0$,
electrons move in a constant potential $V_1=-mv_{F1}^2/2$, and
$V_2=-mv_{F2}^2/2)$ for $z>0$. 
\begin{figure}
\centering
\includegraphics[width=3.5in]{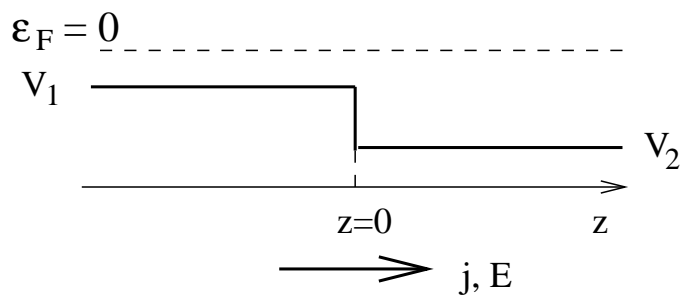}
\caption[Step-like potential experienced by electrons at an 
interface between two metals]{Step-like potential experienced by 
electrons at an interface between two metals.}
\label{two_lrs_pot} 
\end{figure}

Equations~(\ref{eq_mu_1}) and~(\ref{eq_mu_2}) with the reflection
coefficients~(\ref{R_ij}) are solved in order to investigate how the
chemical potential profile and interface resistance change as the
functions of the potential barrier height $V_{2}/V_{1}$, and the amount of
diffuse scattering at the interface. In Fig.~\ref{mu_two}, the forms of
the chemical potential for $V_{2}/V_{1}=2$, and different values of the
diffuse scattering parameter $S$ are shown. As in the case of the same
metal on both sides of the boundary, the chemical potential varies within
several mean-free paths of the interface. Resistance measured far from the
interface, $R_1$, is bigger then that measured right at the interface,
$R_0$, for interfaces where diffuse scattering dominates ($S$ is close to
$0$). $R_1$ is smaller then $R_0$ for the interfaces where specular
scattering dominates ($S$ is close to $1$). Unlike the case of two
identical layers, the resistance of the step-like barrier decreases as the
amount of diffuse scattering decreases.~\cite{Zh_L}
\begin{figure}
\centering
\includegraphics[width=3.5in]{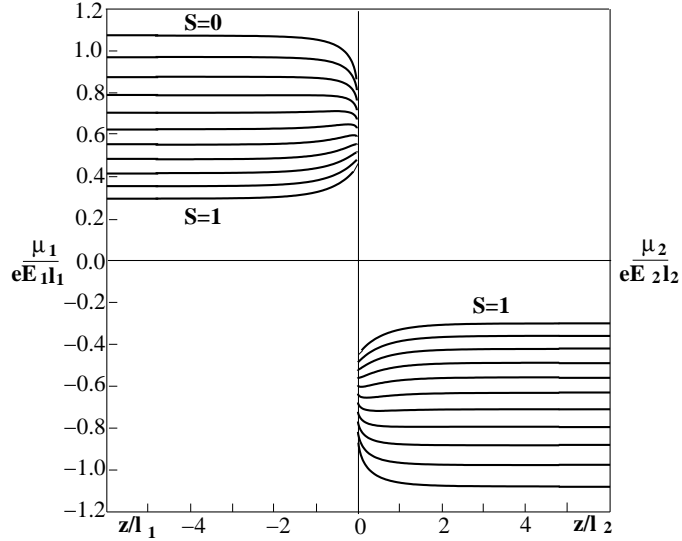}
\caption[Chemical potential profile in two-layered system with
$V_{2}/V_{1}=2$ for different amount of diffuse scattering at the
interface]{Chemical potential profile in two-layered system for different
values of diffuse scattering $S=0$, $0.1$, $\dots$, $1$ at the interface, 
normalized with respect to potential drop $eE_il_i$ within a mean-free 
path $l_i$ of each metal, so that the plots are anti-symmetrical around 
$z=0$.}
\label{mu_two}
\end{figure}

Figure~\ref{R_ball} demonstrates the effect of diffuse scattering in the
bulk of the layers on the interface resistance. In the case of a
completely specular interface, $S=1$, the variation of interface
resistance as the function of the barrier height $V_2/V_1$ is shown in
comparison with that obtained by Barnas and Fert in the system of
ballistic layers.~\cite{BF_prb,BF_jmmm} In the presence of diffuse
scattering in the bulk of the layers, both $R_0$ and $R_1$ are less than
the interface resistance obtained for purely ballistic transport. In the
diffusive layer, electrons reflected off the barrier can be scattered back
on the barrier, and again have a chance to be transmitted through, making
the interface resistance smaller.
\begin{figure}
\centering
\includegraphics[width=\textwidth]{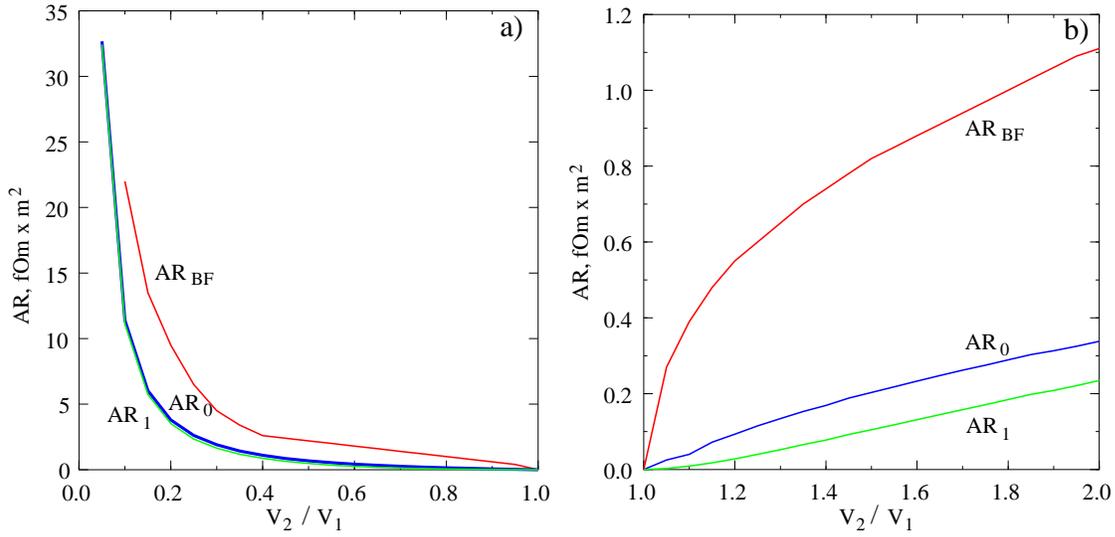}
\caption[Effect of the diffuse scattering in the bulk of the layers on the
interface resistance in two-layered system]{Effect of the diffuse
scattering in the bulk of the layers on the interface resistance in  
two-layered system. Red line is the Barnas and Fert result for the sheet
interface resistance, $AR_{BF}$, adapted from Refs.~\cite{BF_prb,BF_jmmm},
for the ballistic transport in the bulk of the layers and specular
scattering at the interface; blue and green lines are resistances $AR_0$
and $AR_1$ for diffuse transport in the bulk of the layers and
specular scattering at the interface, $S=1$, as the functions of the
electron potentials ratio a) $V_2/V_1 < 1$, b) $V_2/V_1 > 1$.}
\label{R_ball}
\end{figure} 
 
Figure~\ref{one_all} summarizes the behavior of the interface resistance
as the amount of diffuse scattering at the interface, and the height of
the potential barrier experienced by the electrons at the interface
change. Arbitrary units for the resistance are used, since the
calculations are performed irrespective of the particular metals. Note
that the ratio of the electron Fermi velocities in the layers
$v_{F2}/v_{F1}$ is used as a variable, not the ratio of the electron
potentials $V_2/V_1$, as in the Fig.~\ref{R_ball}. In
Figs.~\ref{one_all}a)-b), the variation of the interface resistance as the
function of the amount of diffuse scattering at the interface for the
system with $v_{F2}/v_{F1}=2$ is shown. The resistance $R_0$ measured
directly at the interface is larger than the resistance $R_1$ measured far
from it, and $R_0$ grows as the interface becomes less diffusive ($S$
approaches $1$), while $R_1$ decreases slightly, so that the absolute
value of the difference $R_1-R_0$ grows, reaching about 70\% if calculated
relative to $R_0$, and 120\% if calculated relative to $R_1$ for a
completely specular interface. In Figs.~\ref{one_all}c)-f), the variation
of the interface resistance as the function of $v_{F2}/v_{F1}$ for
different values of $S$ is shown. In the absence of the potential barrier
($v_{F2}/v_{F1}=1$), both $R_0$ and $R_1$ increase as the amount of
diffuse scattering at the interface increase ($S$ approaches $0$), since
only diffuse scattering is responsible for the resistance in this case. As
the height of the barrier grows ($v_{F2}/v_{F1}$ becoming $>1$ or $<1$),
specular reflection starts to influence the interface resistance, and the
values of both $R_0$ and $R_1$ are closer to each other for different
values of $S$ than they were in the absence of the potential barrier.
Values of the potential barrier heights exist where the interface
resistance is almost independent of the amount of diffuse scattering at
the interface. These values are different for $R_0$ and $R_1$. The
percentage difference between $R_1$ and $R_0$ shows no variation with the
barrier height in the case of completely diffuse interface ($S=0)$, since
$v_{F2}/v_{F1}$ only defines the scale of the interface resistance in this
case. As the amount of diffuse scattering at the interface decreases, both
$(R_1-R_0)/R_0$ and $(R_1-R_0)/R_1$ show a strong dependence on the
potential barrier height, reaching maximum values of several tens percent
for small barriers. This fact is of particular interest, because small
barrier heights are realized in Fe-Cr multilayers for minority electrons
and Co-Cu multilayers for majority electrons. These systems will be
discussed further below.
\begin{figure}
\centering
\includegraphics[width=\textwidth]{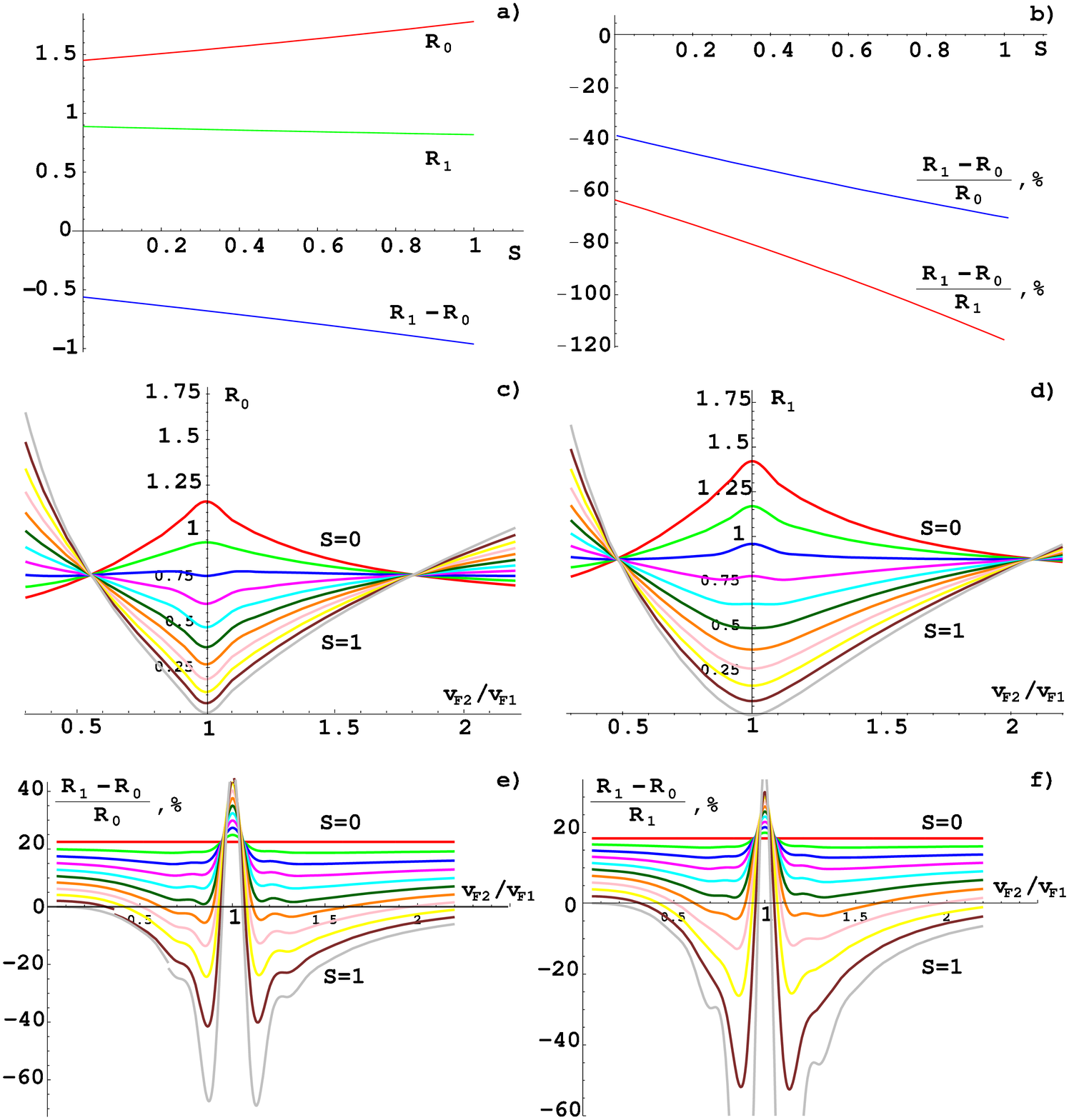}
\caption[Summary of the results for the two-layered system]
{Summary of the results for the two-layered system: 
a) resistances $R_0$, $R_1$, and the difference $R_1-R_0$ as the functions 
of the diffuse scattering $S$ at the interface for $v_{F2}/v_{F1}=2$, 
b) percentage differences $R_1-R_0$ relative to $R_0$ and $R_1$ as the 
functions of $S$ for $v_{F2}/v_{F1}=2$,  
c)-d) resistances $R_0$ and $R_1$ as the functions of the electrons Fermi 
velocities ratio $v_{F2}/v_{F1}$ for different values of the diffuse 
scattering 
$S$ at the interface. In the plots a)-d), arbitrary units for the 
resistance are used. 
e)-f) percentage differences $R_1-R_0$ relative to $R_0$ and $R_1$ as the 
functions of the electrons Fermi velocities for different values of $S$.} 
\label{one_all}
\end{figure}

\subsection{Anticipation of the breakdown of the resistors-in-series 
model}
The results of the previous section show that the resistance of the
interface deduced from the voltage drop measured far from the interface
($R_1$) should not be interpreted as the interface resistance that would
be obtained by measuring the voltage drop directly at the interface
($R_0$), as they may differ significantly. As seen from
Figs.~\ref{iden_met}a) and~\ref{mu_two}, the chemical potential varies
within the electron mean-free path from the interface. Hence, in the
multilayered system with layers thicknesses of the order or less than the
mean-free path, the chemical potential can be expected not to reach its
asymptotic value within a layer. The breakdown of the resistors-in-series
model can be anticipated, which means that while the resistance of the
whole structure still can be found by adding the resistances due to the
bulk of the layers and due to the interfaces, neither the resistance $R_1$
nor $R_0$ of the interface between the semi-infinite layers should be
considered as the resistance of the interface between the layers which are
thinner than the mean free path due to the scattering indigenous to them.
In other words, even if one maintains the same interface between layers,
its contribution to the resistance of a multilayered structures will
depend on the thickness of the adjacent layers, or, more precisely, on the
ratio of the thickness to the mean-free path. Its contribution will be
constant, $R_1$, only if the interfaces are far enough from each other for
the chemical potential to level off.

In the following section, the prediction about the resistors-in-series
model breakdown and interface resistance dependence on the ratio of the
layer thickness to the electron mean-free path in this layer will be 
tested in the systems consisting of three and five layers.

\section{Three and five layers}
In this section, systems consisting of three and five layers of different
metals are considered.~\cite{ShpLZh_mrs_01} Figure~\ref{thr_fv_pot} shows
the step-like potentials experienced by electrons in such systems. In the
three-layered system (Fig.~\ref{thr_fv_pot}a)), electrons move in a
constant potential $V_1=-mv^2_{F1}/2$ at $z<-d$ and $z>d$, and
$V_2=-mv^2_{F2}/2$ at $-d<z<d$. In the five-layered system
(Fig.~\ref{thr_fv_pot}b)), electrons move in a constant potential
$V_1=-mv^2_{F1}/2$ at $z<-3d$, $-d<z<d$, and $z>3d$, and
$V_2=-mv^2_{F2}/2$ at $-3d<z<-d$ and $d<z<3d$. The structures with only
two possible values of the electron Fermi velocity are chosen for
simplicity; further below, when the realistic Fe-Cr and Co-Cu systems will
be considered,this restriction will be removed, and the electron Fermi
velocity will depend not only on the material but on the magnetization
direction in the magnetic layer.
\begin{figure}
\centering
\includegraphics[width=\textwidth]{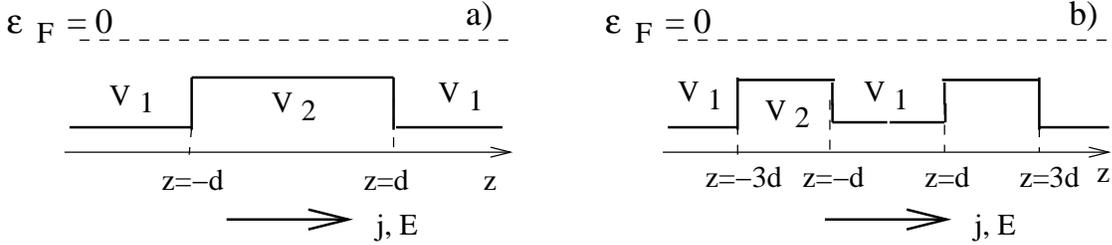}
\caption[Step-like potential experienced by electrons in a 
multilayered system]{Step-like potential experienced by electrons in a) 
a three-layered system, b) a five-layered system.}
\label{thr_fv_pot}
\end{figure}

Equations~(\ref{eq_mu3_1})-(\ref{eq_B3})  
and~(\ref{eq_mu5_1})-(\ref{eq_C5}) with the reflection
coefficients~(\ref{R_ij}) are solved in order to investigate how the
chemical potential profile changes when the thickness of the layers
changes and becomes less than the electron mean-free path in the layers.
I find the resistance due to the interfaces as the function of the
layers thickness relative to the electron mean-free path, for different
values of diffuse scattering $S$ at the interfaces, and investigate
the deviation from the resistors-in-series model as the barrier height
$v_{F2}/v_{F1}$ changes, for different values of $S$.

Figure~\ref{mu_many} shows the chemical potential profiles for three- and
five-layered systems with $S=0.5$; this represents interfaces where there
is a 50$\%$ probability that the electrons are scattered diffusely and
50$\%$ specularly. In the case where the thicknesses of the inner layers
are much larger than an electron mean free path in these layers
(Fig.~\ref{mu_many}a),b)), the chemical potential approaches constant
values in the inner layers. This means that the interfaces are independent
of each other, and the interface resistance of the N-interface system is
found to be N times the resistance of one single interface $R_1$,
independently of the layer thicknesses as long as they are larger than the
mean free path. Because of the diffuse scattering processes in the inner
layers electrons lose their memory of the scattering at an interface and
come to the next interface as if they were propagating in an infinite
layer. In the second case, where the layer thicknesses are much smaller
then the electron mean free path (Fig.~\ref{mu_many}c),d)), electrons
retain a memory of the scattering from a previous interface; in this case
the chemical potential does not approach a constant value, and different
interfaces interact with each other, so that the interface resistance
depends on the distance between the interfaces.
\begin{figure}
\centering
\includegraphics[width=\textwidth]{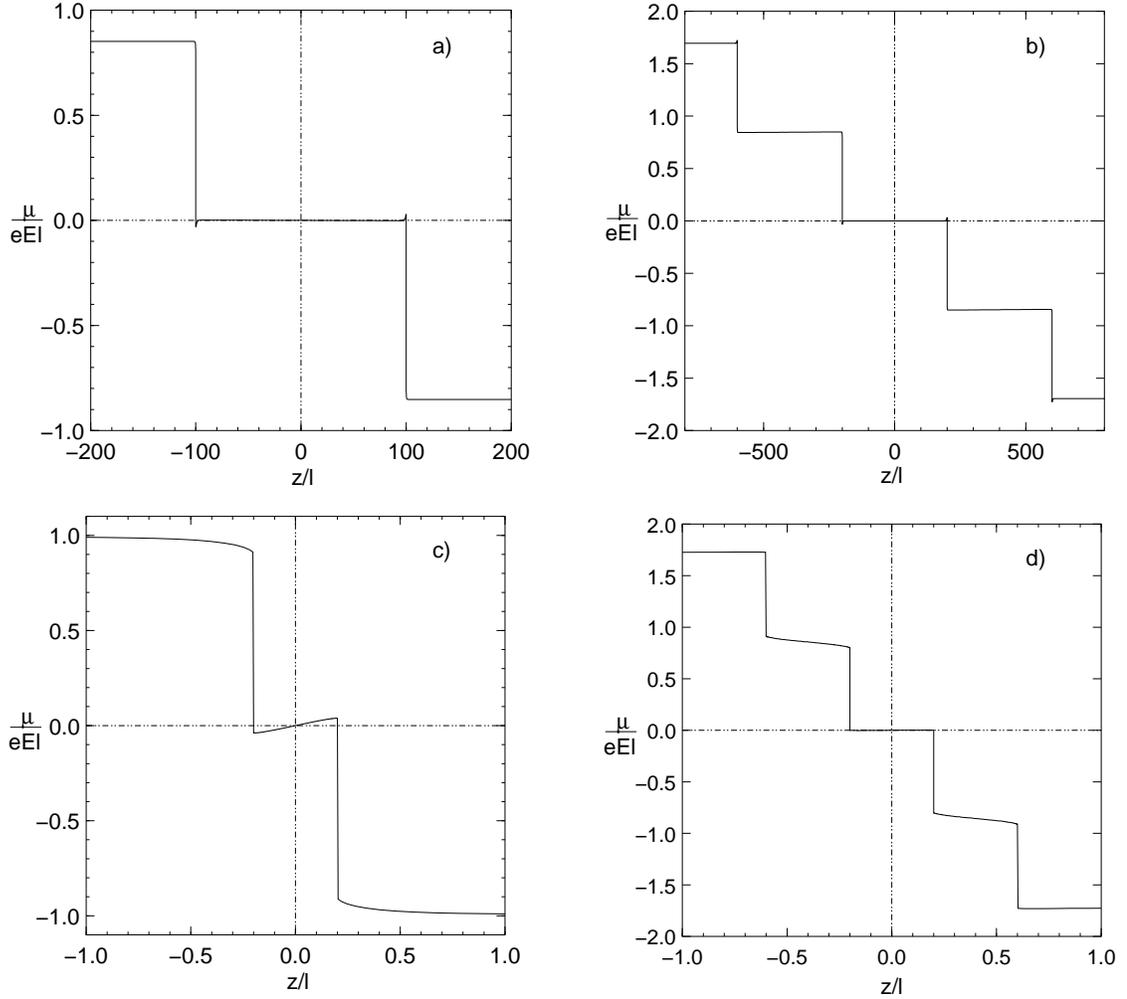}
\caption[Chemical potential profiles in the multilayered systems]
{Chemical potential profiles for $S=0.5$, normalized with respect to
potential drop $eEl$ within {\it each} layer: a) in the three-layered
system with the middle layer thickness much larger than the electron
mean-free path in the layer, b) in the five-layered system with the
middle layers thicknesses much larger than the electron mean-free path in
the layers, c) in the three-layered system with the middle layer
thickness much smaller than the electron mean-free path in the layer, d)
in the five-layered system with the middle layers thicknesses much
smaller than the electron mean-free path in the layers.} 
\label{mu_many}
\end{figure}

Figure~\ref{many_all} summarizes the behavior of the interface
resistance in the multilayers as the thickness of the layers, amount of
diffuse scattering at the interfaces, and the height of the potential
barriers experienced by electrons at the interfaces change. Arbitrary
units for resistance are used, as in the Fig.~\ref{one_all}, since the
calculations are performed irrespective of the particular metals.
Figure~\ref{many_all}a) shows how the resistance due to two interfaces,
$R_2$, in the trilayer system depends on the thickness $2d$ of the
middle layer relative to the electron mean free path $l_2$ in this
layer, for different amounts of diffuse scattering at the interfaces,
and in the case where the electron Fermi velocity in the middle layer is
twice smaller than the electron Fermi velocity in the outer layers,
i.e., $v_{F1}=2v_{F2}$. When the thickness of the middle layer is large
($2d/l_2\gg 1$), $R_2$ is independent of $2d$, and it is twice the
resistance of one single interface $R_1$. As the ratio $2d/l_2$
decreases, interface resistance in the trilayer system changes. For
completely diffuse interfaces ($S=0$) interface resistance decreases as
$2d/l_2$ decreases. If a small amount of specular scattering is present
at the interfaces, interface resistance at first increases significantly
as $2d/l_2$ decreases, and then decreases abruptly when $2d/l_2$
becomes less than 0.2. As the amount of specular scattering at the
interfaces grows ($S$ reaches 1), the change in the interface resistance
as $2d/l_2$ decreases becomes smaller.
\begin{figure}
\centering
\includegraphics[width=\textwidth]{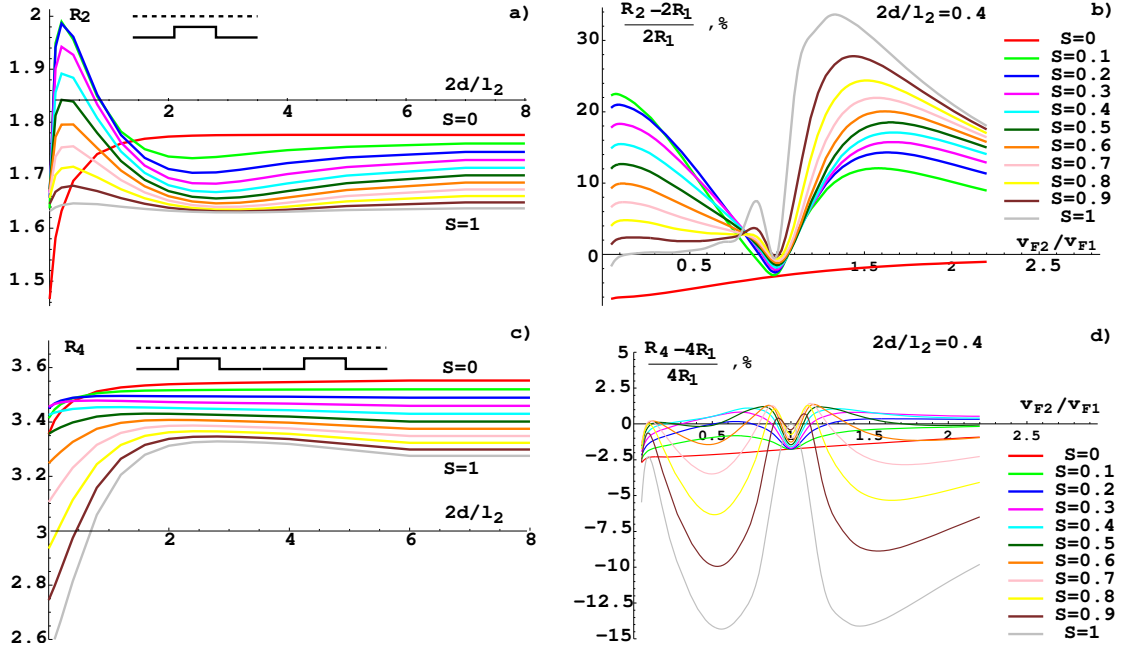}
\caption[Summary of results for the three-layered and five-layered 
systems]
{Summary of results for the three-layered and five-layered systems:
a) resistance due to the interfaces $R_2$ of the three-layered system as
a function of the middle layer thickness to the electron mean-free path
in this layer ratio $2d/l_2$ for different values of diffuse scattering
$S$ at the interfaces and for the electron Fermi velocities ratio
$v_{F2}/v_{F1}=0.5$,
b) percentage difference between $R_2$ and twice the resistance of the
single interface between two semi-infinite layers $2R_1$ relative to
$2R_1$ as a function of the electron Fermi velocities ratio
$v_{F2}/v_{F1}$ for different values of $S$ at the interfaces and for
$2d/l_2=0.4$,
c) resistance due to the interfaces $R_4$ of the five-layered system as a
function of $2d/l_2$ for different values of $S$ at the interfaces and
for $v_{F2}/v_{F1}=0.5$,
d) percentage difference between $R_4$ and four times the resistance of
the single interface between two semi-infinite layers $4R_1$ relative to
$4R_1$ as a function of $v_{F2}/v_{F1}$ for different values of $S$ at
the interfaces and for $2d/l_2=0.4$, In the plots a) and c), arbitrary
units for the resistance are used.}
\label{many_all}
\end{figure}

In the Fig.~\ref{many_all}b), the difference between the resistance due to
the interfaces in the three-layered system $R_2$ and twice the resistance
of the single interface between two semi-infinite layers $2R_1$ relative to
$2R_1$ as a function of the electron Fermi velocities ratio $v_{F2}/v_{F1}$
for different values of $S$ at the interfaces and for $2d/l_2=0.4$ is
shown. This difference represents the deviation from the
resistors-in-series model. In the absence of the barrier
($v_{F2}/v_{F1}=1$), resistance exists due to diffuse scattering, and
$(R_2-2R_1)/2R_1=0$ for completely specular interfaces only. The deviation
from the resistors-in-series model is the largest for the completely
specular interfaces ($S=1$) and $v_{F2}/v_{F1}=1.4$. The shape of the
potential barrier (whether the electrons experience a step up or step down
at the interface between the first and the second layers) strongly
influences the $(R_2-2R_1)/2R_1$ difference, as the plots are asymmetrical
around $v_{F2}/v_{F1}=1$, the ratio of the electron Fermi velocities that
represents the absence of the barrier.

Similarly, Fig.~\ref{many_all}c) shows the resistance due to four
interfaces in a five-layered system as a function of the layers
thickness $2d$ relative to the electron mean-free path in the second
layer $l_2$. Unlike in the three-layered system, the variation of the
interface resistance with $2d/l_2$ is less pronounced for diffusive
interfaces ($S$ approaching $0$), and more pronounced for specular
interfaces ($S$ approaching $1$). Figure~\ref{many_all}d) shows the
deviation from the resistors-in-series model in the five-layered system,
the difference between $R_4$ and $4R_1$ relative to $4R_1$ for different
$S$ and $2d/l_2=0.4$. As the number of steps experienced by the
electrons grows from $2$ in the three-layered system to $4$ in the
five-layered system, the actual shape of the barrier seems to play less
important role in defining the $(R_N-NR_1)/NR_1$ difference as a
function of $v_{F2}/v_{F1}$, as the plots for $(R_4-4R_1)/4R_1$ look
symmetrical around $v_{F2}/v_{F1}=1$. As in the three-layered system,
the maximum deviation from the resistors-in-series model is obtained
for completely specular interfaces; it reaches 15\% for
$v_{F2}/v_{F1}=1.6$ and $v_{F2}/v_{F1}=0.6$. 

\section{\label{CoCu_FeCr}Co-Cu and Fe-Cr systems}
In this section the resistance due to the interfaces in the realistic
two-, three-, and five-layered Fe-Cr and Co-Cu systems is studied.
The following cases are considered: 1) a FM$\uparrow$-N
single interface, 2)  FM$\uparrow$-N-FM$\uparrow$($\downarrow$)- a
trilayer with parallel (anti-parallel) magnetization directions in the
ferromagnetic layers, and 3)
FM$\uparrow$-N-FM$\uparrow$($\downarrow$)-N-FM$\uparrow$ - a five-layered
system with parallel (anti-parallel) magnetization directions in the
ferromagnetic layers (see Fig.~\ref{real_struct}). 
\begin{figure}
\centering
\includegraphics[width=5in]{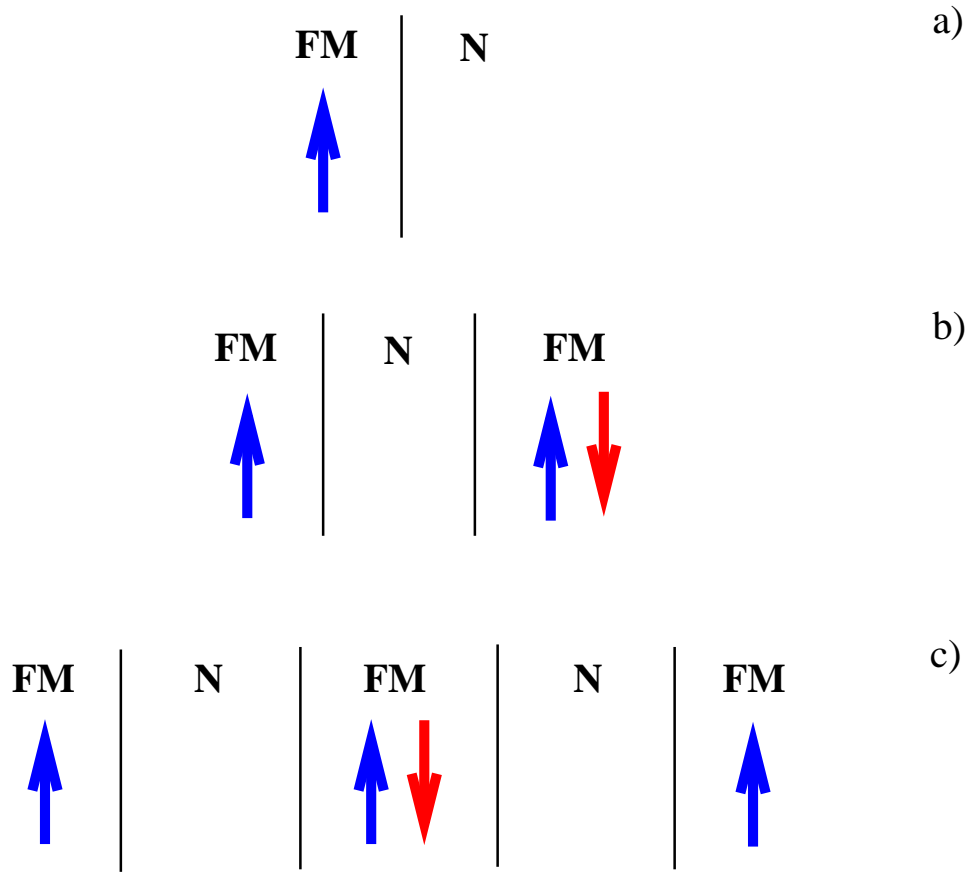}
\caption[FM-N structures]
{a) FM$\uparrow$-N single interface, 
b) FM$\uparrow$-N-FM$\uparrow$($\downarrow$) three-layered system, 
c) FM$\uparrow$-N-FM$\uparrow$($\downarrow$)-N-FM$\uparrow$ five-layered 
system.}
\label{real_struct}
\end{figure}
In the case of the five-layered system the thicknesses of the inner 
layers are taken to be the same, and the magnetization of the middle 
ferromagnetic layer only is changed. Majority and minority electrons 
experience different potential steps at the interfaces between magnetic 
and non-magnetic metals (see Fig.~\ref{real_many}a)).
\begin{figure}
\centering
\includegraphics[width=\textwidth]{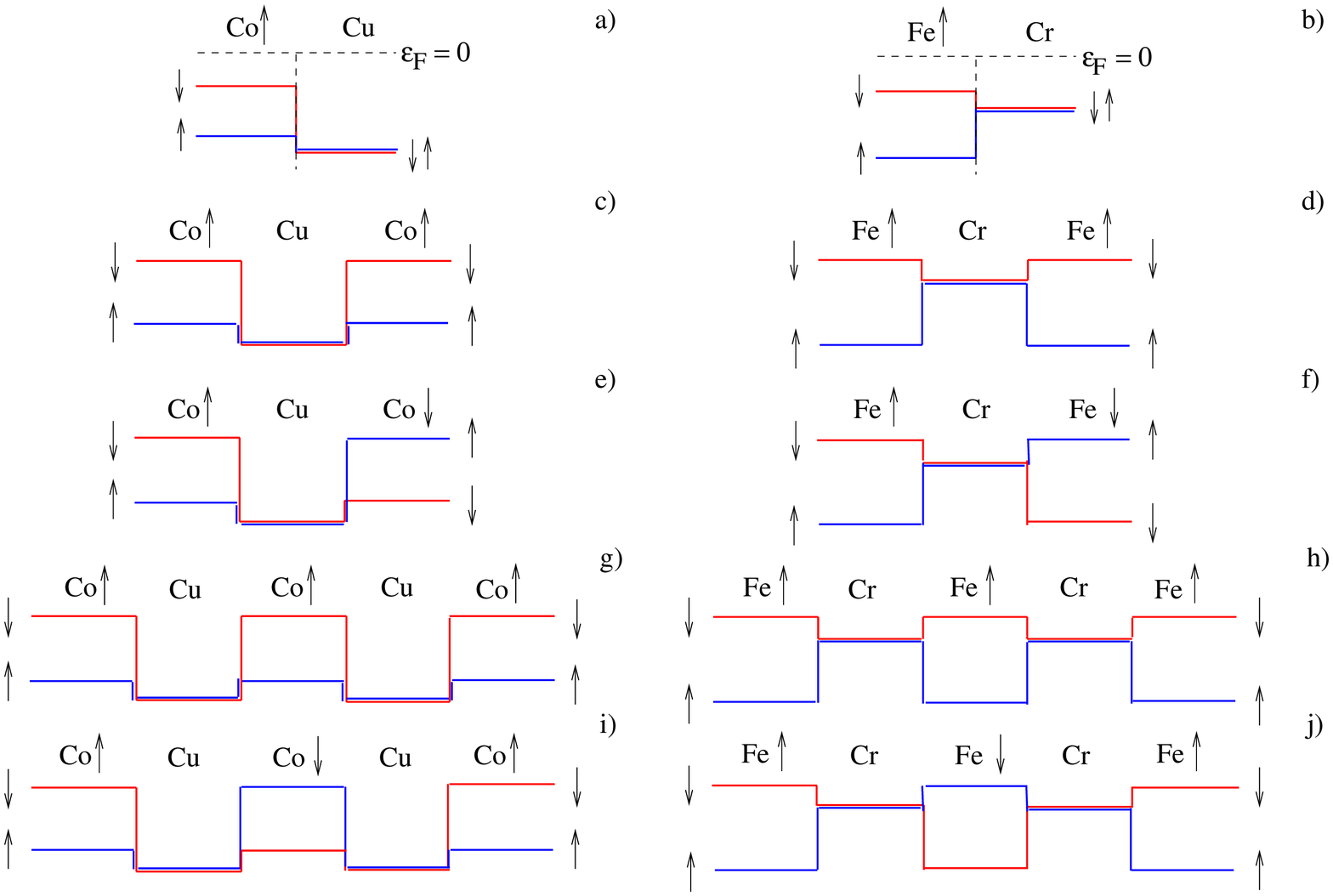}
\caption[Step-like potentials experienced by up and down electrons in the 
Co-Cu and Fe-Cr multilayered structures]
{Step-like potentials experienced by up (blue) and down (red) 
electrons in the
a)-b) Co-Cu and Fe-Cr structures,
c)-d) Co-Cu-Co and Fe-Cr-Fe structures with parallel magnetizations of 
the ferromagnetic layers,
e)-f) Co-Cu-Co and Fe-Cr-Fe structures with anti-parallel magnetizations   
of the ferromagnetic layers,
g)-h) Co-Cu-Co-Cu-Co and Fe-Cr-Fe-Cr-Fe structures with parallel
magnetizations,
i)-j) Co-Cu-Co-Cu-Co and Fe-Cr-Fe-Cr-Fe structures with anti-parallel
magnetizations.}
\label{real_many}
\end{figure}
The ratio of the electron Fermi velocities in the Fe-Cr systems 
$v_{F}(Cr)/v_{F}(Fe)$ is taken to be 0.837 for majority electrons, and 
1.003 for minority electrons.~\cite{Hood_Fal} In the Co-Cu systems the 
ratio $v_{F}(Cu)/v_{F}(Co)$ is calculated to be 1.05 for majority 
electrons, and 1.18 for minority electrons 
(see appendix~\ref{cocu_ratio}).
Interface resistances are calculated by assuming majority and minority 
channels to be independent,~\cite{sdtmn_ch2} so that the total 
resistance is the parallel combination of the resistances of up 
electrons $R_\uparrow$ and down electrons $R_\downarrow$:
\begin{equation}
\label{res_parall}
R=\frac{R_\uparrow R_\downarrow}{R_\uparrow+R_\downarrow}. 
\end{equation}
The step-like potentials experienced by electrons in the two-, three-, and
five-layered Co-Cu and Fe-Cr structures are shown schematically in the
Fig.~\ref{real_many}. In order to find the resistances of up and down
electrons $R_\uparrow$ and $R_\downarrow$ in each case a) through j)  (see
Fig.~\ref{real_many}),
equations~(\ref{eq_mu_1})-(\ref{eq_mu_2}),~(\ref{eq_mu3_1})-(\ref{eq_B3}),
and~(\ref{eq_mu5_1})-(\ref{eq_C5}) were numerically solved with the
reflection coefficients~(\ref{R_ij}) and the appropriate parameters
$v_{F1}$, $v_{F2}$, and $v_{F3}$. I present the correspondence between
$R_\uparrow$, $R_\downarrow$, type of the structure, equations that need to
be solved, and parameters $v_{F1}$, $v_{F2}$, and $v_{F3}$ entering these
equations in the Appendix~\ref{table}. In order to deduce the interface
resistance from the jump of the chemical potential (Eq.~(\ref{real_res})),
the values of the electrons Fermi momentum in the non-magnetic layers are
taken to be $k_{FCu}=1.36 A^{-1}$ for copper, and $k_{FCr}=3.8 A^{-1}$ for
chromium.~\cite{Hood_Fal}

Figures~\ref{cocu_Rofd} and ~\ref{cocu_RNR1} summarize the behavior of the
resistance due to the interfaces in the Co-Cu multilayers with different
magnetization configurations as I change the amount of the diffuse
scattering $S$ at the interfaces and the thickness of the layers $2d$
relative to the electron mean-free path $l_{mfp}$.  
Figure~\ref{cocu_Rofd}a) shows the resistance $R_1$ measured far from the
interface between two semi-infinite layers of Co and Cu, and that measured
right at the interface, $R_0$, as the functions of $S$. For diffuse
interfaces ($S$ is close to zero), $R_1$ is larger than $R_0$. Both $R_1$
and $R_0$ decrease as the amount of diffuse scattering at the interface
decreases ($S$ increases), but $R_1$ decreases faster, and there exists a
value of parameter $S$ when $R_1=R_0$. The resistance $R_0$ becomes larger
than $R_1$ for a specular interface. The relative difference between $R_1$
and $R_0$ (Fig.~\ref{cocu_Rofd}b)) is about 20\% for completely diffuse
interfaces; as the resistances decrease, this difference may reach more
than 100\%.

Figures~\ref{cocu_Rofd}c)-f) show the resistances due to the interfaces
in the multilayers $R_2$ for three-layered systems and $R_4$ for
five-layered systems as the functions of the layer thicknesses relative
to the electron mean-free path for different amounts of diffuse
scattering at the interfaces and different magnetization configurations
(see Fig.~\ref{real_many}). In all four geometries, resistance due to the
interfaces decreases as the amount of diffuse scattering decreases ($S$
changes from $0$ to $1$).  Interface resistance is very small for
completely specular interfaces, due to the very small potential barrier
height that the majority electrons experience at the interface between Co
and Cu, so that the majority channel shunts the current. As seen from the
Figs.~\ref{cocu_Rofd}c)-f), resistance due to the interfaces in the
multilayers do not depend on the thickness of the layers $2d$ relative to
the electron mean-free path $l_{mfp}$ when $2d/l_{mfp}$ is greater than
$1$, and they are found to be equal to the sum of resistances $R_1$ of
the corresponding number of the independent interfaces, $2R_1$ or $4R_1$.
As the thickness of the layers becomes much smaller than the mean-free
path, $2d/l_{mfp}\ll 1$, resistances $R_2$ and $R_4$ start to deviate
from $2R_1$ or $4R_1$, as predicted, indicating the breakdown of the
resistors-in-series model.
 
Figures~\ref{cocu_RNR1}a)-d) show the deviations of the actual
resistances $R_2$ and $R_4$ from the resistors-in-series model results,
$2R_1$ or $4R_1$, relative to $2R_1$ or $4R_1$, as the functions of
$2d/l_{mfp}$ for different amounts of diffuse scattering at the
interfaces and different magnetization configurations. These deviations
are not monotonic with $2d/l_{mfp}$. The difference between $R_N$ and
$NR_1$ is especially large for extreme interfaces, almost completely
specular, $S$ close to $1$, or completely diffuse, $S$ close to $0$. (In
the three-layered systems the plot for $S=1$ is not shown, as it has a
much larger scale than other plots.) It reaches 10-15\% in the
three-layered systems, 20\% in the five-layered systems with the parallel
magnetization directions, and 30\% in the five-layered systems with the
anti-parallel magnetization directions for completely specular
interfaces.

Finally, Figs.~\ref{cocu_RNR1}e)-f) show the difference between actual
resistance due to interfaces $R_N$ ($N=$1, 2, and 4) and the resistance
that would be obtained assuming interfaces to be independent of each
other as a function of the amount of diffuse scattering $S$ at the
interfaces. The thickness $2d$ of the non-magnetic layers is taken to be
small compared to the electron mean free path in the layers
($2d/l_{mfp}$=0.4). The actual resistance is compared with both the
resistance of an independent interface measured right at the interface,
$NR_0$, and that measured far from the interface, $NR_1$. In all
geometries, both $(R_N-NR_1)/NR_1$ and $(R_N-NR_0)/NR_0$ are significant
for either completely diffuse or completely specular interfaces. For the
realistic interfaces where an equal amount of specular and diffuse
scattering is present, $S=0.5$, the difference between the interface
resistance one could measure and the resistance of independent interfaces
measured right at the interfaces, $NR_0$, reaches 10\%, which is
comparable with the experimental error for the interface 
resistance.~\cite{BP_jmmm_99}

Figures~\ref{fecr_Rofd} and~\ref{fecr_RNR1} show the same information as
the figures~\ref{cocu_Rofd} and~\ref{cocu_RNR1} for Fe-Cr systems.  
Similarly to the resistances in the Co-Cu systems, $R_1$ is larger than
$R_0$ at small $S$, both $R_0$ and $R_1$ decrease as the interface becomes
more specular ($S$ approaching $1$), $R_1$ decreasing faster (see
Fig.~\ref{fecr_Rofd}a)). The difference with Co-Cu systems is that $R_1$
becomes equal to $R_0$ for $S$ closer to $1$. The relative difference
between $R_0$ and $R_1$ is between 12 and 18\% for $S$ up to
$0.9$,(Fig.~\ref{fecr_Rofd}b)), after that the difference grows
significantly (not shown). Resistances $R_2$ and $R_4$ for different
magnetization configurations decrease as the amount of diffuse scattering
at the interfaces decreases. In the case of the parallel magnetization
directions both in the three- and five-layered structures, resistances of
the completely specular interfaces are not shown, because they are small
compared to the resistances for $S\neq 1$, due to the very small potential
barrier experienced by minority electrons at the interface between Fe and
Cr (smaller than that for majority electrons at the interface between Co
and Cu). Resistance due to the interfaces depend on the thickness of the
layers relative to the electron mean-free path, $2d/l_{mfp}$, as it
becomes of the order or less then $1$ (Fig.~\ref{fecr_Rofd}c)-f)). As it
is for the Co-Cu systems, the deviation from the resistors-in-series model
for Fe-Cr structures is substantial for extreme interfaces, with $S$ close
to $0$ or $1$ (Fig.~\ref{fecr_RNR1}a)-d)), but for the interfaces with
$S=0.5$, the difference between $R_N$ and $NR_0$ reaches 20$\%$ for the
systems with parallel magnetic configuration and more then 15$\%$ for the
systems with anti-parallel magnetic configuration
(Fig.~\ref{fecr_Rofd}e)).  This exceeds experimental error, which is about
10$\%$ for the interface resistance and about 15$\%$ for the bulk
resistance~\cite{BP_jmmm_99}.
\begin{figure} 
\centering
\includegraphics[width=\textwidth]{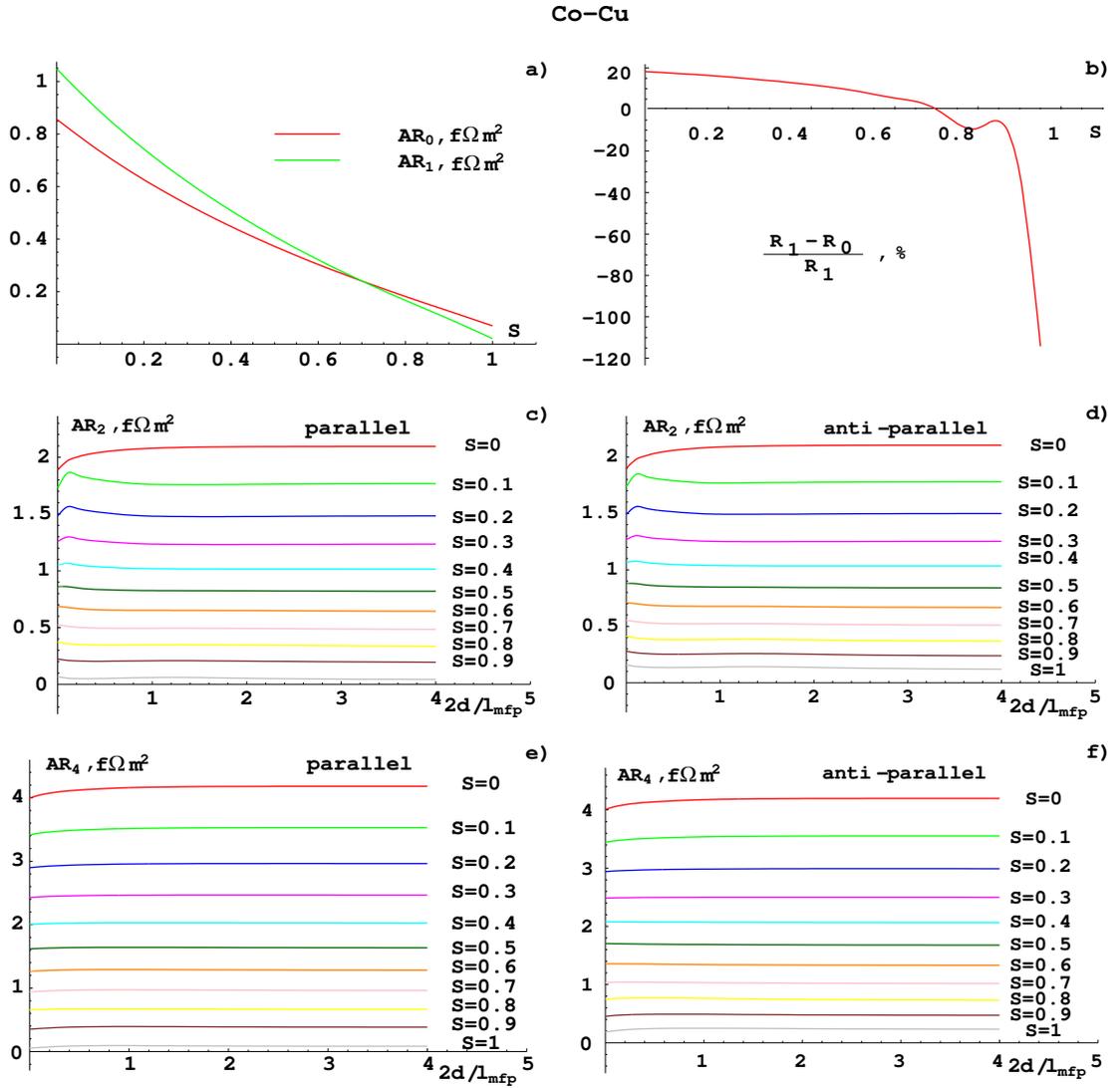}
\caption[Resistances due to the interfaces in the Co-Cu systems]
{Resistances due to the interfaces in the Co-Cu systems: 
a) sheet resistance measured far from the interface, $AR_1$, and that 
measured at
the interface, $AR_0$, in the two-layered systems as the functions of the
amount of diffuse scattering $S$ at the interface, b) the difference
between $R_1$ and $R_0$ relative to $R_1$ in the two-layered systems as a
function of $S$, c)-d) resistance $R_2$ as a function of the inner layer
thickness relative to the mean-free path in the three-layered systems for
different values of the diffuse scattering $S$ and different
magnetization configurations, e)-f) resistance $R_4$ as a function of the
inner layers thicknesses relative to the mean-free path in the
five-layered systems for different values of $S$ and different
magnetization configurations.} 
\label{cocu_Rofd} 
\end{figure}
\begin{figure} 
\centering
\includegraphics[width=\textwidth]{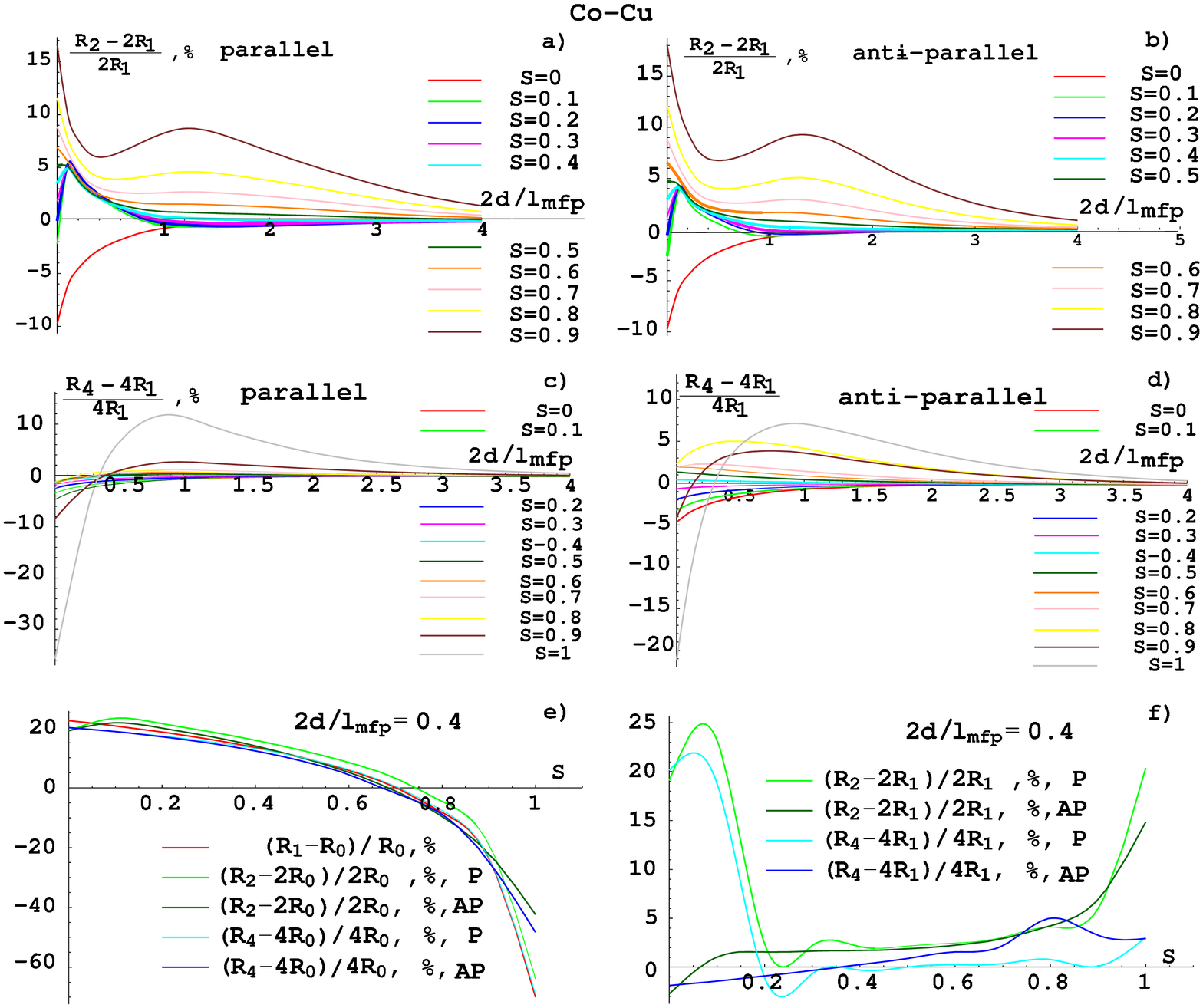}
\caption[Deviations from the resistors-in-series model in the Co-Cu
systems] {Deviations from the resistors-in-series model in the Co-Cu
systems: a)-b) the difference between the resistance $R_2$ and twice the
resistance of a single interface $2R_1$ relative to $2R_1$ as a function
of the inner layer thickness $2d$ relative to the mean-free path
$l_{mfp}$ in the three-layered system for different values of $S$ and
different magnetization configurations, c)-d) the difference between the
resistance $R_4$ and four times the resistance of a single interface,
$4R_1$, relative to $4R_1$ as a function of $2d$ relative to $l_{mfp}$ in
the five-layered system for different values of $S$ and different
magnetization configurations, e) the difference between the resistances
$R_N$, with $N=1$, $2$, $4$ and $N$ times the resistance $R_0$, $NR_0$, of 
a single independent interface measured right at the interface, relative to
$NR_0$, in the multilayers with $2d \ll l_{mfp}$ as a function of $S$ for
different magnetization configurations, f) the difference between the
resistances $R_N$, with $N=2$, $4$ and $N$ times the resistance $R_1$ of
a single independent interface measured far from the interface, 
$NR_1$, relative to $NR_1$, in the multilayers with $2d\ll l_{mfp}$ as a 
function of $S$ for different magnetization configurations. } 
\label{cocu_RNR1}
\end{figure} 
\begin{figure} 
\centering
\includegraphics[width=\textwidth]{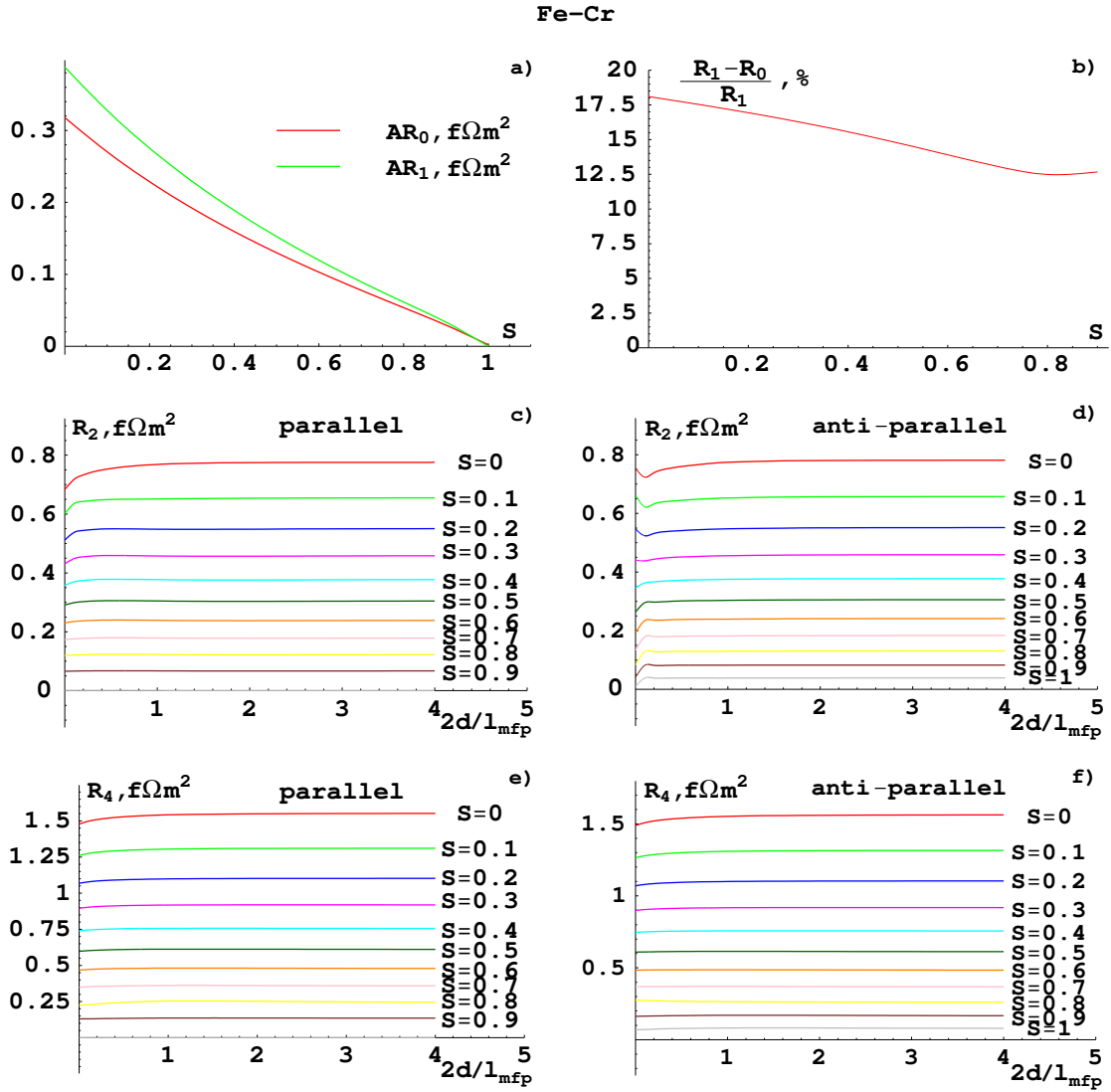}
\caption[Resistances due to the interfaces in the Fe-Cr systems]
{Resistances due to the interfaces in the Fe-Cr systems: 
a) sheet resistance measured far from the interface, $AR_1$, and that
measured at the interface, $AR_0$, in the two-layered systems as the
functions of the amount of diffuse scattering $S$ at the interface,
b) the difference between $R_1$ and $R_0$ relative to $R_0$ in the
two-layered systems as a function of $S$ (note the different $S$-axis 
limit),
c)-d) resistance $R_2$ as a function of the inner layer thickness relative
to the mean-free path in the three-layered systems for different values of
the diffuse scattering $S$ and different magnetization configurations,
e)-f) resistance $R_4$ as a function of the inner layers thicknesses
relative to the mean-free path in the five-layered systems for different
values of $S$ and different magnetization configurations.}
\label{fecr_Rofd} 
\end{figure} 

\begin{figure} 
\centering
\includegraphics[width=\textwidth]{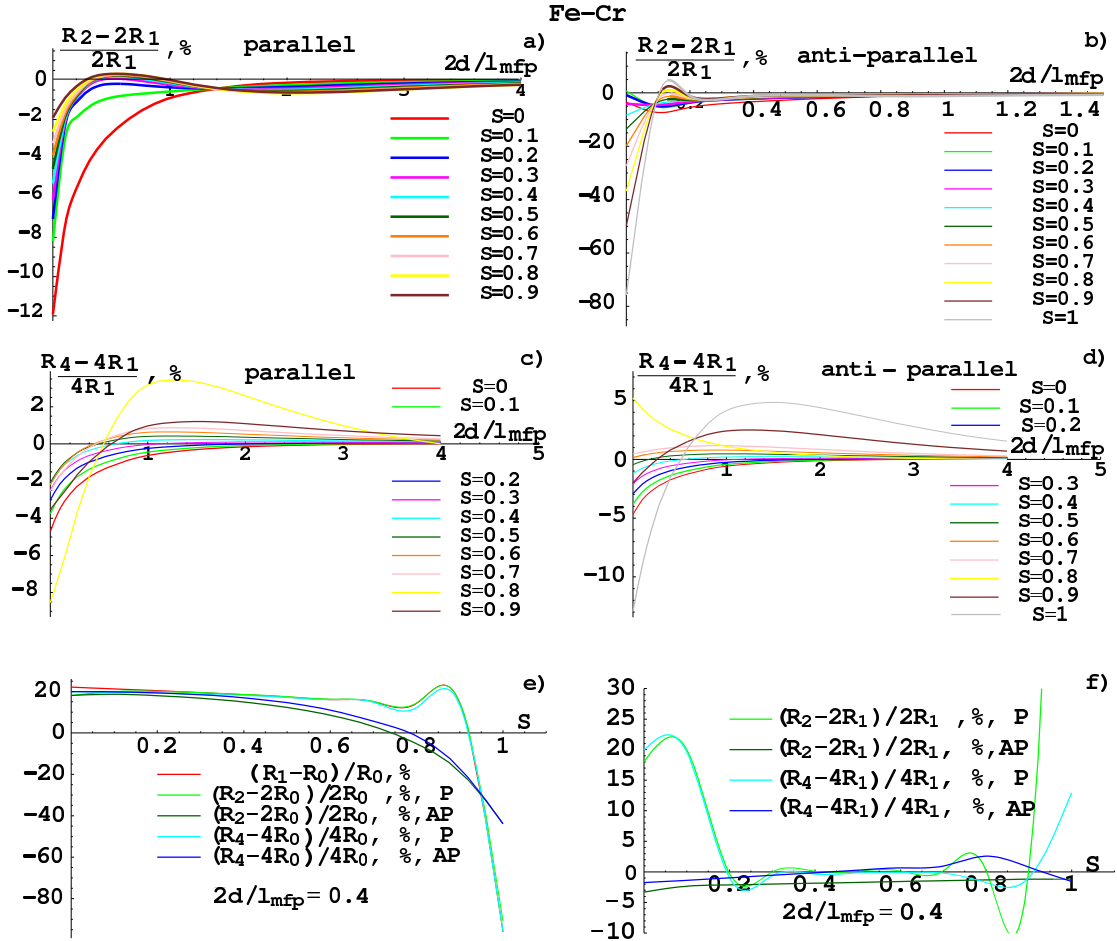}
\caption[Deviations from the resistors-in-series model in the Fe-Cr
systems] {Deviations from the resistors-in-series model in the Fe-Cr
systems: a)-b) the difference between the resistance $R_2$ and twice the
resistance of a single interface $2R_1$ relative to $2R_1$ as a function of
the inner layer thickness $2d$ relative to the mean-free path $l_{mfp}$ in
the three-layered system for different values of $S$ and different
magnetization configurations, c)-d) the difference between the resistance
$R_4$ and four times the resistance of a single interface, $4R_1$, relative
to $4R_1$ as a function of $2d$ relative to $l_{mfp}$ in the five-layered
system for different values of $S$ and different magnetization
configurations, e) the difference between the resistances $R_N$, with
$N=1$, $2$, $4$ and $N$ times the resistance $R_0$ of a single independent
interface measured right at the interface, $NR_0$, relative to $NR_0$, in
the multilayers with $2d \ll l_{mfp}$ as a function of $S$ for different
magnetization configurations, f) the difference between the resistances
$R_N$, with $N=2$, $4$ and $N$ times the resistance $R_1$ of a single
independent interface measured far from the interface, $NR_1$, relative to
$NR_1$, in the multilayers with $2d\ll l_{mfp}$ as a function of $S$ for
different magnetization configurations.  } 
\label{fecr_RNR1} 
\end{figure}

\chapter{\label{chap_swt_thr} Spin-torques - theory}

In the following two chapters, a mechanism of the magnetization
switching of magnetic multilayers driven by a spin-polarized current is
studied. First, I present a spin transfer model in which the equation of
motion of the spin accumulation is solved in order to derive the torque
acting on the background magnetization. Next, a system of two thick
magnetic layers separated by a non-magnetic spacer is considered. The
analytical expressions for the spin-current and spin-accumulation
distributions in the system, and spin-torque and effective field acting
on the local magnetization in the ferromagnetic layers are obtained, and
I discuss the significance of the result for the spin-current at the
interface between layers in the system with magnetization directions
close to anti-parallel. Finally, the correction to the CPP
magnetoresistance due to the interaction of the electron spins and
background magnetization is calculated. In the next chapter, I present
the results for a realistic multilayered structure.

\section{\label{rev_form}Review of formalism} 
The linear response of the current to the electric field in the
$x$-direction for diffusive transport, Eq.~(\ref{curr_lin_resp}), can be
written in a spinor form as~\cite{ZhLF_prl_02}
\begin{equation}
\label{spin_cur_gen}
\hat{\jmath}(x)=\hat{C}E(x)-\hat{D}\frac{\partial \hat{n}}{\partial x},
\end{equation}
where $E(x)$ is the electric field, $\hat{j}$, $\hat{C}$, $\hat{D}$, and 
$\hat{n}$ are the $2\times 2$ matrices representing the current, the 
conductivity, the diffusion constant, and the charge and spin 
accumulations at a given position. The diffusion constant and the 
conductivity are related via the Einstein relation 
$\hat{C}=e^2\hat{N}(\epsilon_F)\hat{D}$ for a degenerate metal, where 
$\hat{N}(\epsilon_F)$ is the density of states at the Fermi level. In 
general, these matrices can be expressed in terms of the Pauli spin 
matrix $\mbox{\boldmath$\sigma$}$ as
\begin{equation}
\left\{
\begin{array}{l}
\label{cdNnj}
\hat{C}=C_0{\hat I}+\mbox{\boldmath$\sigma$}\cdot{\bf C},\nonumber \\
\hat{D}=D_0{\hat I}+\mbox{\boldmath$\sigma$}\cdot{\bf D},\nonumber \\
\hat{N}=N_0{\hat I}+\mbox{\boldmath$\sigma$}\cdot{\bf N}, \\
\hat{n}=n_0{\hat I}+\mbox{\boldmath$\sigma$}\cdot{\bf m},\nonumber \\
\hat{j}=j_e{\hat I}+\mbox{\boldmath$\sigma$}\cdot{\bf j}_m,\nonumber
\end{array}
\right.
\end{equation}
where $2n_0$ is the charge accumulation, $2n_0=n_\uparrow+n_\downarrow$,
and ${\bf m}$ is the spin-accumulation, $2|{\bf
m}|=n_\uparrow-n_\downarrow$, $n_\uparrow$ is the number of electrons
with the spins parallel to the principal quantization axis,
$n_\downarrow$ is the number of electrons with the spins anti-parallel
to the principal quantization axis, and, similarly, electric current
$2j_e=j_\uparrow+j_\downarrow$, magnetization current $2|{\bf 
j}_m|=j_\uparrow-j_\downarrow$. Inserting Eqs.~(\ref{cdNnj}) in 
Eq.~(\ref{spin_cur_gen}), the electric current $j_e$ and magnetization 
current ${\bf j}_m$ can be written as
\begin{equation}
\label{je}
j_{e}\equiv{\rm Re}({\rm Tr}\hat{j})=2C_0E(x)-
2D_0\frac{\partial n_0}{\partial x}-
2{\bf D}\cdot\frac{\partial{\bf m}}{\partial x},
\end{equation}
\begin{equation}
\label{jm}
{\bf j}_{m}\equiv{\rm Re}{\rm Tr}(\mbox{\boldmath$\sigma$}{\hat j})=
2{\bf C}E(x)-2{\bf D}\frac{\partial n_0}{\partial x}
-2D_0\frac{\partial{\bf m}}{\partial x},
\end{equation}
where the units $e=\mu_B=1$ are chosen for the notation convenience. The
first term on the right hand side of Eq.~(\ref{jm}) is the contribution
to the spin polarized current from the electric field, the third term is
the contribution from the spin accumulation, which is present in a
magnetically inhomogeneous structure. The second term, the contribution
coming from the charge accumulation, will be neglected, since its
characteristic lengthscale, of the order of several angstroms, is of the
order of magnitude smaller than other relevant lengthscales.  As will be
shown below, in magnetic multilayers the contribution of the third term
to the spin current can dominate over the first. Comparing Eq.~(\ref{je})
with the expression~(\ref{curr_lin_resp}) for the electric current, the
parameters $C_0$ and $D_0$ can be identified with
$C_0=\frac{1}{2}\sigma=\frac{k_F^3}{6\pi^2}\frac{e^2\tau}{m}$,
$D_0=\frac{1}{2}D=\frac{1}{6}v_F^2\tau=\frac{1}{6}v_F\lambda_{mfp}$,
where $\lambda_{mfp}\equiv v_F\tau$ is the electron mean-free path.

For a transition metal ferromagnet, the spin-polarization parameter for
the conductivity $\beta$, the diffusion constant $\beta^\prime$, and for
the density of states $\beta^{\prime\prime}$ can be defined as ${\bf
C}=\beta C_0{\bf M}_d$, ${\bf D}=\beta^\prime D_0{\bf M}_d$, and ${\bf
N}=\beta^{\prime\prime}N_0{\bf M}_d$, where ${\bf M}_d$ is the unit
vector to represent the direction of the local (background)
magnetization. Similarly to the spin-polarization parameters for the
interface, $\gamma$, $\gamma^\prime$, and $\gamma^{\prime\prime}$,
introduced in the Appendix~\ref{app_bound_cond}, spin-polarization
parameters for the bulk are related as
$\beta^\prime=(\beta-\beta^{\prime\prime})/(1-\beta\beta^{\prime\prime})$
(compare with Eq.~(\ref{gam_prime})). The experimental value for $\beta$
can be found in the Ref.~\cite{BP_jmmm_99}, and the value of
$\beta^{\prime\prime}$ can be found from the bulk densities of states of
up and down electrons, $N_\uparrow$ and $N_\downarrow$, as
$\beta^{\prime\prime}=(N_\uparrow-N_\downarrow)/(N_\uparrow+N_\downarrow)$.
Inserting the expressions for ${\bf C}$ and ${\bf D}$ in Eqs.~(\ref{je})
and~(\ref{jm}), and eliminating the electric field and charge density,
the following expression for the spin-current is obtained:
\begin{equation}
\label{jm_fin}
{\bf j}_m=\beta j_e{\bf M}_{d}
-2D_0\left[\frac{\partial{\bf m}}{\partial x}
-\beta\beta^\prime{\bf M}_d({\bf M}_d\cdot\frac{\partial{\bf m}}{\partial x})
\right].  
\end{equation}

The equation of motion for the spin accumulation may be obtained by
generalizing the continuity equation~(\ref{cont}) to the systems with
both spin-up and spin-down electrons, and including the possibility of
spin-flip scattering. The continuity equation for the spin-up electrons
density $n_\uparrow$ (spin-down electrons density $n_\downarrow$) has
the form
$$
\frac{\partial n_{\uparrow(\downarrow)}}{\partial t}+
\frac{\partial j_{\uparrow(\downarrow)}}{\partial 
x}=-\frac{n_{\uparrow(\downarrow)}-n_{\downarrow(\uparrow)}}{\tau_{sf}},
$$
where $\tau_{sf}$ is the spin-flip relaxation time of the conduction 
electrons, and, for the spin-accumulation,
$$
\frac{\partial|{\bf m}|}{\partial t}+\frac{\partial|{\bf j}_m|}{\partial 
x}=-\frac{2|{\bf m}|}{\tau_{sf}}.
$$
Introducing the term $J/\hbar{\bf m}\times{\bf M}_d$ (see
Chap.~\ref{chap_intro}), the equation of motion for ${\bf m}$ may be
written as
\begin{equation}
\label{full_cont_m}
\frac{\partial{\bf m}}{\partial t}+\frac{\partial{\bf j}_m}{\partial x}
+\frac{J}{\hbar}{\bf m}\times{\bf M}_d=-\frac{2{\bf m}}{\tau_{sf}}.  
\end{equation}
Upon placing the expression for the magnetization current
Eq.~(\ref{jm_fin}) in Eq.~(\ref{full_cont_m}) one finds the following
equation of motion for the spin-accumulation vector:
\begin{eqnarray}
\label{full_diff_eq}
\frac{1}{2D_0}\frac{\partial{\bf m}}{\partial t}-
\frac{\partial^2{\bf m}}{\partial x^2}+\beta\beta^{\prime }{\bf M}_d
\left({\bf M}_d\cdot\frac{\partial^2{\bf m}}{\partial x^2}\right) 
&+&\frac{{\bf m}}{\lambda_{sf}^2}+
\frac{{\bf m}\times{\bf M}_d}{\lambda_J^2} \nonumber \\
&=&-\frac{1}{2D_0}\frac{\partial}{\partial x}(\beta j_e{\bf M}_d),  
\end{eqnarray}
where $\lambda _{sf}\equiv\sqrt{D_0\tau_{sf}}$ and
$\lambda_J\equiv\sqrt{2\hbar D_0/J}$. The term on the right hand side of
the time dependent diffusion equation Eq.~(\ref{full_diff_eq}) for the
spin accumulation is the source term; it is this term that drives the
accumulation.~\cite{ZhL_prb_02} Following the conventional treatment of
magnetic multilayers for current perpendicular to the plane of the
layers, one can assume that the magnetization is uniform throughout a
layer, and changes discontinuously between the
layers.~\cite{Cambl_prb,VF_prb} If one looks for the steady-state
solutions of the equation~(\ref{full_diff_eq}), the first term on the
left-hand side is zero, and the electric current $j_e$ is constant
throughout the multilayer; therefore one can see that the source term is
confined to interfaces between the layers, and can be taken into account
by appropriate boundary conditions. The boundary conditions at the
diffusive interfaces between two layers with noncollinear magnetizations
are addressed in detail in Appendix~\ref{app_bound_cond}. It is shown
that while the spin-accumulation experiences a jump at the interfaces,
proportional to the interface resistance, the spin-current is continuous
across the diffusive interface, and no spin-torque is created at the
interface. The specular scattering at the interfaces can also be taken
into account within the semiclassical framework presented in this work,
as it is done, for example, in Chap.~\ref{chap_res_thr}; this involves
considering each ${\bf k}$ vector separately which for a noncollinear
multilayer is a cumbersome problem, and it will not be addressed here.
Below, the equations for the spin-accumulation and the effect of the
spin-accumulation on the local magnetization in the bulk of the layers
are discussed.

The equation of motion for the local magnetization ${\bf M}_d$ is the 
Landau-Lifshitz-Gilbert equation  
\begin{equation}
\label{LLG}
\frac{d{\bf M}_d}{dt}=-\gamma_0{\bf M}_d\times({\bf H}_e+J{\bf m})
+\alpha{\bf M}_d\times\frac{d{\bf M}_d}{dt},  
\end{equation}
where $\gamma_0$ is the gyromagnetic ratio, ${\bf H}_{e}$ is the
magnetic field including the contributions from the external field,
anisotropy and magnetostatic field, the additional effective field
$J{\bf m}$ is due to coupling between the local moments and the spin
accumulation, and the last term is the Gilbert damping term. Note that
the characteristic time scale for the local moment is
$\gamma_0^{-1}H_e$, of the order of nanoseconds for a magnetic
field of 0.1 T, while the time scale of the spin accumulation is of the
order of $\tau_{sf}$ and $h/J$, i.e., of the order of picoseconds.
Therefore, the background magnetization can be assumed to be fixed on
the time scale of the spin accumulation, and the steady state solution
of the equation~(\ref{full_diff_eq}) can be considered.

By separating the spin accumulation into longitudinal (parallel to the
{\it local} moment) and transverse (perpendicular to the {\it local}
moment) modes, ${\bf m}_{||}$ and ${\bf m}_\perp$, and looking for the
steady state solutions, equation~(\ref{full_diff_eq}) can be written as
\begin{equation}
\label{eq_m_long}
\frac{\partial^2{\bf m}_{||}}{\partial x^2}-
\frac{{\bf m}_{||}}{\lambda_{sdl}^2}=0, 
\end{equation}
where $\lambda_{sdl}=\sqrt{1-\beta\beta ^\prime}\lambda_{sf}$, and 
\begin{equation}
\label{eq_m_trans} 
\frac{\partial^2{\bf m}_\perp}{\partial x^2}-
\frac{{\bf m}_\perp}{\lambda_{sf}^2}-
\frac{{\bf m}_\perp\times{\bf M}_d}{\lambda_J^2}=0.
\end{equation}
Note that the terms longitudinal and transverse refer to the
magnetization in the individual layers, i.e., they are {\it locally }
defined and have no global meaning throughout a noncollinear multilayer. 

The longitudinal accumulation ${\bf m}_{||}$ decays at the length scale of
the spin diffusion length $\lambda_{sdl}$, while the transverse spin
accumulation ${\bf m}_{\perp }$ decays as $\lambda_J$ if one assumes
$\lambda_J\ll\lambda_{sf}$. This assumption is valid for cobalt, for
example. where the spin-diffusion length $\lambda_{sdl}$ has been measured
to be about 60 nm, $\beta$ is 0.5, $\lambda_{mfp}$ is about 6
nm,~\cite{BP_jmmm_99} $v_F=0.6\cdot 10^6$ m/s, $J=0.3$
eV.\cite{antr_unpub} I estimate $\beta^\prime$ to be about 0.9 using the
bulk densities of states for up and down electrons of 2.42 states/(atom
Ry) and 13.42 states/(atom Ry)~\cite{Wang}, so that $\lambda_{sf}=80$ nm,
$\lambda_J=2$ nm, $\lambda_J\ll\lambda_{sf}$, and the transverse spin
accumulation has a much shorter length scale compared to the longitudinal
one. However, for permalloy, $\lambda_{sdl}$ has been measured to be about
5 nm, $\beta$ is 0.7,~\cite{BP_jmmm_99} $\beta^\prime$ is estimated to be
about 0.95 using the bulk densities of states for up and down electrons of
2.4 states/Ry and 15 states/Ry~\cite{Mazin,Nadg}, so that $\lambda_{sf}$
is about 9 nm. Taking the typical diffusion constant of a metal to be
$D_0=10^{-3} $m$^2$/s, and $J=0.1$ eV,~\cite{Coop}, $\lambda_J$ can be
estimated to be 4 nm, comparable with $\lambda_{sf}$. The majority of the
results presented in this work are obtained in the limit
$\lambda_J\ll\lambda_{sf}$, so that they are applicable to cobalt. The
structure consisting of two semi-infinite permalloy layers divided by a
nonmagnetic spacer is considered separately below.

As seen from Eq.~(\ref{LLG}), the longitudinal spin accumulation has no
effect on the local moment; therefore, one can re-write Eq.~(\ref{LLG})
in terms of the transverse spin accumulation only by replacing ${\bf m}$
by ${\bf m}_{\perp}$:
\begin{equation}
\label{LLG_perp}
\frac{d{\bf M}_d}{dt}=
-\gamma_0{\bf M}_d\times({\bf H}_e+J{\bf m}_\perp)
+\alpha{\bf M}_d\times\frac{d{\bf M}_d}{dt}.
\end{equation} 
In order to describe the transverse accumulation, it is convenient to 
introduce 
an auxiliary vector ${\bf A}$ such that $J{\bf m}_\perp={\bf 
A}\times{\bf M}_d$.~\cite{ZhLF_prl_02} If one considers a system with 
two noncollinear ferromagnetic layers, the spin accumulation in one layer depends on 
the orientation of the other. Supposing that the above equation is 
used for the layer F1 (see Fig.~\ref{pic_multi}), the local 
magnetization of this layer is labeled as ${\bf M}_d^{(1)}$, and the 
magnetization of the other layer is labeled as ${\bf M}_d^{(2)}$. 
Without loss of generality, the vector ${\bf A}$ can be written as a 
linear combination of the vector ${\bf M}_d^{(2)}$, and the vector 
perpendicular to ${\bf M}_d^{(2)}$, for example, ${\bf A}=a{\bf 
M}_d^{(2)}-b{\bf M}_d^{(2)}\times{\bf M}_d^{(1)}$, so that 
the two components of the accumulation in 
the plane perpendicular to ${\bf M}_d^{(1)}$ are written as 
\begin{equation}
\label{def_a_b}
J{\bf m}_\perp=a{\bf M}_d^{(2)}\times{\bf M}_d^{(1)}
+b{\bf M}_d^{(1)}\times({\bf M}_d^{(2)}\times{\bf M}_d^{(1)}),
\end{equation}
where $a$ and $b$ are determined by geometric details of the
multilayer. By placing this form of the accumulation in the equation of
motion for the background magnetization, Eq.~(\ref{LLG_perp}), one finds
the following equation of motion for $\bf{M}_{d}^{(1)}$:
\begin{equation}
\label{LLG_fin}
\frac{d{\bf M}^{(1)}_d}{dt}=
-\gamma_0{\bf M}^{(1)}_d\times({\bf H}_e+b{\bf M}^{(2)}_d)
-\gamma_0a{\bf M}^{(1)}_d\times({\bf M}_d^{(2)}\times{\bf M}_d^{(1)})
+\alpha{\bf M}_d\times\frac{d{\bf M}_d}{dt}.
\end{equation}
Thus the transverse spin accumulation produces two effects
simultaneously: the term $b{\bf M}_d^{(1)}\times{\bf M}_d^{(2)}$ is the
torque due to an "effective field" $b{\bf M}_d^{(2)}$, and the other
is $a{\bf M}_d^{(1)}\times({\bf M}_d^{(2)}\times{\bf M}_d^{(1)})$ is
called the "spin torque" predicted in the
Refs.~\cite{Sl_jmmm_96,Sl_jmmm_99,Sl_jmmm_02,Berg_prb_96, Berg_jap_01}.
The first term produces a precessional motion about $\bf{M}_{d}^{(1)}$;
in this sense it acts {\it as if} the spin current creates a magnetic
field on $\bf{M}_{d}^{(1)}$ (see Fig.~\ref{pic_motion}). The second term
acts so as to increase or decrease the angle between $\bf{M}_{d}^{(1)}$
and $\bf{M}_{d}^{(2)}$; also, it acts so as to assist or oppose the
damping term in Eq.~(\ref{LLG}). Note that both the spin-torque and
effective field terms appear on an equal footing in the
equation~(\ref{LLG_fin}), as both are related to the transverse spin
accumulation.
\begin{figure}
\centering
\includegraphics[width=5in]{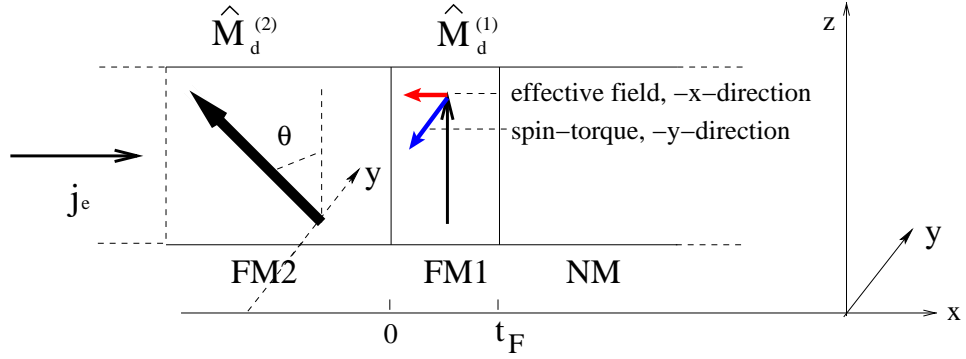}
\caption[Direction of the effective field and spin-torque acting on a 
thin FM layer background magnetization]{Direction of the effective field 
(red vector) and spin-torque (blue vector) acting on a
thin FM layer background magnetization ${\bf M}_d^{(1)}={\bf e}_z$, if 
the magnetization of the thick layer is
${\bf M}_d^{(2)}=\cos\theta{\bf e}_z-\sin\theta{\bf e}_y$.
}
\label{pic_motion}
\end{figure}

The torque transmitted by the current to the background may also be
determined from the following considerations. The torque $\tau$ acting on
the background angular momentum ${\bf M}_d$ is the time derivative of
${\bf M}_d$, $\tau=d{\bf M}_d/dt$. The {\it total} angular momentum of
the system, which is the sum of the electron angular momentum ${\bf m}$
and the background angular momentum ${\bf M}_d$ is conserved, so that 
${\bf m}+{\bf M}_d=const$. The torque acting on the background 
magnetization can be written in terms of the electron spin accumulation 
and spin current as 
$$
\tau=-\frac{d{\bf m}}{dt}=-\frac{\partial{\bf m}}{\partial t}-
\frac{\partial{\bf j}_m}{\partial x},
$$
so that the torque transmitted by a steady-state current is given by the
{\it gradient} of the spin current. Integrating this over a layer with
the thickness $d$ which absorbs the momentum, one finds
\begin{equation}
\label{tot_torque}
\Delta\tau=-\int_0^d\frac{\partial{\bf j}_m}{\partial x}dx=
{\bf j}_m(0)-{\bf j}_m(d).
\end{equation}
The total torque absorbed by the layer is the difference between the spin
current values at the boundaries of the layer. 

\section{Two ferromagnetic layers} 
In order to illustrate the method developed in 
Sec.~\ref{rev_form} to determine the effective field and spin-torque
acting on the ferromagnetic layer, I choose a simplified system where I
can analytically derive the spin accumulation, spin-current, effective
field, and spin-torque. I consider a system consisting of two thick
ferromagnetic layers with the magnetizations ${\bf
M}_d^{(1)}=\cos\theta{\bf e}_z+\sin\theta{\bf e}_y$ and ${\bf
M}_d^{(2)}=\cos\theta{\bf e}_z-\sin\theta{\bf e}_y$, where ${\bf x}$ is
the direction of the electric current (see Fig.~\ref{pic_two}),
separated by a nonmagnetic spacer layer so that there is no exchange
coupling between ferromagnetic layers, and backed by nonmagnetic layers,
or leads. The nonmagnetic spacer layer thickness is considered to be
small compared to the spin-diffusion length in the nonmagnetic metal
($\lambda_{sdl}^N\sim 600$ nm), so that both spin accumulation and spin
current are constant in the spacer, and I will use both terms "spacer"
and "interface between magnetic layers" interchangeably. The thickness
of the ferromagnetic layers $t_F$ is large enough so that one can
neglect the reflections from the outer boundaries of the layers. The
difference between this system and the system considered in the
Ref.~\cite{ZhLF_prl_02} is that in the latter, the second layer is
assumed to be half metallic, so that the current is fully
spin-polarized, while here this approximation is abandoned.
\begin{figure}
\centering
\includegraphics[width=5in]{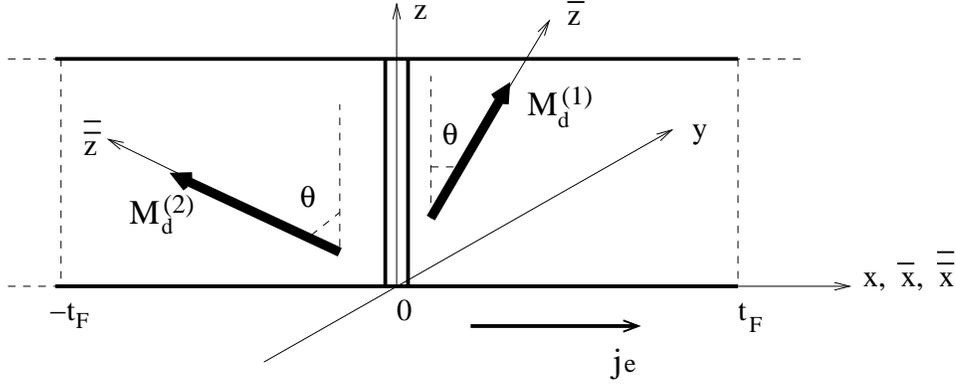}
\caption[System of two thick ferromagnetic layers]{System of two thick
ferromagnetic layers separated by a spacer layer and backed by
nonmagnetic layers, or leads. Magnetization of the first (right) layer is
${\bf M}_d^{(1)}=\cos\theta{\bf e}_z+\sin\theta{\bf e}_y$, magnetization
of the left (second) layer is ${\bf M}_d^{(2)}=\cos\theta{\bf
e}_z-\sin\theta{\bf e}_y$. ($x,y,z$) is the global coordinate system,
($\bar x,\bar y,\bar z$) and (${\bar{\bar x}},{\bar{\bar y}},{\bar{\bar
z}}$) are the local coordinate systems connected to the local
magnetizations in the first and second layers.}
\label{pic_two}
\end{figure}

I solve Eqs.~(\ref{eq_m_long}) and~(\ref{eq_m_trans}) to find the
longitudinal and transverse spin accumulation in both ferromagnetic
layers, and use Eq.~(\ref{jm_fin}) to find the spin-current. In the
local coordinate system ($\bar x,\bar y,\bar z$) connected to the local
magnetization of the right layer, the spin-accumulation and spin current
take the following form:
\begin{equation}
\label{m_loc_1}
\left\{
\begin{array}{l}
m_{\bar z}^{(1)}=G_1\exp(-x/\lambda_{sdl}) \nonumber \\
m_{\bar x}^{(1)}=2{\rm Re}(G_2\exp(-x/l_+))\\
m_{\bar y}^{(1)}=2{\rm Im}(G_2\exp(-x/l_+)), \nonumber
\end{array}
\right.
\end{equation}
\begin{equation}
\label{jm_loc_1}
\left\{
\begin{array}{l}
j_{m,{\bar z}}^{(1)}=\beta 
j_e+\frac{2D_0(1-\beta\beta^\prime)}{\lambda_{sdl}}G_1\exp(-x/\lambda_{sdl})
\nonumber \\ \\
j_{m,{\bar x}}^{(1)}=4D_0{\rm Re}\left(G_2\frac{\exp(-x/l_+)}{l_+}\right) 
\\ \\
j_{m,{\bar y}}^{(1)}=4D_0{\rm Im}\left(G_2\frac{\exp(-x/l_+)}{l_+}\right), 
\end{array}
\right.
\end{equation}
where $G_1$ and $G_2$ are constants of integration, and 
\begin{equation}
\label{l+}
\frac{1}{l_+}=\sqrt{\frac{1}{\lambda_{sf}^2}-\frac{i}{\lambda_J^2}}
\approx\frac{1-i}{\sqrt 2\lambda_J}
\end{equation}
when $\lambda_J\ll\lambda_{sf}$. Similar expressions for the local 
spin-accumulation and spin-current are written for the second layer. 

To determine the constants of integration, the boundary conditions at
the interface between two ferromagnetic layers have to be invoked. The
boundary conditions are discussed in detail in
Appendix~\ref{app_bound_cond}. At a diffusive interface, the spin-current
is conserved, and the spin-accumulations to the left and to the right of
the interface in the global ($x,y,z$) coordinate system are related as 
follows:
\begin{eqnarray}
\label{bc_m}
\left\{
\begin{array}{c}
\nonumber
m_x(0-)-m_x(0+)=2j_{m,x}(0)r\frac{\lambda_J}{\sqrt 2D_0} \\ \\
m_y(0-)-m_y(0+)=2j_{m,x}(0)\left(\cos^2\theta+
\frac{\sin^2\theta}{(1-\gamma\gamma^\prime)}\right)r\frac{\lambda_J}{\sqrt 
2D_0} \\ \\ \nonumber 
m_z(0-)-m_z(0+)=-\gamma j_e\cos\theta r\frac{\lambda_J}{\sqrt 
2(1-\gamma\gamma^\prime)D_0}+
2j_{m,x}(0)\left(\sin^2\theta+\frac{\cos^2\theta}{(1-\gamma\gamma^\prime)}\right)
r\frac{\lambda_J}{\sqrt 2D_0}, 
\end{array}
\right.
\end{eqnarray} 
where $r$ is proportional to the interface resistance $AR_I$ (see 
Eq.~(\ref{res})), $\gamma$ and $\gamma^\prime$ are the spin polarization 
parameters for the conductivity and the diffusion constant at the 
interface (see Eq.~(\ref{gam_gam_prime})). Note that the 
spin-accumulation is conserved across the interface if there is no 
diffuse scattering at the interface ($AR_I=0$). Vectors in the 
local and global coordinate systems are related via the rotation matrix 
(see Eq.(\ref{rot_matr})), and the constants of integration are
\begin{equation}
\label{G1}
G_1=-\frac{\beta j_e}{\sqrt 2\lambda_J J}\frac{\hbar a_0^3}{e\mu_B}\times
\frac{\sin^2\theta(1+{\bar r})+\frac{{\bar r}(1-\gamma/\beta)}
{1-\gamma\gamma^\prime}\cos^2\theta}
{\cos^2\theta(1+\frac{{\bar r}\lambda}{1-\gamma\gamma^\prime})+
\lambda\sin^2\theta(1+{\bar r})} 
\end{equation}
\begin{equation}
\label{G2}
G_2=\frac{\beta j_e}{2\sqrt 2\lambda_J J}\frac{\hbar a_0^3}{e\mu_B}\times
\frac{(1+{\bar r}t\lambda)\sin 2\theta}
{\cos^2\theta(1+\frac{{\bar r}\lambda}{1-\gamma\gamma^\prime})+
\lambda\sin^2\theta(1+{\bar r})}(1-i),
\end{equation}
where $\lambda=(1-\beta\beta^\prime)\lambda_J/\sqrt 2\lambda_{sdl}$,
$t=\gamma/\beta(1-\gamma\gamma^\prime)$, ${\bar r}=\sqrt
2AR_Ie^2N_0^I(\epsilon_F)D_0(1-\gamma^{\prime\prime
2})/\lambda_J(1-\gamma\gamma^\prime)$, $AR_I$ is the resistance of the
interface between two ferromagnetic layers, $\beta$, $\beta^\prime$ are
spin-polarization parameters for conductivity and diffusion constant in
the bulk of the layers, $\gamma$, $\gamma^\prime$, $\gamma^{\prime\prime
}$ are spin-polarization parameters for conductivity, diffusion constant,
and electron density of states at the interface; $N_0^I(\epsilon_F)$ is
the density of states at the interface at Fermi energy (see
Appendix~\ref{app_bound_cond}). The expressions for the spin-accumulation
in the right layer are
\begin{eqnarray}
\label{m_glob_1}
\left\{
\begin{array}{l}
m_x^{(1)}=2e^{-\frac{x}{\sqrt 2\lambda_J}}{\rm Re}G_2[\cos\frac{x}{\sqrt 
2\lambda_J}+\sin\frac{x}{\sqrt 2\lambda_J}] \\ 
m_y^{(1)}=-2e^{-\frac{x}{\sqrt 2\lambda_J}}{\rm Re}G_2[\cos\frac{x}{\sqrt
2\lambda_J}-\sin\frac{x}{\sqrt 
2\lambda_J}]\cos\theta+G_1e^{-\frac{x}{\lambda_{sdl}}}\sin\theta \\
m_z^{(1)}=2e^{-\frac{x}{\sqrt 2\lambda_J}}{\rm Re}G_2[\cos\frac{x}{\sqrt
2\lambda_J}-\sin\frac{x}{\sqrt
2\lambda_J}]\sin\theta+G_1e^{-\frac{x}{\lambda_{sdl}}}\cos\theta. 
\end{array}
\right.
\end{eqnarray}
In the left layer they are 
\begin{eqnarray}
\label{m_glob_2}
\left\{
\begin{array}{l}
m_x^{(2)}(x)=m_x^{(1)}(-x)  \nonumber \\
m_y^{(2)}(x)=m_y^{(1)}(-x) \\ 
m_z^{(2)}(x)=-m_z^{(1)}(-x). \nonumber
\end{array}
\right.
\end{eqnarray}
The spin-current in the layers can be found from Eq.~(\ref{jm_fin}). The
expressions for the components of ${\bf j}_m$ are complicated, and I
will only present the values of the $x-$, $y-$, and $z-$components of
${\bf j}_m$ at $x=0$, in the nonmagnetic spacer, in the absence of the
scattering at the interface. Due to the symmetry of the problem, only
the $z$-component is non zero, so that
\begin{equation}
\label{jm_x0}
{\bf j}_m(x=0)=\frac{\beta j_e\cos\theta}
{\cos^2\theta+\lambda\sin^2\theta}{\bf e}_z,
\end{equation}
where $\lambda=\sqrt{1-\beta\beta^\prime}\lambda_J/\sqrt
2\lambda_{sf}$, and it is of the order of 0.02 for cobalt. The spin 
current at the interface reaches its maximum value of 
\begin{equation}
\label{j_m_max}
j_{m,max}(x=0)=\frac{\beta j_e\cos\theta^\star}{2\lambda}
\end{equation}
when the angle between the local magnetizations is $2\theta^\star$, where 
\begin{equation}
\label{theta_star}
\cos2\theta^\star=-\frac{1-3\lambda}{1-\lambda},
\end{equation}  
and $2\theta^\star$ is close to $\pi$ when $\lambda\ll 1$. The magnitude
of the spin-current is enhanced by a large factor of $\lambda^{-1}$
compared to the bare spin-current $\beta j_e\cos\theta$.  This
enhancement comes from the interplay between longitudinal and transverse
accumulations; it is the result of the global nature of the spin-current
even though the transverse components of the spin current and
accumulation are absorbed within a region of several $\lambda_J$ of the
interface. While the total spin torque acting on a symmetric two-layered
structure is zero, the result of Eq.~(\ref{jm_x0}) can serve as an
indication that in the non-symmetric structure when one layer is pinned
and the other free; the spin-torque acting on the free layer, and,
hence, the angular momentum transferred to it, exceeds the transverse
component of the bare portion of the incoming spin-current by
$1/2\lambda$. This prediction will be tested in the next chapter, where
the non-symmetric three-layered system is discussed.

The spin-torque $a_1$ and the effective field $b_1$ acting on the
ferromagnetic layer can be found from the following considerations.  From
Eq.~(\ref{m_loc_1}), the transverse spin-accumulation in the right layer
can we written as
\begin{equation}
\label{m_perp_G}
{\bf m}_\perp^{(1)}={\bf m}_{\bar x}^{(1)}{\bf e}_{\bar x}+
{\bf m}_{\bar y}^{(1)}{\bf e}_{\bar y}=
2{\rm Re}(G_2\exp(-x/l_+)){\bf e}_{\bar x}+2{\rm Im}(G_2\exp(-x/l_+)){\bf 
e}_{\bar y},
\end{equation}
where ${\bf e}_{\bar x}$ and ${\bf e}_{\bar y}$ are unit vectors in the
{\it local} coordinate system. By noticing that $G_2$ can be written as
$\tilde{G_2}\sin 2\theta$, ${\bf M}^{(2)}_d\times{\bf M}^{(1)}_d=-\sin
2\theta{\bf e}_{\bar x}$, ${\bf M}^{(1)}_d\times({\bf M}^{(2)}_d\times{\bf
M}^{(1)}_d)= -\sin 2\theta{\bf e}_{\bar y}$, one can write ${\bf
m}_\perp^{(1)}$ as
\begin{eqnarray}
\label{m_perp_Gtilde}
{\bf m}_\perp^{(1)}&=&-2{\rm Re}({\tilde G_2}\exp(-x/l_+)){\bf 
M}^{(2)}_d\times{\bf M}^{(1)}_d \nonumber \\
&-&2{\rm Im}({\tilde G_2}\exp(-x/l_+)){\bf M}^{(1)}_d\times({\bf 
M}^{(2)}_d\times{\bf
M}^{(1)}_d).
\end{eqnarray}
One can see that the form of the transverse spin-accumulation given by 
Eq.~(\ref{m_perp_Gtilde}) is exactly the form used in the definition of 
$a$ and $b$ (Eq.(\ref{def_a_b})):
\begin{equation}
\label{def_a_b_rep}   
J{\bf m}_\perp=a{\bf M}_d^{(2)}\times{\bf M}_d^{(1)}
+b{\bf M}_d^{(1)}\times({\bf M}_d^{(2)}\times{\bf M}_d^{(1)}).
\end{equation}
To obtain the coefficients $a$ and $b$, one averages the 
spin-accumulation~(\ref{m_perp_Gtilde}) over the thickness $t_F$ of the 
right layer. The spin-torque and effective field {\it per unit length} 
acting on the ferromagnetic layer take the form
\begin{equation}
\label{a_result}
a_1=-
\frac{\beta j_e}{2\sqrt 2\lambda_J}\frac{\hbar a_0^3}{e\mu_B}\times
\frac{1-\cos(\frac{t_F}{\sqrt 2\lambda_J})e^{-\frac{t_F}{\sqrt 
2\lambda_J}}}{t_F/(\sqrt 2\lambda_J)}\times
\frac{(1+{\bar r}t\lambda)}
{\cos^2\theta(1+\frac{{\bar r}\lambda}{1-\gamma\gamma^\prime})+
\lambda\sin^2\theta(1+{\bar r})},
\end{equation}
\begin{equation}
\label{b_result}
b_1=
\frac{\beta j_e}{2\sqrt 2\lambda_J}\frac{\hbar a_0^3}{e\mu_B}\times
\frac{\sin(\frac{t_F}{\sqrt 2\lambda_J})e^{-\frac{t_F}{\sqrt
2\lambda_J}}}{t_F/(\sqrt 2\lambda_J)}\times
\frac{(1+{\bar r}t\lambda)}
{\cos^2\theta(1+\frac{{\bar r}\lambda}{1-\gamma\gamma^\prime})+
\lambda\sin^2\theta(1+{\bar r})}.
\end{equation}
Note that in order to get the true torque and field acting on the 
ferromagnetic layers, the coefficients $a$ and $b$ has to be multiplied by 
$\sin2\theta$ (see Eq.~(\ref{def_a_b_rep})).

Both spin-torque $a\sin2\theta$ and effective field $b\sin2\theta$ reach 
their maximum value when $\theta=\theta^{\star\star}$ where
\begin{equation}
\label{theta_star_star}
\cos2\theta^{\star\star}=-\frac{1-\lambda}{1+\lambda}
\end{equation}
in the absence of the interface resistance; we see that 
$\theta^{\star\star}$ is close to $90^\circ$. 

In the second layer, the spin-torque $a_2$ is equal to $-a_1$, and the
effective field $b_2$ is equal to $-b_1$ due to the symmetry of the
problem.

In Fig.~\ref{m_j_all} I present the spin accumulation and spin current
distribution in the FM-Sp-FM structure with $\lambda_J=4$ nm,
$\lambda_{sdl}=60$ nm in the presence of the interface resistance
$AR_I=0.5$ f$\Omega\cdot$m$^2$;~\cite{BP_jmmm_99} the diffusion constant
is taken to be 10$^{-3}$ m$^2$/s. Both ${\bf m}$ and ${\bf j}_m$ are
constant in the spacer, so I don't show the spacer in these plots. As
follows from Eq.~(\ref{m_glob_1}), the characteristic length scale of the
x-component of the accumulation, perpendicular to the plane of the layer
magnetizations, is $\lambda_J$. {\it Global} y- and z- components of spin
accumulation are defined by both transverse and longitudinal {\it local}
components of the accumulation, so for them, two length scales may be
distinguished: one of the order of $\lambda_J$ for a more rapid change of
the spin accumulation, and the longer one, of the order of the
spin-diffusion length $\lambda_{sdl}$. Only the z-component of the spin
accumulation is discontinuous in the presence of the interface resistance
due to the symmetry of the problem. Far from the interfaces,
$x\gg\lambda_J$ and $x\ll-\lambda_J$, the spin current is collinear with
background magnetization ${\bf M}_d^{(1)}$ or ${\bf M}_d^{(2)}$,
approaching its bare values ${\bf j}_m=\beta j_e{\bf M}_d^{(1)}$ or ${\bf
j}_m=\beta j_e{\bf M}_d^{(2)}$. As follows from the
equation~(\ref{jm_x0}), a large amplification of z-component of the
spin-current occurs within a distance of several $\lambda_J$ from the
interface.
\begin{figure}
\centering
\includegraphics[width=\textwidth]{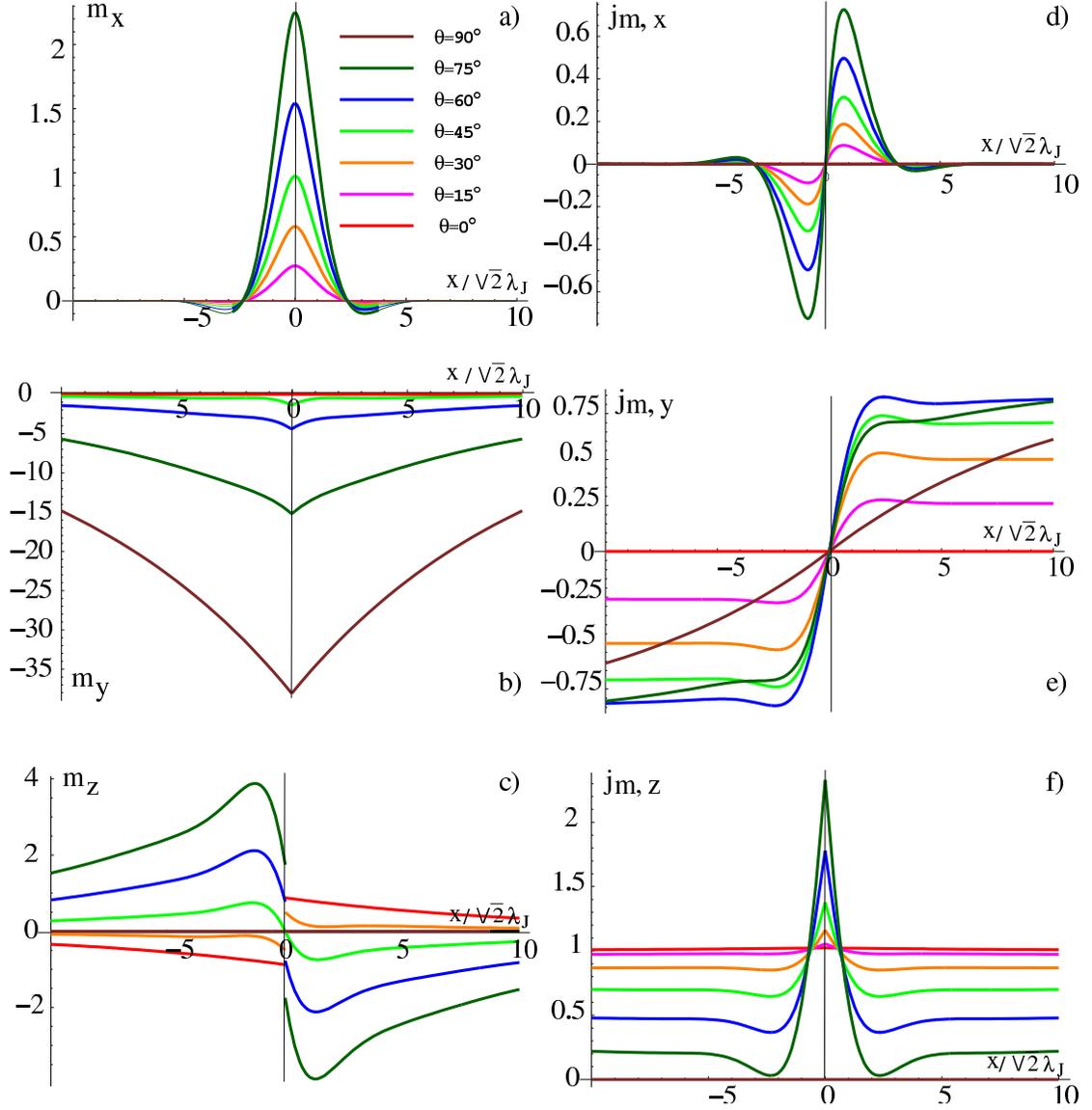}
\caption[Spin accumulation and spin current distribution in two-layered 
system]{$x$-, $y$-, and
$z$-components of the spin-accumulation ${\bf m}/(\beta
j_e/\sqrt 2 \lambda_J J)(\hbar a_0^3/e\mu_B)$ (a-c) and spin-current ${\bf  
j}_m/\beta j_e$ (d-f) distribution in FM-Sp-FM system in the presence 
of interface resistance for $\lambda_J=4$ nm, $\lambda_{sdl}=60$ nm and 
different angles $\theta$.}
\label{m_j_all}   
\end{figure}

In Fig.~\ref{a_b_all} I show the spin-torque and effective field per
unit length acting on the first ferromagnetic layer as a function of the
thickness of the layers $t_F$ relative to $\sqrt 2\lambda_J$.  One can
see that while the torque and field terms $a$ and $b$
(Fig.~\ref{a_b_all}a) and c)) are largest for $\theta=90^\circ$, they
don't act on the background magnetization, since $\sin2\theta=0$. The
true torque and field $a\sin2\theta$ and $b\sin2\theta$
(Fig.~\ref{a_b_all}b) and d)) are largest for $\theta$ close to
$90^\circ$ (see Eq.~(\ref{theta_star_star})). It is worth noting that
while the effective field decreases much faster then the spin torque as
the thickness of the layers increases, at their maximum, they have the
same magnitude, so they indeed have to appear on an equal footing in the
equation of motion for the local magnetization.
\begin{figure}
\centering
\includegraphics[width=\textwidth]{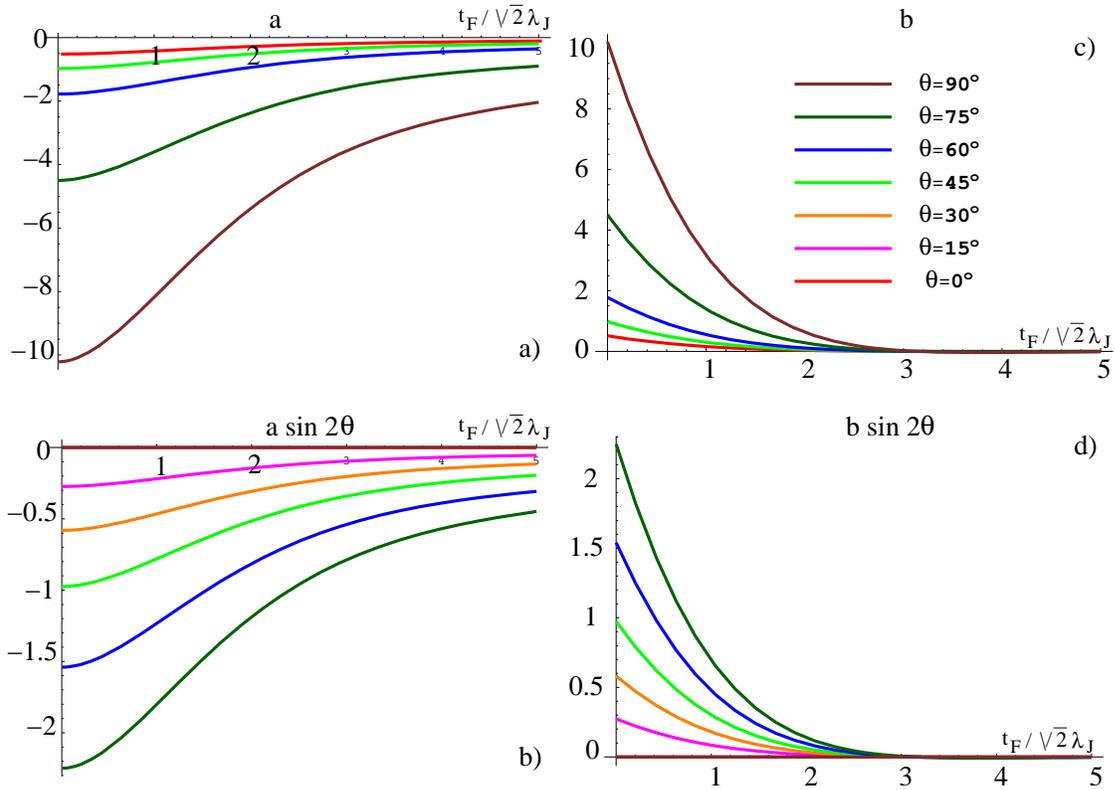}
\caption[Spin-torque and effective field acting on the FM 
layer in FM-Sp-FM system as a function of the layer thickness]{Spin-torque 
$a/\beta j_e(\hbar a_0^3/e\mu_B)$  and (b) $a\sin2\theta/\beta j_e(\hbar 
a_0^3/e\mu_B)$, (c) effective field $b/\beta j_e(\hbar a_0^3/e\mu_B)$  
and (d) $b\sin2\theta/\beta j_e(\hbar 
a_0^3/e\mu_B)$ acting on the FM layer in FM-Sp-FM system as a 
function of $t_F/\sqrt 2\lambda_J$ in the presence of interface 
resistance for $\lambda_J=4$ nm, $\lambda_{sdl}=60$ nm and different 
angles $\theta$.}
\label{a_b_all}
\end{figure}

\section{Correction to CPP resistance} 
The normalized angular dependence of the resistance can be defined 
as~\cite{Pratt_priv}
\begin{equation}
\label{r_norm_def}
R_{norm}=\frac{R(2\theta)-R(0)}{R(2\pi)-R(0)},
\end{equation}
and the data was fit to
\begin{equation}
\label{r_norm_xi}
R_{norm}=\frac{1-\cos^2(\theta)}{1+\chi\cos^2(\theta)}.
\end{equation}

I calculate the angular dependence of the CPP resistance based on the 
spin currents that are found using the diffusion equation for the spin 
accumulation in noncollinear structures as follows. 

The resistivity of a system is a proportionality coefficient between
electric field and electric current, $E=\rho j_e$. From Eq.~(\ref{je}),
electric current can be written in terms of the longitudinal spin
accumulation ${\bf m}_{||}$ as
\begin{equation}
\label{je_mparr}
j_e=2C_0E-2D_0\beta^\prime\frac{\partial m_{||}}{\partial x},
\end{equation}
where in the right layer $m_{||}\equiv m_{\bar 
z}=G_1\exp{(-x/\lambda_{sdl})}$, in the left layer  
$m_{||}\equiv m_{\bar{\bar z}}=-G_1\exp{(x/\lambda_{sdl})}$ (see 
Eq.~(\ref{m_loc_1})), $G_1$ is defined by Eq.~(\ref{G1}). The 
resistivity of the first (right) and the second (left) layers then 
takes the form:
\begin{equation}
\label{rho}
\rho_{1,2}=\frac{1-\frac{2D_0\beta^\prime}{\lambda_{sdl}}
\frac{G_1}{j_e}\exp{(\mp x/\lambda_{sdl})}}{2C_0}.
\end{equation}
The sheet resistance of the whole system can be found by integrating the 
resistivity over the $x$-coordinate from $-t_F$ to $t_F$, and adding the 
interface resistance $AR_I$; I find
\begin{equation}
\label{AR}
AR(\theta)=\frac{t_F}{C_0}\left(1+
\frac{\beta\beta^\prime\lambda}{(1-\beta\beta^\prime)}
\frac{(1+{\bar r})(1-\gamma\gamma^\prime)\sin^2\theta+
{\bar r}(1-\gamma/\beta)\cos^2\theta}
{(1-\gamma\gamma^\prime+{\bar r}\lambda)\cos^2\theta+
\lambda(1+{\bar r})(1-\gamma\gamma^\prime)\sin^2\theta}\right)+AR_I,
\end{equation}
where $\theta$ is half of the angle between magnetizations. The 
parameter $\chi$ then takes the following form:
\begin{equation}
\label{chi}
\chi=-1+\frac{1}{\lambda(1+{\bar r})}
+\frac{{\bar r}}{(1+{\bar r})(1-\gamma\gamma^\prime)},
\end{equation} 
where $\lambda=(1-\beta\beta^\prime)\lambda_J/(\sqrt 2\lambda_{sdl})$,
${\bar r}=\sqrt 2AR_Ie^2N_0^I(\epsilon_F)D_0(1-\gamma^{\prime\prime
2})/\lambda_J(1-\gamma\gamma^\prime)$, $e$ is the electron charge, $N_0^I$
is the density of states at the interface, $D_0$ is the diffusion constant
in the bulk, $\gamma$, $\gamma^\prime$, $\gamma^{\prime\prime}$ are the
spin-polarization parameters for the conductivity, diffusion constant, and
density of states at the interface (see Appendix~\ref{app_bound_cond}).
For cobalt, I estimate $\chi$ to be about 60 if I take $\lambda_J=2$ nm,
$\lambda_{sdl}=60$ nm, $D_0=10^{-3}$m$^2$/s, $\beta=0.5$, $\gamma=0.75$,
$N_\uparrow=2.42$ states/(atom Ry), $N_\downarrow=13.42$ states/(atom Ry)
for the bulk, $N_\uparrow=$1.83 states/(atom Ry), $N_\downarrow=17.75$
states/(atom Ry) for the interface (with Cu), so that $\beta^\prime=0.9$,
$\gamma^\prime=0.97$, $N_0^I(\epsilon_F)=6.5\cdot$10$^{28}$ 
states/m$^3$eV. To my knowledge, there is no experimental data for the 
value of $\chi$ for cobalt, so my result can not be compared with the 
experimental one. Below, I present a calculation for permalloy, for which 
an experimental value of $\chi$ exists.\cite{Pratt_priv} 

\section{Permalloy}
For permalloy, the approximation $\lambda_J\ll\lambda_{sf}$ is not valid, 
so one has to use the exact expression for $l_+$ (see Eq.~(\ref{l+})),
\begin{equation}
\label{l+_2}
\frac{1}{l_+}=\sqrt{\frac{1}{\lambda_{sf}^2}-\frac{i}{\lambda_J^2}}=
\frac{c-id}{\sqrt 2\lambda_J}
\end{equation}
where 
$c=\sqrt{(\lambda_J/\lambda_{sf})^2+\sqrt{(\lambda_J/\lambda_{sf})^4+1}}$, 
$d=\sqrt{-(\lambda_J/\lambda_{sf})^2+\sqrt{(\lambda_J/\lambda_{sf})^4+1}}$. 
Note that for cobalt $c\approx d\approx 1$.
In the absence of the interface resistance, I obtain the 
following analytical expressions for spin-accumulation distribution, 
spin-torque, and effective field:
\begin{eqnarray}
\label{m_glob_perm} 
\left\{
\begin{array}{l}
m_x^{(1)}=2e^{-\frac{cx}{\sqrt 2\lambda_J}}[{\rm Re}G_2\cos\frac{dx}{\sqrt 
2\lambda_J}-{\rm Im}G_2\sin\frac{dx}{\sqrt 2\lambda_J}] \nonumber \\
m_y^{(1)}=-2e^{-\frac{cx}{\sqrt 2\lambda_J}}[{\rm 
Re}G_2\cos\frac{dx}{\sqrt
2\lambda_J}+{\rm Im}G_2\sin\frac{dx}{\sqrt
2\lambda_J}]\cos\theta+G_1e^{-\frac{x}{\lambda_{sdl}}}\sin\theta \\
m_z^{(1)}=2e^{-\frac{cx}{\sqrt 2\lambda_J}}[{\rm Re}G_2\cos\frac{dx}{\sqrt
2\lambda_J}+{\rm Im}G_2\sin\frac{dx}{\sqrt
2\lambda_J}]\sin\theta+G_1e^{-\frac{x}{\lambda_{sdl}}}\cos\theta, 
\nonumber   
\end{array}
\right.
\end{eqnarray}
\begin{eqnarray}
\label{a_perm}
a=&-&\frac{J}{c^2+d^2}
\frac{1-\cos(\frac{dt_F}{\sqrt 2\lambda_J})e^{-\frac{ct_F}{\sqrt 
2\lambda_J}}}{t_F/(\sqrt 2\lambda_J)}(c{\rm Re}\tilde G_2-d{\rm 
Im}\tilde G_2) \nonumber \\ 
&-&\frac{J}{c^2+d^2}
\frac{\sin(\frac{dt_F}{\sqrt 2\lambda_J})e^{-\frac{ct_F}{\sqrt
2\lambda_J}}}{t_F/(\sqrt 2\lambda_J)}(d{\rm Re}\tilde G_2+c{\rm 
Im}\tilde G_2),
\end{eqnarray}
\begin{eqnarray}
\label{b_perm}   
b&=&\frac{J}{c^2+d^2}(c{\rm Re}\tilde G_2-d{\rm Im}\tilde G_2)
\frac{\sin(\frac{t_F}{\sqrt 2\lambda_J})e^{-\frac{t_F}{\sqrt 
2\lambda_J}}}{t_F/(\sqrt 2\lambda_J)} \nonumber \\
&-&\frac{J}{c^2+d^2}(d{\rm Re}\tilde G_2+c{\rm Im}\tilde G_2)
\frac{1-\cos(\frac{dt_F}{\sqrt 2\lambda_J})e^{-\frac{ct_F}{\sqrt
2\lambda_J}}}{t_F/(\sqrt 2\lambda_J)},
\end{eqnarray}
where
\begin{equation}
\label{G1_perm}
G_1=-\frac{\beta j_e\sqrt 2}{\lambda_J J}\frac{\hbar 
a_0^3}{e\mu_B}\times
\frac{c\sin^2\theta}{(c^2+d^2)\cos^2\theta+2c\lambda\sin^2\theta}
\end{equation}
\begin{equation}
\label{G2_perm}
G_2=\tilde{G_2}\sin 2\theta=-\frac{\beta j_e}{\sqrt 2\lambda_J 
J}\frac{\hbar a_0^3}{e\mu_B}\times
\frac{(d-ic)\sin 2\theta}{(c^2+d^2)\cos^2\theta+2c\lambda\sin^2\theta}.
\end{equation}
The parameter $\chi$ (Eq.~\ref{r_norm_xi}) for permalloy takes the form
\begin{equation}
\label{chi_perm}
\chi=-1+\frac{c^2+d^2}{2c\lambda},
\end{equation}
where  
\begin{eqnarray}
c&=&\sqrt{(\lambda_J/\lambda_{sf})^2+\sqrt{(\lambda_J/\lambda_{sf})^4+1}}
 \nonumber \\
&=&\sqrt{(1-\beta\beta^\prime)(\lambda_J/\lambda_{sdl})^2
+\sqrt{(1-\beta\beta^\prime)^2(\lambda_J/\lambda_{sdl})^4+1}}, \nonumber 
\end{eqnarray}
$$
c^2+d^2=2\sqrt{(\lambda_J/\lambda_{sf})^4+1}=
2\sqrt{(1-\beta\beta^\prime)^2(\lambda_J/\lambda_{sdl})^4+1},
$$
$$
\lambda=\frac{(1-\beta\beta^\prime)}{\sqrt 
2}\frac{\lambda_J}{\lambda_{sdl}}.
$$
Taking $\lambda_J$ to be 4 nm, $\lambda_{sdl}$ to be 5 nm, $\beta=0.7$,
$\beta^\prime=0.95$ (see Sec.~\ref{rev_form}), I estimate $\chi$ to
be about 3.8. This number has to be compared with the experimentally
obtained $\chi=1.17$.~\cite{Pratt_priv} While the theoretical value of
$\chi$ is more than three times larger than the experimental one, it is
calculated without taking into account the resistance of the interface
between permalloy layers; the presence of he interface resistance leads
to a smaller value of $\chi$ (see Eq.~(\ref{chi})).

\chapter{\label{chap_swt_rslt} Spin-torques in the thin FM - thick FM - 
NM structure}

In this chapter, I consider a realistic pillar-like multilayered
structure used for current induced reversal of a magnetic
layer~\cite{kat_prl_00,alb_apl_00,fert_apl_01} consisting of a thick
ferromagnetic layer, whose primary role is to polarize the current, and
whose magnetization is pinned, a thin ferromagnetic layer that is to be
switched, a nonmagnetic spacer layer so that there is no interlayer
exchange coupling between the thick and thin layers, and nonmagnetic
layer or lead in the back of the thin magnetic layer,
Fig.~\ref{pic_multi_6}. The lead in the back of the thick magnetic layer
does not have to be considered, because the thickness of the left
ferromagnetic layer is taken to exceed the spin diffusion length
$\lambda_{sdl}$ in this layer; the spin polarization of the current is
dictated by the thick magnetic layer. As in the symmetrical structure
considered in the previous chapter, the thickness of the nonmagnetic
spacer layer $t_{Sp}$ is taken to be small compared to the
spin-diffusion length in the nonmagnetic metal ($\lambda_{sdl}^N\sim
600$ nm), so that both spin accumulation and spin current are constant
in the spacer, and all the results are obtained in the approximation
that $t_{Sp}=0$. The majority of the results for the spin-current,
spin-accumulation, spin-torque, and effective field acting on the local
magnetization in the thin magnetic layer are obtained numerically using
the Mathematica$^{\rm TM}$ software,~\cite{math} and presented in
Sec.~\ref{num_res}. Next, I derive an analytical expression for the
transverse spin-current at the interface between two ferromagnetic
layers within an approximation that the thickness of the thin FM layer
$t_F$ is much smaller than the spin-flip length $\lambda_{sf}$, but much
larger than the characteristic length scale of the transverse spin
accumulation $\lambda_J$. Finally, I study the angular dependence of the
resistance of the structure described above.
\begin{figure} 
\centering
\includegraphics[width=4.5in]{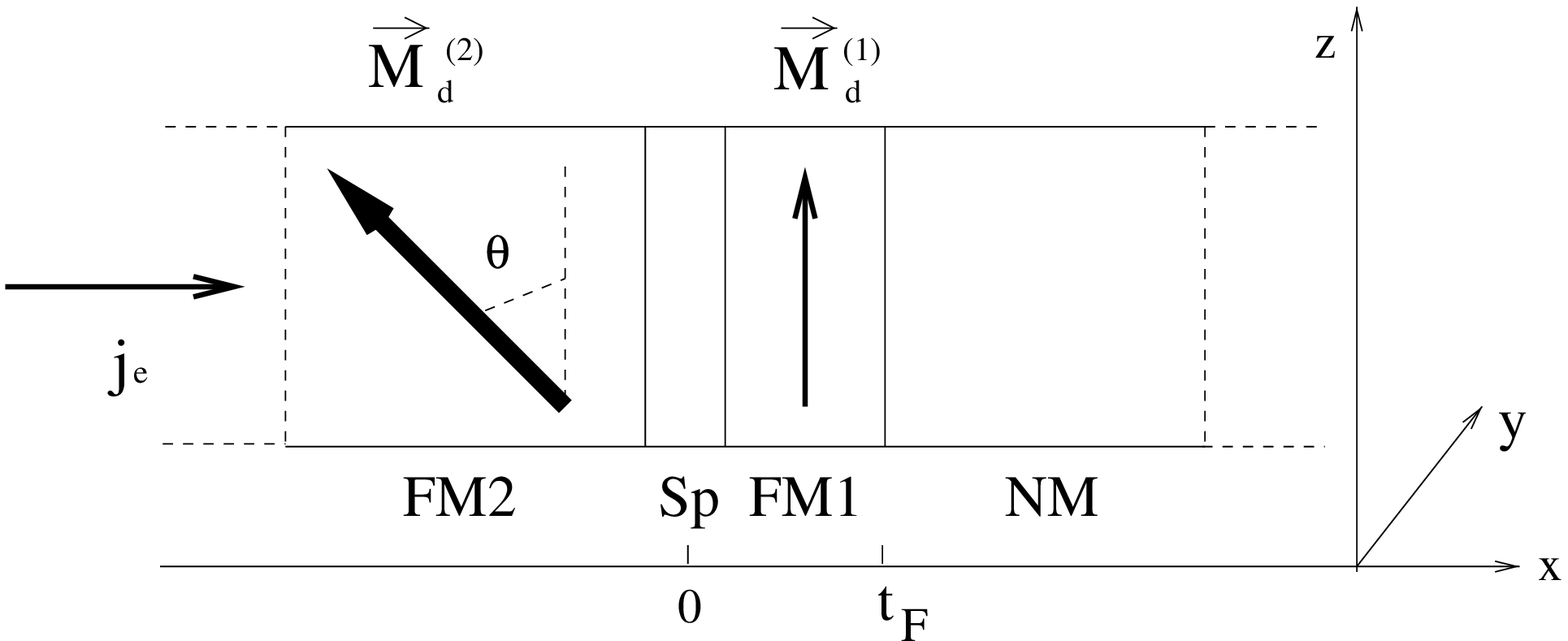}
\caption[Multilayered pillar-like structure used for current induced
reversal of a magnetic layer] {Multilayered pillar-like structure used for
current induced reversal of a magnetic layer. FM2 is a thick ferromagnetic
layer with the thickness exceeding $\lambda^F_{sdl}$ and local
magnetization ${\bf M}_d^{(2)}=\cos\theta{\bf e}_z-\sin\theta{\bf e}_y$,
Sp is a thin nonmagnetic spacer, FM1 is a thin ferromagnetic layer with 
the thickness $t_F$ and local magnetization ${\bf M}_d^{(1)}={\bf e}_z$, 
and NM is a nonmagnetic back layer.} 
\label{pic_multi_6} 
\end{figure}


\section{\label{num_res} Numerical results}

In order to find the spin accumulation and spin current distributions in
the structure depicted in Fig~\ref{pic_multi_6}, I solve 
Eqs.~(\ref{eq_m_long}) and~(\ref{eq_m_trans}) in the thick
ferromagnetic layer, thin ferromagnetic layer, and nonmagnetic back layer,
and find the spin current using Eq.~(\ref{jm_fin}) (see
appendix~\ref{app_sol_three}). In Appendices~\ref{app_bound_cond}
and~\ref{app_sol_three} I derive the boundary conditions on the
accumulation and current at the interface between the thick and thin
ferromagnetic layers, and at the interface between the thin magnetic layer
and the nonmagnetic layer; with these one can determine the spin current
across the entire structure, and consequently the spin torque and
effective field acting on the thin magnetic layer. In the absence of
specular scattering at the interfaces, spin current is continuous across
the whole system; the spin accumulation is discontinuous across the
diffusive interfaces, the discontinuity being proportional to the
interface resistance, and continuous across the interface with zero
resistance. I assume that there is neither specular nor diffuse scattering
at the interface between the thin ferromagnetic layer and nonmagnetic
layer, and that there is no specular scattering at the interface between
ferromagnetic layers. I obtain all the results both in the absence and in
the presence of the diffuse scattering at this interface. Without further
simplifications I am unable to give analytic expressions for the
accumulation and current across the multilayer, and I present the
numerical results for these quantities as well as the torque and field
they create. In all the plots below, the diffusion constant is taken to be 
10$^{-3}$ m$^2$/s in the magnetic layers, and 5$\cdot$10$^{-3}$ m$^2$/s in 
the nonmagnetic layers, and the spin diffusion length in the nonmagnetic 
layers $\lambda_{sdl}^N$ is taken to be 600 nm. The amount of diffuse 
scattering at the interface $AR_I$, the spin polarization parameters for 
the conductivity $\gamma$ at the interface and $\beta$ in the bulk of 
magnetic layers $\beta$ are taken from the experimental data on CPP-MR, 
$AR_I(1-\gamma^2)=0.5$~fOm$\cdot$m$^2$, $\gamma=0.75$, 
$\beta=0.5$.~\cite{BP_jmmm_99} Other spin polarization parameters are 
obtained using the results of the {\it ab-initio} calculations of the 
density of states for up and down electrons.~\cite{Wang}

In Figs.~\ref{sw_atot}-\ref{sw_btotR} I show the {\it total} (not per
unit length as in the previous chapter for a symmetrical structure) spin
torque $a\sin\theta t_F$ and effective field $b\sin\theta t_F$ as a
function of the thickness of the thin magnetic layer $t_{F}$ which is
being switched for different combinations of $\lambda_J$,
$\lambda_{sdl}$, $AR_I$, and for different angles $\theta$ between the
magnetization directions in the layers. Two values of $\lambda_J$ are
chosen for illustration, $\lambda_J=4$ nm, which is comparable to the
mean free path $\lambda_{mfp}=6$ nm, and a smaller value of
$\lambda_J=1$ nm; $\lambda_{sdl}$ is taken to be 60 nm and 30 nm. One
can see that while the spin torque rapidly increases for small but
finite $t_F\approx\lambda_J$ and then gradually levels off
(Figs.~\ref{sw_atot}-\ref{sw_atotR}), the effective field is largest
about $t_F\approx 0.5\lambda_J$ and then decreases toward zero with
$t_F$ (Figs.~\ref{sw_btot}-\ref{sw_btotR}); this can be understood as
follows. When the thickness of the thin layer $t_F$ is much smaller than
$\lambda_J$, the spin accumulation in the thin layer is the same as that
of the thick layer at the interface, and its direction is parallel to
that of the magnetization of the thick layer ${\bf M}_d^{(2)}$ (note
that I do not consider torques created directly at the interface between
the ferromagnetic layers); therefore, as follows from the definition of
$a$ and $b$, $J{\bf m}_\perp=a{\bf M}_d^{(2)}\times{\bf M}_d^{(1)}+b{\bf
M}_d^{(1)}\times({\bf M}_d^{(2)}\times{\bf M}_d^{(1)})$
(Eq.~(\ref{def_a_b})), only the effective field $b$ can exist, since
only the second term has a component along ${\bf M}_d^{(2)}$. As $t_F$
increases, the spin accumulation in the thin magnetic layer rotates away
from ${\bf M}_d^{(2)}$ and develops a transverse component, parallel to
${\bf M}_d^{(2)}\times{\bf M}_d^{(1)}$, i.e., a spin torque $a$
develops.  When $t_F$ becomes larger than $\lambda_J$, the spin
accumulation is further rotated in the thin layer, out of the plane of
${\bf M}_d^{(2)}$ and ${\bf M}_d^{(1)}$, and thus the component of the
spin accumulation in the plane of the magnetizations decreases rapidly,
i.e., the effective field diminishes faster than the torque. As $t_F$
increases further there are no additional contributions to either the
field and torque because they represent effects that are centered at the
interface with the spacer layer and averaged over the entire thickness
of the thin magnetic layer. Although the effective field is negligible
compared to the torque in the limit of large $t_{F}$, it is noteworthy
that at its maximum the field $b$ is at least as large as $a$, so that
both spin torque and effective field should enter the equation of motion
of the background magnetization, Eq.~(\ref{LLG}). One also notes that
while the torque and field terms $a$ and $b$ are largest for
$\theta=180^\circ$ and~$0^\circ$, they do not act on the background
magnetization because $\sin\theta=0$. The largest effects are found for
$\theta$ close to $180^\circ$ (see below). Comparing the
figures~\ref{sw_atot} and~\ref{sw_atotR}, \ref{sw_btot}
and~\ref{sw_btotR}, one can see that interface resistance decreases both
spin torque and effective field, and causes the spin torque to achieve
saturation for slightly smaller $t_F$. Comparing the diagrams a) and c),
b) and d) at each of the Figs.~\ref{sw_atot}-\ref{sw_btotR}, one notices
that when the spin diffusion length is reduced from $\lambda_{sdl}=60$
nm to $\lambda_{sdl}=30$ nm, the spin torque and effective field are
reduced. Comparing the diagrams a) and b), c) and d) at each of the
Figs.~\ref{sw_atot}-\ref{sw_btotR}, one can see that when the spin
transfer length is reduced from $\lambda_J=4$ nm to $\lambda_J=1$ nm,
the spin torque achieves saturation for smaller $t_F$ and the effective
field is increased and peaks for lower $t_F$.
\begin{figure}
\includegraphics[width=\textwidth]{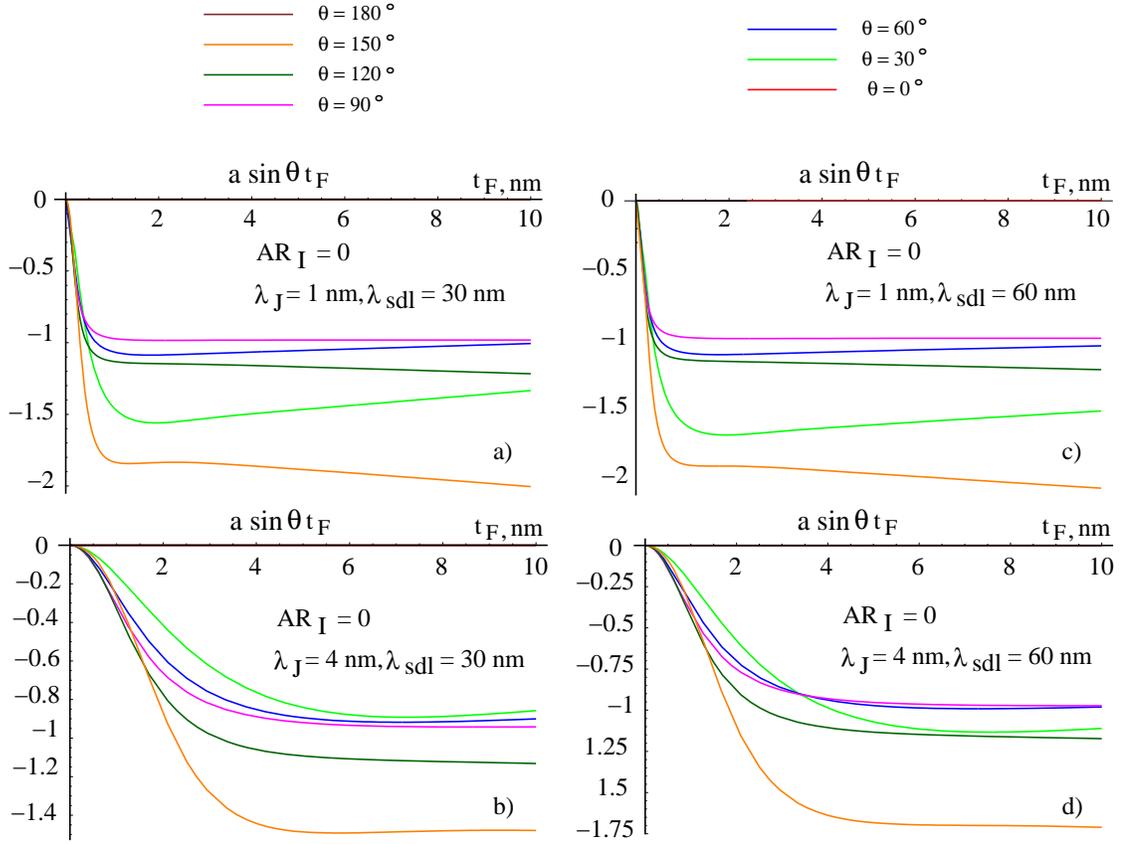}
\caption[Total torque acting on the thin FM layer in the 
FM-Sp-FM-NM system as a function of the layer thickness for different 
$\lambda_J$ and $\lambda_{sdl}^F$ 
and zero interface resistance]{Total torque 
$a\sin\theta t_F/\beta j_e(\hbar a_0^3/e\mu_B)$ acting on the thin FM 
layer in the 
FM-Sp-FM-NM system as a function of the layer thickness for different 
$\lambda_J$ 
and $\lambda_{sdl}$ and zero interface resistance.}
\label{sw_atot}
\end{figure}

\begin{figure}
\includegraphics[width=\textwidth]{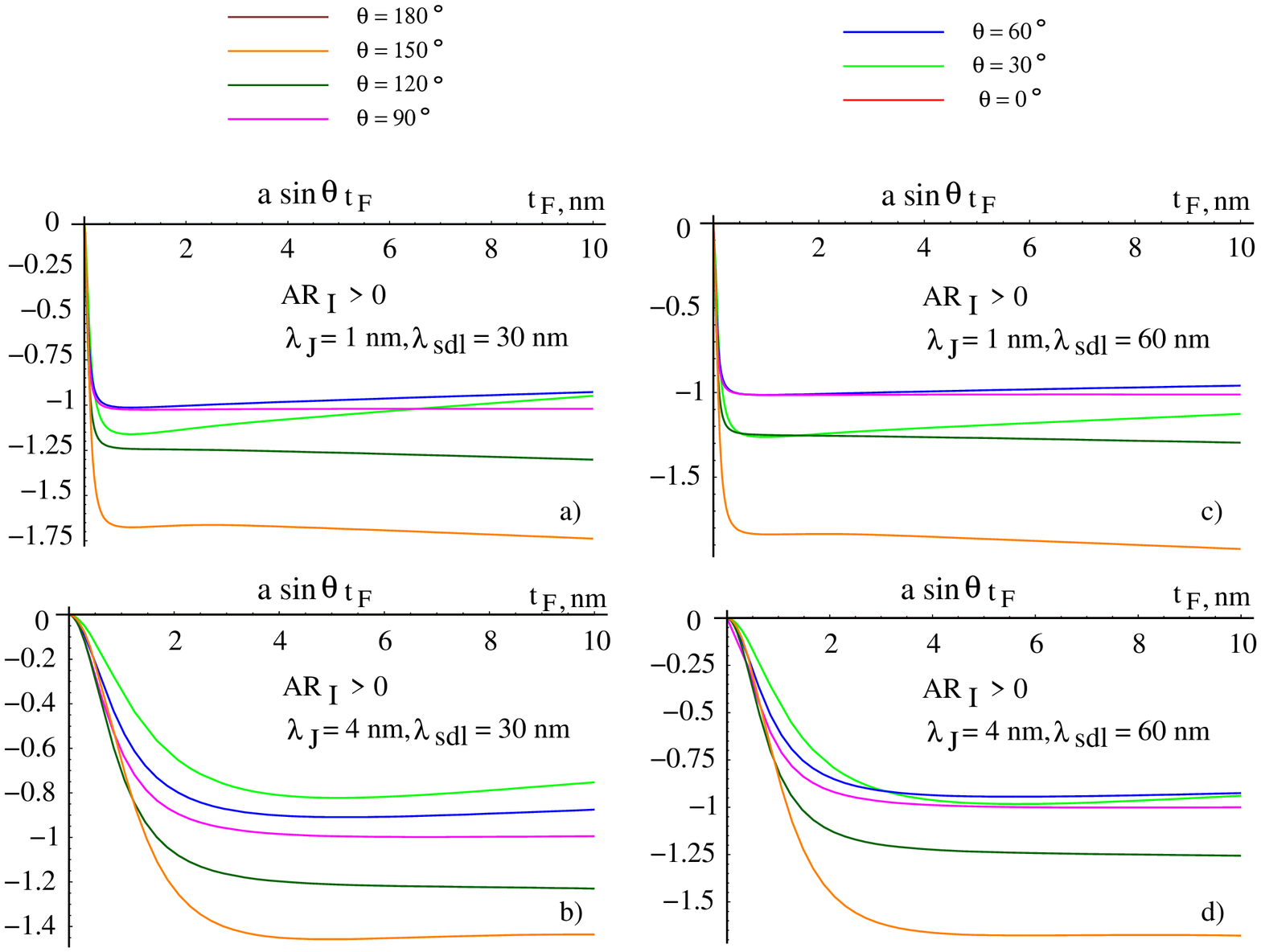}
\caption[Total torque acting on the thin FM layer in the 
FM-Sp-FM-NM system as a function of the layer thickness for different 
$\lambda_J$ and $\lambda_{sdl}^F$
and non-zero interface resistance]{Total torque  $a\sin\theta t_F/\beta 
j_e(\hbar a_0^3/e\mu_B)$ acting on the thin FM 
layer in the FM-Sp-FM-NM system as a function of the layer thickness  
for different $\lambda_J$
and $\lambda_{sdl}$ and non-zero interface resistance.}
\label{sw_atotR}
\end{figure}

\begin{figure}  
\includegraphics[width=\textwidth]{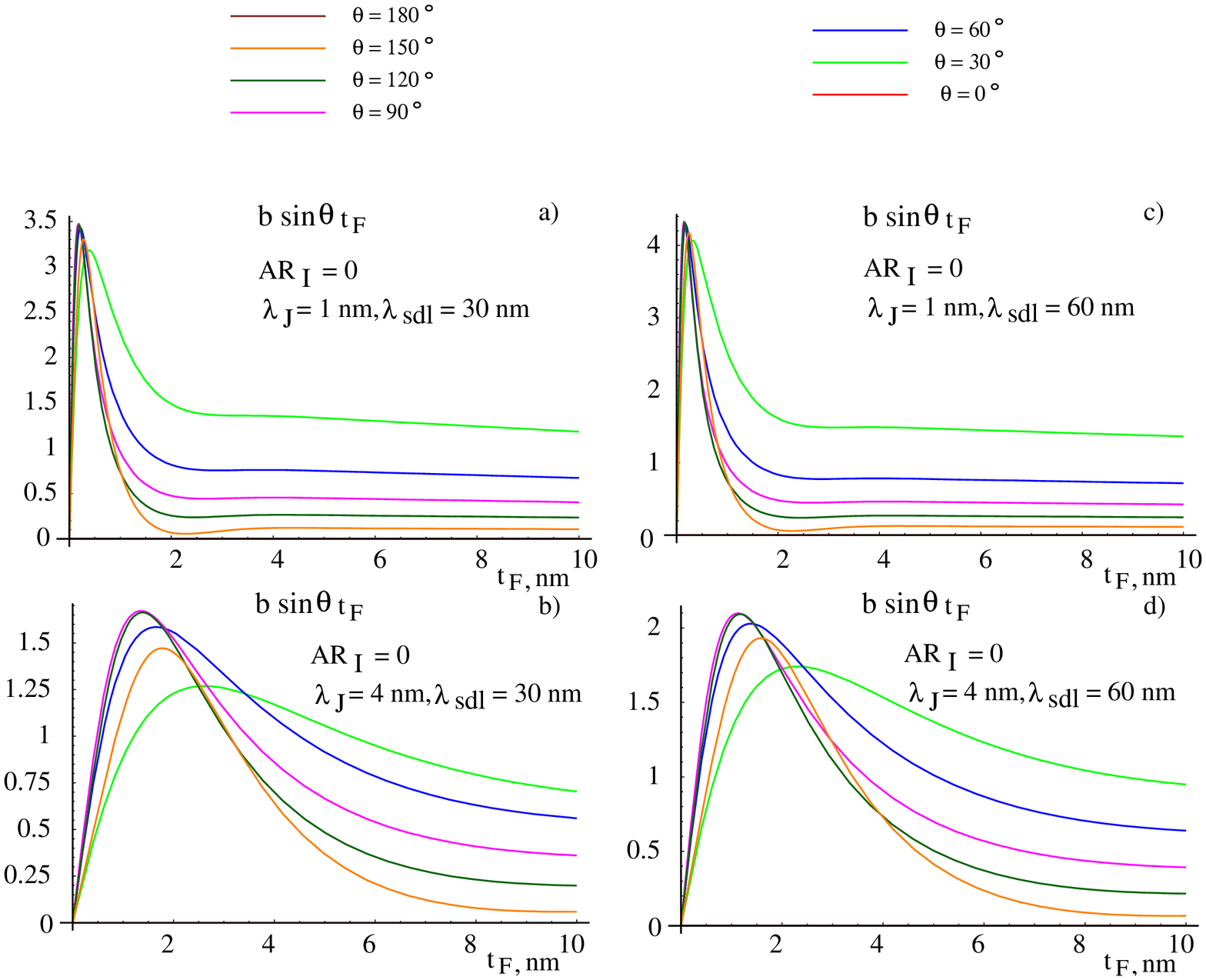}
\caption[Total effective field acting on the thin FM layer in the 
FM-Sp-FM-NM system as a function of the layer thickness for 
different $\lambda_J$ and 
$\lambda_{sdl}^F$ and zero interface resistance]{Total effective field
$b\sin\theta t_F/\beta j_e(\hbar a_0^3/e\mu_B)$ acting on the thin FM 
layer in the 
FM-Sp-FM-NM system as a function of the layer thickness for different 
$\lambda_J$ and $\lambda_{sdl}$ and zero interface resistance.}  
\label{sw_btot}
\end{figure}

\begin{figure}  
\includegraphics[width=\textwidth]{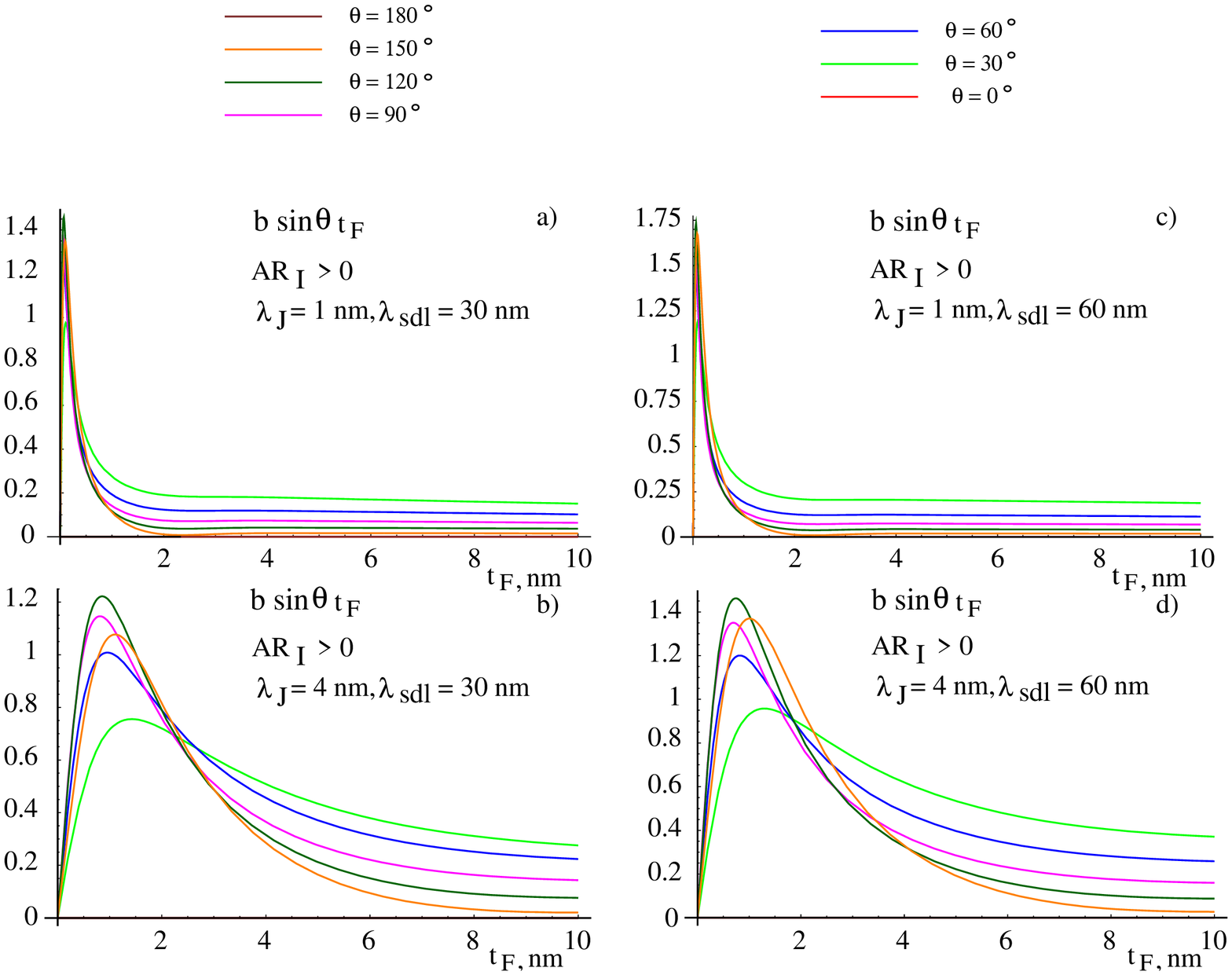}
\caption[Total effective field acting on the thin FM layer in the 
FM-Sp-FM-NM system as a function of the layer thickness for 
different $\lambda_J$ and
$\lambda_{sdl}^F$ and non-zero interface resistance]{Total effective 
field $b\sin\theta t_F/\beta j_e(\hbar a_0^3/e\mu_B)$ acting on the 
thin FM layer in the 
FM-Sp-FM-NM system as a function of the layer thickness for different
$\lambda_J$ and $\lambda_{sdl}$ and non-zero interface resistance.}  
\label{sw_btotR}
\end{figure}

In Figs.~\ref{mj_3}-\ref{mjR_15} I show the spin accumulation and spin
current distribution in the thick FM-thin FM-NM system both in the
absence and in the presence of the interface resistance for
$\lambda_J=4$ nm, $\lambda_{sdl}=60$ nm, for different angles $\theta$
between magnetization directions in the layers, and for two different
thicknesses $t_F$ of the thin ferromagnetic layer. I choose $t_F=3$ nm,
smaller than $\lambda_J$, but where the spin torque $a\sin\theta t_F$
starts to saturate, and $t_F=15$ nm, much larger than $\lambda_J$, but
smaller than $\lambda_{sdl}$.  The magnetization in the thin layer ${\bf
M}_d^{(1)}$ is taken as the global $z$ axis; the current is along the
$x$ axis which is along the growth direction of the multilayer, and the
$y$ direction is perpendicular to the other two. In these global axes,
longitudinal and transverse in the thin magnetic layer refers to the
directions $z$ and $x-y$; however for the thick layer, whose
magnetization ${\bf M}_d^{(2)}$ is at an angle $\theta$ relative to
${\bf M}_d^{(1)}$, the global $y$ and $z$ axes do not define what is
meant by longitudinal and transverse in this layer.
 
Comparing the figures~\ref{mj_3} and~\ref{mjR_3}, \ref{mj_15}
and~\ref{mjR_15}, one notices that the diffuse scattering at the interface
between the ferromagnetic layers produces sizable discontinuities in the
accumulation at the interface (see Appendices~\ref{app_bound_cond}
and~\ref{app_sol_three}), but it does not lead to an increased spin 
torque (see Figs.~\ref{sw_atot}-\ref{sw_btotR}); this result is different 
from the ones obtained within other models.

One can see that far from the interface $x\ll -\lambda_J$ the
accumulation and current in the thick layer are collinear with
background magnetization ${\bf M}_d^{(2)}$, i.e., referred to its local
axes they are longitudinal with no transverse components, and the spin
current approaches its bare value ${\bf j}_{m}\rightarrow \beta
j_{e}{\bf M }_d^{(2)}$ (see Eq.~(\ref{jm_fin})), as one expects in the
bulk of a ferromagnet. Even though one still has a longitudinal spin
accumulation in the region $-\lambda_{sdl}\ll x\ll -\lambda_J$ its
gradient is small compared to that of the transverse accumulation which
makes large contributions to the spin current in the region $x >
-\lambda_J$. This is clear from the plots for $m_x$ and $j_{m,x}$ which
go to zero, while $m_y\rightarrow m\sin\theta$, $j_{m,y}\rightarrow
\beta j_e\sin\theta$, $m_z\rightarrow m\cos\theta$, and
$j_{m,z}\rightarrow \beta j_e\cos\theta$. For example, for
$\theta=90^\circ$, $j_{m,y}\rightarrow 1$, in units of $\beta j_{e}$,
while $j_{m,z}\rightarrow 0$ in the thick ferromagnetic layer far from
the interface. In the nonmagnetic layer to the right, $x > t_F$, the
spin current is polarized along the magnetization direction in the thin
magnetic layer ${\bf M}_d^{(1)}$, and there is no, or very little, spin
current in the $x-y$ direction. 

Surprisingly, there is a huge enhancement of the transverse ($x-y$)  
components of the spin current in the region about $x=0$
(diagrams~d)~and~e) in Figs.~\ref{mj_3}-\ref{mjR_15}). Since the torque
transmitted to the thin magnetic layer is the difference between the
transverse component of the spin current at the boundaries of this layer
(see Eq.~(\ref{tot_torque})), it is greatly amplified compared to what one
would find neglecting the spin accumulation. The thick magnetic layer to
the left $x<0$ is pinned so that the enhanced torques acting in the region
of the interface do not produce any rotation. The $z$ or longitudinal
component of the incoming spin current is not absorbed by the thin
magnetic layer as there is no transfer of spin angular momentum along this
direction ($t_F\ll\lambda_{sdl}$), see the diagram~f)~in
Figs.~\ref{mj_3}-\ref{mjR_15}. The slight decrease in $j_{m,z}$ is due to
the spin flip scattering in the magnetic layer which is characterized by
$\lambda_{sdl}=60$ nm. The much slower decrease in $j_{m,z}$ in the
nonmagnetic layer, $x>t_F$, comes from the spin flip scattering in the
nonmagnetic layer whose $\lambda_{sdl}^N=600$ nm.

The large enhancement of the transverse spin currents can be understood as
follows. At the interface between the thick and thin ferromagnetic layers,
the spin accumulation has to adjust to the new magnetization direction.
Both the longitudinal and transverse components of ${\bf m}$ have to
experience the change of the same order, but the distance over which the
transverse component is absorbed, $\lambda_J$, is much smaller than that
for the longitudinal accumulation, $\lambda_{sdl}$. Therefore the gradient
of the transverse accumulation about $x=0$ is large and as it is the
gradient that contributes to the spin current, Eq.~(\ref{jm_fin}), one
finds an amplification of the transverse components of the spin current at
the interface. In the next section I present a quantitative estimate for 
the enhancement. 

It is worth noting that since the transport in the spin transfer region is
assumed to be diffusive (see Chap.~\ref{chap_intro}), so that the length
scale of ${\bf m}_\perp$ and ${\bf j}_{m\perp}$ is $\lambda_J$, the
transverse component of the spin current is absorbed over the {\it entire}
thin layer, if its thickness is comparable with $\lambda_J$ (diagrams~d)
and~e)~in Figs.~\ref{mj_3} and~\ref{mjR_3}). For larger thicknesses $t_F$ 
(Figs.~\ref{mj_15} and~\ref{mjR_15}), the transverse spin current in the 
thin magnetic layer indeed goes to zero before reaching the interface 
with the nonmagnetic back layer $x=t_F$.The reason for the different 
behavior in the thin layer arises from its confined geometry, i.e., the 
reflections from the thin magnetic/nonmagnetic back layer interface create 
the patterns observed for the transverse spin currents.

\begin{figure}  
\includegraphics[width=\textwidth]{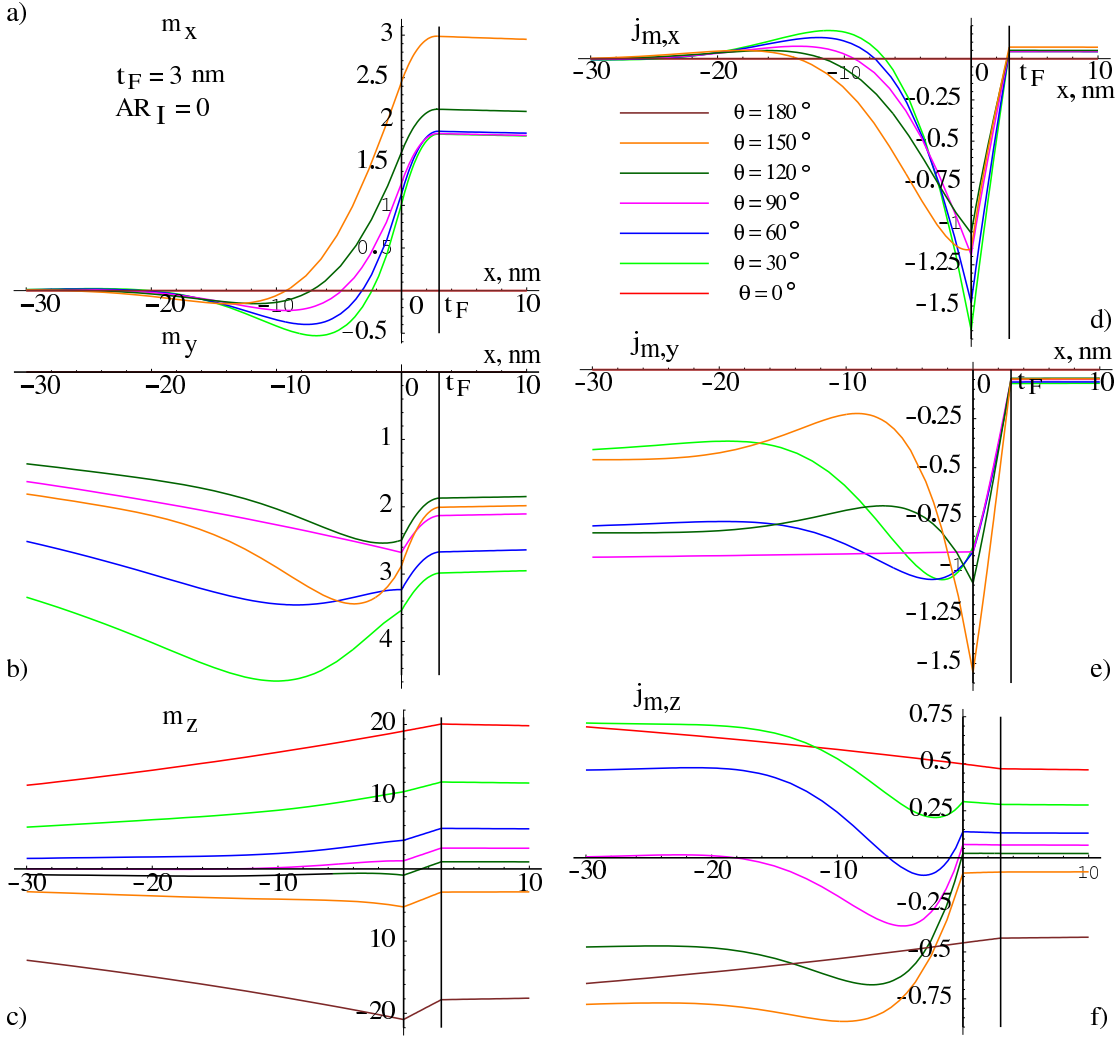}
\caption[Spin-accumulation and spin-current distribution in the 
FM-Sp-FM-NM system with the thin FM layer thickness $t_F=3$ nm and 
zero interface resistance]{$x$-, $y$-, and
$z$-components of the spin-accumulation ${\bf m}/(\beta
j_e/\sqrt 2 \lambda_J J)(\hbar a_0^3/e\mu_B)$ (a-c) and spin-current ${\bf 
j}_m/\beta j_e$ (d-f) distribution in the FM-Sp-FM-NM system with 
$\lambda_J=4$ nm, $\lambda_{sdl}=60$ nm, the thin FM layer thickness  
$t_F=3$ nm and zero interface resistance.}  
\label{mj_3}
\end{figure}

\begin{figure}  
\includegraphics[width=\textwidth]{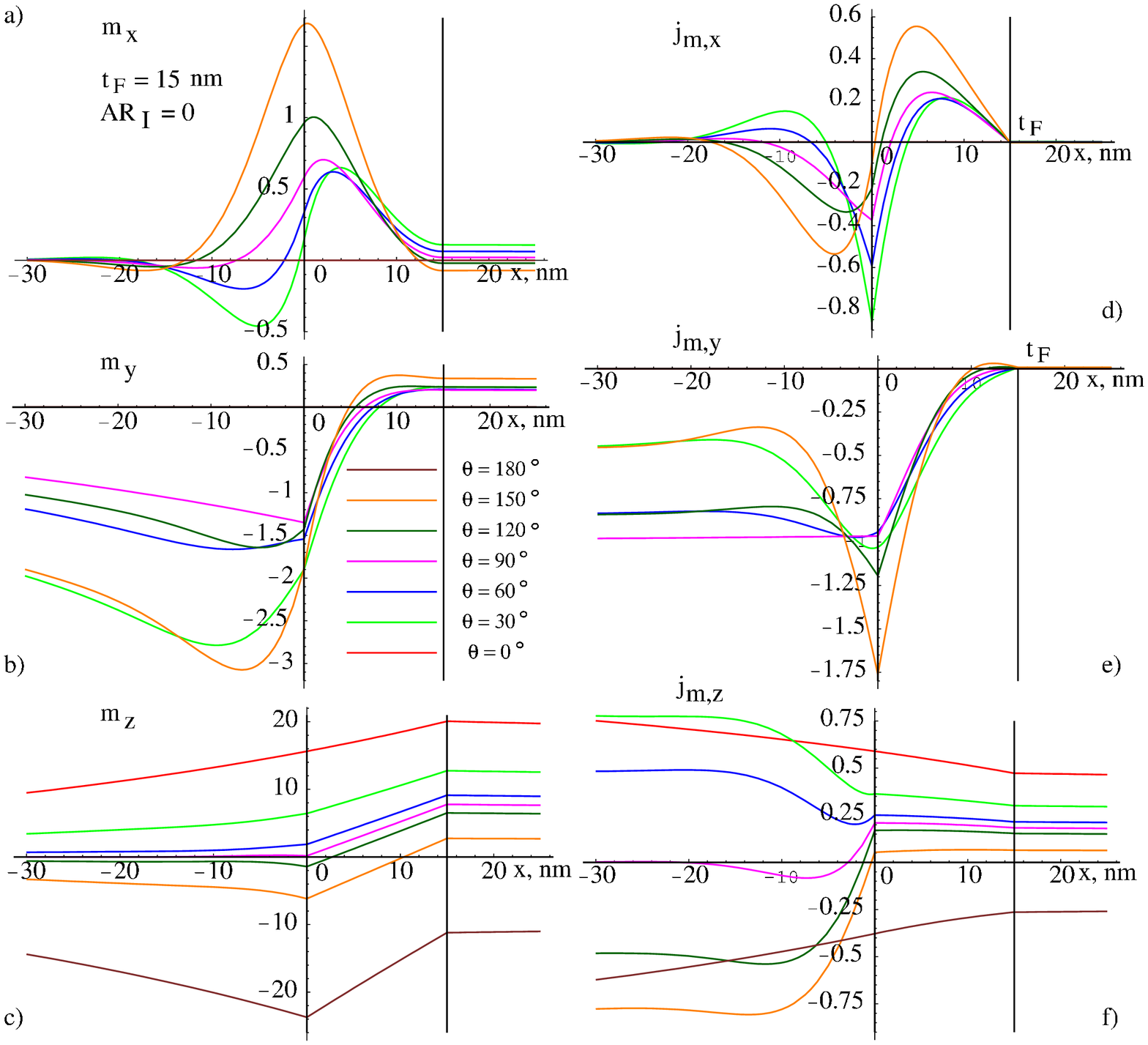}
\caption[Spin-accumulation and spin-current distribution in the 
FM-Sp-FM-NM system wight the thin FM layer thickness $t_F=15$ nm and 
zero interface resistance]{$x$-, $y$-, and 
$z$-components of the spin-accumulation ${\bf m}/(\beta 
j_e/\sqrt 2 \lambda_J J)(\hbar a_0^3/e\mu_B)$ (a-c) and spin-current ${\bf 
j}_m/\beta j_e$ (d-f) distribution in the FM-Sp-FM-NM system with 
$\lambda_J=4$ nm, $\lambda_{sdl}=60$ nm, the thin FM layer thickness 
$t_F=15$ nm and zero interface resistance.}  
\label{mj_15}
\end{figure}

\begin{figure}  
\includegraphics[width=\textwidth]{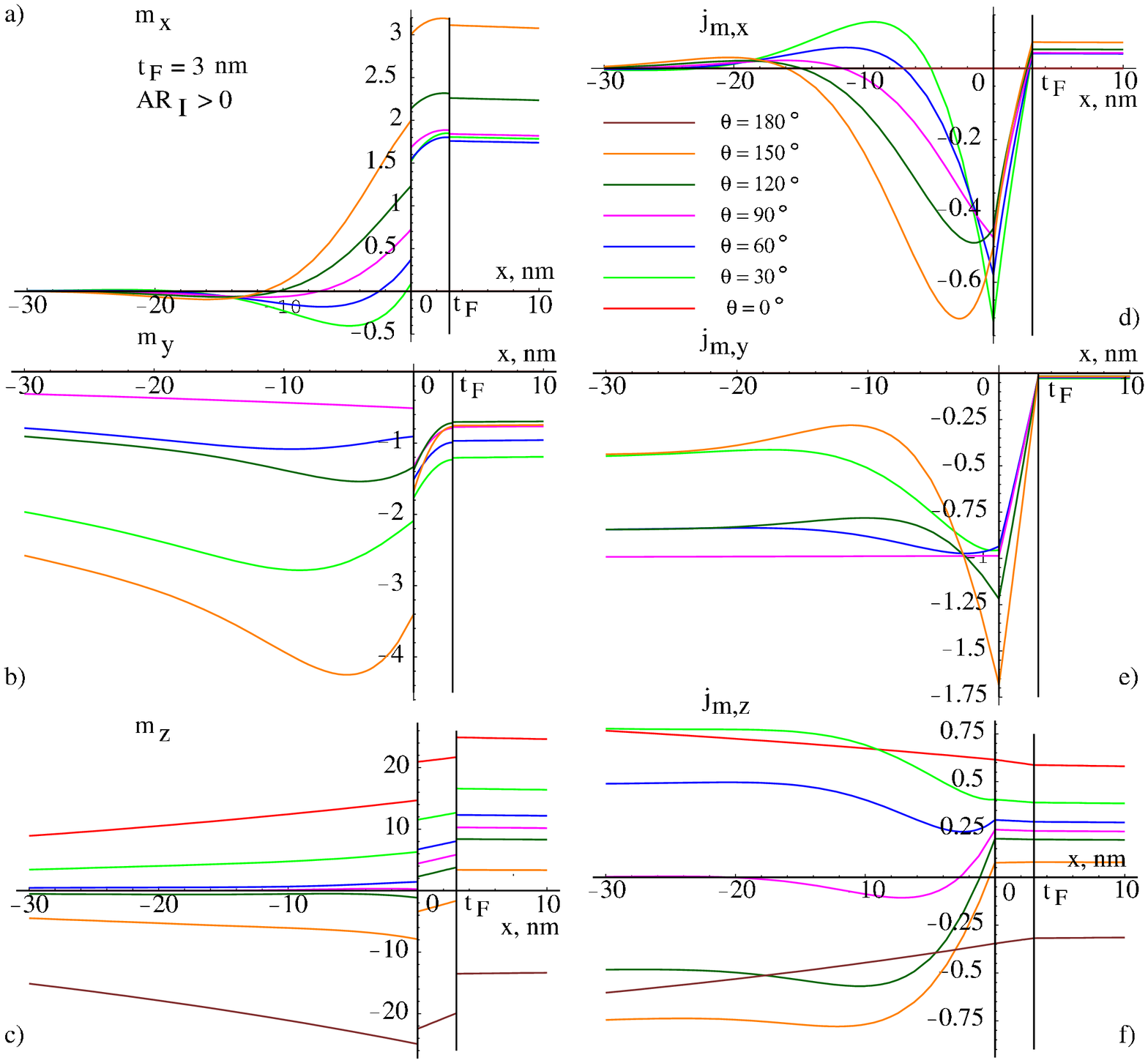}
\caption[Spin-accumulation and spin-current distribution in the 
FM-Sp-FM-NM system whit the thin FM layer thickness $t_F=3$ nm and 
non-zero interface resistance]{$x$-, $y$-, and
$z$-components of the spin-accumulation ${\bf m}/(\beta
j_e/\sqrt 2 \lambda_J J)(\hbar a_0^3/e\mu_B)$ (a-c) and spin-current ${\bf 
j}_m/\beta j_e$ (d-f) distribution in the FM-Sp-FM-NM system with 
$\lambda_J=4$ nm, $\lambda_{sdl}=60$ nm, the thin FM layer thickness  
$t_F=3$ nm and non-zero interface resistance.}  
\label{mjR_3}
\end{figure}

\begin{figure}  
\includegraphics[width=\textwidth]{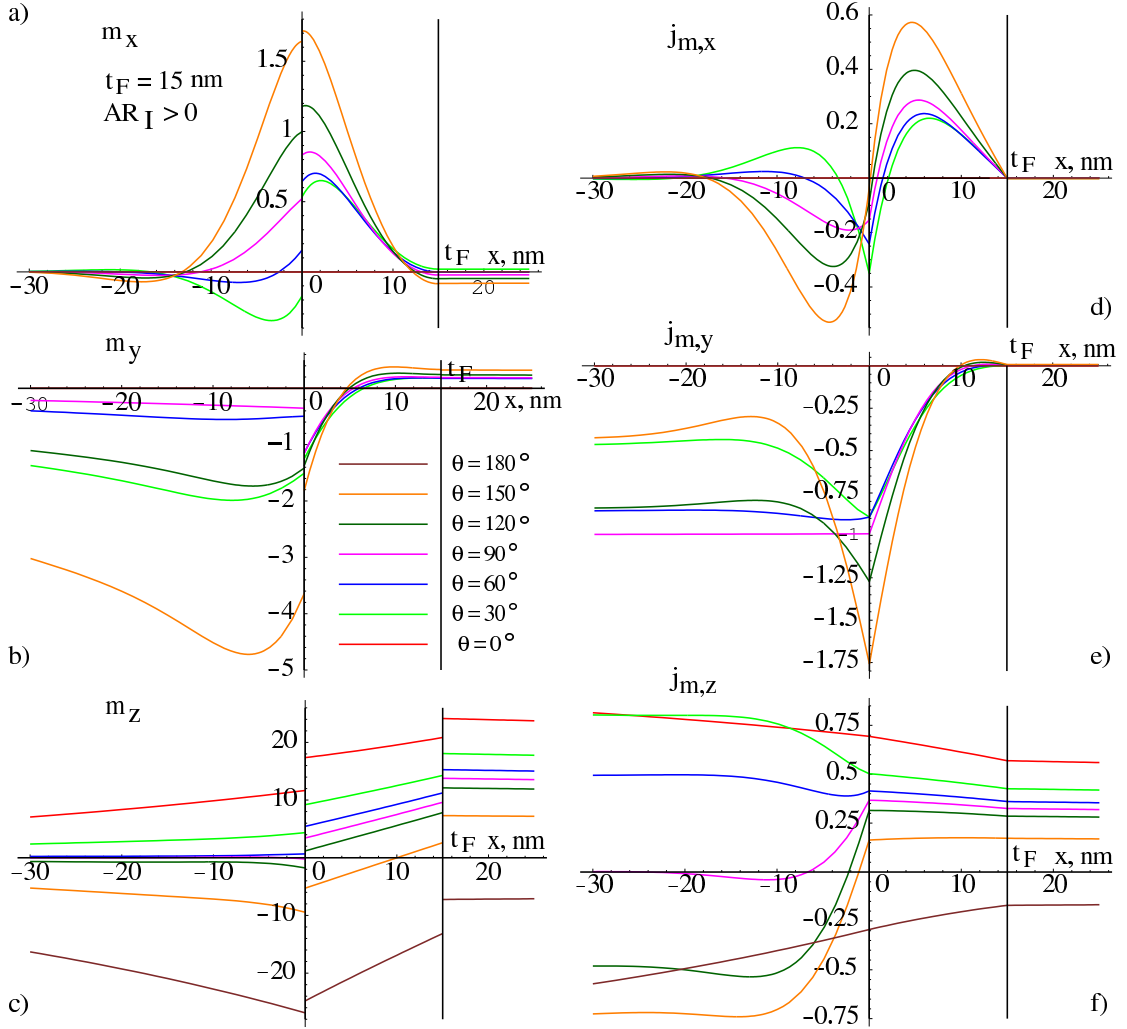}
\caption[Spin-accumulation and spin-current distribution in the 
FM-Sp-FM-NM system Whit the thin FM layer thickness $t_F=15$ nm and 
non-zero interface resistance]{$x$-, $y$-, and
$z$-components of the spin-accumulation ${\bf m}/(\beta
j_e/\sqrt 2 \lambda_J J)(\hbar a_0^3/e\mu_B)$ (a-c) and spin-current ${\bf 
j}_m/\beta j_e$ (d-f) distribution in the FM-Sp-FM-NM system with 
$\lambda_J=4$ nm, $\lambda_{sdl}=60$ nm, the thin FM layer thickness  
$t_F=15$ nm and non-zero interface resistance.}  
\label{mjR_15}
\end{figure}


\section{\label{enhanc_3} Spin current at the interface between 
ferromagnetic layers}

As seen from Figs.~\ref{mj_3}-\ref{mjR_15}, the spin current at the
interface between the ferromagnetic layers is significantly amplified
compared with its bare value $\beta j_e{\bf M}_d$ for large angles
$\theta$ between the magnetization directions in the layers. In this
section, I derive an analytical expression for the amplification within 
a simplifying approximation $\lambda_J\ll t_F\ll\lambda_{sf}$.

The spin accumulation distribution in the thin layer takes the form (see 
appendix~\ref{app_sol_three}):
\begin{eqnarray}
\label{m_x_13}
m_x^{(1)}&=&2\sin\theta e^{-{\tilde x}}[{\rm Re}G_5\cos{\tilde x}-
{\rm Im}G_5\sin{\tilde x}] \nonumber \\
&+&2\sin\theta e^{{\tilde x}-2\xi}
[({\rm Re}G_6\cos\xi-{\rm Im}G_6\sin\xi)\cos({\tilde x}-\xi) \\
&+&({\rm Im}G_6\cos\xi+{\rm Re}G_6\sin\xi)\sin({\tilde x}-\xi)] \nonumber ,
\end{eqnarray}
\begin{eqnarray}
\label{m_y_13} 
m_y^{(1)}&=&2\sin\theta e^{-{\tilde x}}[{\rm Im}G_5\cos{\tilde x}+
{\rm Re}G_5\sin{\tilde x}] \nonumber \\
&+&2\sin\theta e^{{\tilde x}-2\xi}
[({\rm Im}G_6\cos\xi+{\rm Re}G_6\sin\xi)\cos({\tilde x}-\xi) \\
&-&({\rm Re}G_6\cos\xi-{\rm Im}G_6\sin\xi)\sin({\tilde x}-\xi)] \nonumber ,
\end{eqnarray}
\begin{eqnarray}
\label{m_z_13}
m_z^{(1)}=G_3e^{-x/\lambda_{sdl}}+G_4e^{(x-t_F)/\lambda_{sdl}},
\end{eqnarray}
where ${\tilde x}=x/\sqrt 2\lambda_J$, $\xi=t_F/\sqrt 2\lambda_J$,  
$G_3$, $G_4$, $G_5$ and $G_6$ are constants of integration found via 
the boundary conditions (see Appendix~\ref{app_bound_cond}). The spin 
current is calculated as (see Eq.~(\ref{jm_fin}))
\begin{eqnarray}
\label{jm_13}
j_{m,x}^{(1)}&=&-\frac{1}{2}\frac{\partial m_x^{(1)}(x)}{\partial x}, 
\nonumber \\
j_{m,y}^{(1)}&=&-\frac{1}{2}\frac{\partial m_y^{(1)}(x)}{\partial x}, \\
j_{m,z}^{(1)}&=&\beta j_e-\frac{1}{2}(1-\beta\beta^\prime)\frac{\partial 
m_z^{(1)}(x)}{\partial x}. \nonumber
\end{eqnarray}
Using the approximation $\lambda_J\ll t_F\ll\lambda_{sf}$, the
transverse spin-current at the interface between ferromagnetic layers
$x=0$ in the absence of the interface resistance may be written as
\begin{eqnarray}
\label{jm_trans_13}
j_{m,x}^{(1)}(x=0)&=&({\rm Re}G_5+{\rm Im}G_5)\sin\theta/\sqrt 2\lambda_J, 
\\
j_{m,y}^{(1)}(x=0)&=&-({\rm Re}G_5-{\rm Im}G_5)\sin\theta/\sqrt 
2\lambda_J,
\end{eqnarray}
or, after substituting $G_5$,
\begin{equation}
\label{j_enh}
{\bf j}_{m,\perp}^{(1)}(x=0)=-\frac{\beta 
j_e\sin\theta}{2(\cos^2\theta/2+\lambda\sin^2\theta/2)}{\bf e}_y,
\end{equation}
where $\lambda=\sqrt{1-\beta\beta^\prime}\lambda_J/\sqrt 2\lambda_{sf}$. 
Up to the difference in notation, this expression is the same as the one
obtained in the previous chapter for the spin current at the interface
between two infinite ferromagnetic layers, Eq.~(\ref{jm_x0}). The
magnitude of the spin current at the interface is shown in the
Fig.~\ref{enhance} (red line) in comparison with the bare value of the
transverse current, $\beta j_e\sin\theta$ (blue line).
\begin{figure}
\centering
\includegraphics[width=4 in]{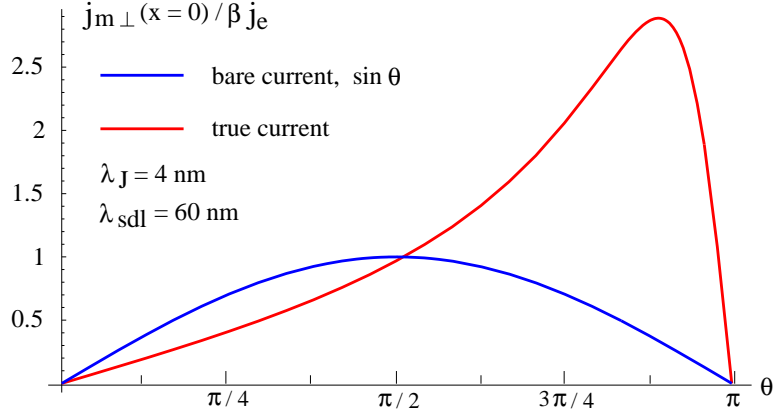}
\caption[True spin current at the interface between the thin and the 
thick ferromagnetic layers in comparison with the bare transverse 
current as a function of the angle between the magnetization directions 
in the layers]
{True spin current at the interface between the thin and the thick 
ferromagnetic layers (red line) in comparison with the bare transverse 
current (blue line) as a function of the angle between the 
magnetization directions 
in the layers for $\lambda_J=4$ nm, $\lambda_{sdl}=60$ nm.}
\label{enhance}
\end{figure}
One can see that for the angles $\theta$ close to $\pi$, the magnitude
of the transverse spin current, and, hence, the spin torque acting on
the thin layer is enhanced compared to its bare value of $\beta
j_e\sin\theta$, reaching its largest value of
\begin{equation}
\label{jm_max_3}
j_{m,\perp max}^{(1)}(x=0)=\frac{\beta j_e\sin\theta(1+\lambda)}{4\lambda}
\end{equation}
for the angle $\theta^\star$ such that
\begin{equation}
\label{theta_star_3}
\cos\theta^\star=-\frac{1-\lambda}{1+\lambda}.
\end{equation}
This huge enhancement, by a factor of $1/4\lambda$, where $\lambda$ is
of the order of 0.03 for cobalt, comes from the interplay between
longitudinal and transverse accumulation; it is the result of the global
nature of the spin current even though the transverse components of the
spin current and accumulation are absorbed within a region $\lambda_J$
of the interface. 


\section{\label{gmr} Resistance}

I calculate the resistance of the structure depicted on
Fig.~\ref{pic_multi_6} using the expression for the resistivity in each
layer which follows from Eq~(\ref{je_mparr}) for the electric 
current:
\begin{equation}
\label{rho_3}
\rho=\frac{1}{2C_0}+\frac{D_0\beta^\prime}{j_eC_0}\frac{\partial 
m_{||}}{\partial x}
\end{equation}
The sheet resistance of the system $AR$ takes the following form:
\begin{equation}
\label{AR_tot}
AR=AR_I+\frac{L}{2C_0}+\frac{\beta\beta^\prime}{2\sqrt 
2j_eC_0}[G_1+(G_4-G_3)(1-e^{-t_F/\lambda_{sdl}})],
\end{equation} 
where $AR_I$ is the resistance due to the interfaces, $L$ is the total
length of the system, $G_1$, $G_3$, and $G_4$ are the constants of
integration, defined in the appendix~\ref{app_sol_three}. 
Figure~\ref{threegmr} 
shows the normalized resistance $R_{norm}=(R(\theta)-R(0))/(R(\pi)-R(0))$ 
as a function of the angle $\theta$ between the magnetization directions 
in the layers for $\lambda_J=4$~nm, $\lambda_{sdl}=60$~nm, and several 
values of the thin layer thickness $t_F$.
\begin{figure}
\centering
\includegraphics[width=5 in]{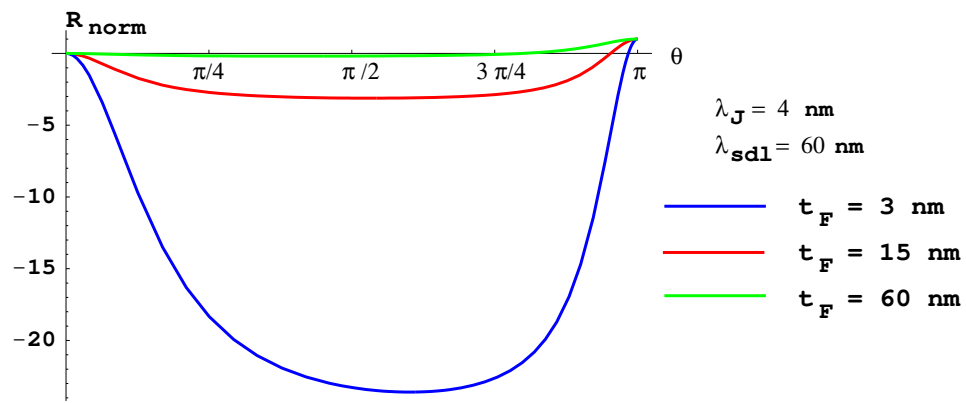}
\caption[Normalized resistance of the thick FM-Sp-thin FM-NM structure 
as a function of the angle between the magnetization directions in the 
layers]{Normalized resistance $R_{norm}$ as  a function of the angle 
$\theta$ between the magnetization directions in the layers in the 
absence of the interface resistance for $\lambda_J=4$~nm, 
$\lambda_{sdl}=60$~nm, and several values of the thin layer thickness $t_F$.}
\label{threegmr}
\end{figure}
One can see that the normalized resistance is a non-monotonic function of
the angle between the magnetization directions in the layers (compare 
with the reference~\cite{manschot}). This is a consequence of the angular 
momentum transfer from the polarized current to the background 
magnetization, and the effect it has on the voltage drop across the 
multilayer.

\chapter{\label{chap_conc} Conclusion}

Chapters~\ref{chap_res_thr} and~\ref{chap_res_rslt} address the problem
of finding the resistance due to the interfaces in the multilayered
metallic structures. In Chap.~\ref{chap_res_thr}, the general method
of solving the semiclassical linearized Boltzmann equation in CPP
geometry is developed, allowing one to obtain the integral equations for
the chemical potentials everywhere in the multilayers in the presence of
both specular and diffuse scattering at the interfaces, and diffuse
scattering in the bulk of the layers. In Chap.~\ref{chap_res_rslt},
the numerical results for the chemical potential and interface
resistance in the multilayered metallic systems are presented. The
variation of the chemical potential within a mean-free path of the
interface between two semi-infinite metallic layers is investigated. The
difference between the potential drop associated with the interface {\it
far} from it and that {\it at} the interface depends on the amount of
diffuse scattering at the interface and can change sign. In the case of
the same metal at both sides of the boundary, the different forms of
diffuse scattering are classified according to whether the measured
resistance is bigger or smaller than that across the interface. For
certain forms of diffuse scattering, the chemical potential remains
constant outside the immediate interfacial region. Interface resistance
in the presence of the diffuse scattering in the bulk of the layers is
less than that calculated in the ballistic transport regime. The
resistance due to interfaces in the metallic structures consisting of
three and five layers is calculated. If one maintains the same interface
between layers, its contribution to the total resistance of the system
depends on the thickness of the adjacent layers. This contribution is
constant only if the distance between interfaces is larger than the
electron mean free path. As the thickness of the layers becomes smaller
than the electron mean free path in these layers, scattering from
interfaces start to interact with each other. The error incurred by
neglecting the dependence of the interface resistance on layer thickness
depends on the amount of diffuse scattering at interfaces and the height
of the potential step at the interface. It reaches 15-20$\%$ in Fe-Cr
and Co-Cu systems with equal amounts of specular and diffuse scattering
at interfaces.  In order to observe this effect one should investigate
systems where interface resistance dominates total resistance. Hence,
the thickness of the layers should be small. Also, by introducing
additional scatterers, one should be able to change the ratio of
thickness to mean free path without changing the thickness of the
layers.


In Chaps.~\ref{chap_swt_thr} and~\ref{chap_swt_rslt} a mechanism for the
magnetization switching that is driven by spin-polarized current in
noncollinear magnetic multilayers is studied.  I find the spin-torque due
to the bulk of the magnetic layers and the diffuse scattering at
interfaces for a number of layered structures; the spin transfer that
occurs at interfaces is self-consistently determined by embedding it in
globally diffusive transport calculations. In Chap.~\ref{chap_swt_thr} a
system consisting of two thick ferromagnetic layers divided by a
nonmagnetic spacer is considered. The spin accumulation and the spin
current distributions, as well as the spin torques acting on the magnetic
layers are found analytically. I obtain an analytical expression for the
CPP magnetoresistance that takes into account the effect of the transfer
of the spin angular momentum from the polarized current on the voltage
drop across the multilayer, and estimate a parameter of the angular
dependence of the CPP-MR that gives a correction to the simple
$\cos^2(\theta/2)$ dependence of the resistance for cobalt and permalloy.
For the latter, this parameter can be compared with the experimental
value. While there is a difference between the theoretical and
experimental values, the theoretical value is obtained without taking into
account the interface resistance, which presence leads to a smaller
discrepancy.  In Chap.~\ref{chap_swt_rslt} a realistic magnetic multilayer
consisting of a thick magnetic layer, a thin magnetic layer that is to be
switched, a nonmagnetic spacer layer, and nonmagnetic layers or leads on
the backs of the magnetic layers; the current is perpendicular to the
plane of the layers. I find the spin-torques acting on the thin magnetic
layer, spin accumulation and current profiles in the whole structure for
different angles between the magnetization directions in the layers and
different thicknesses of the thin layer. The spin angular momentum
transferred from the polarized current to the background magnetization of
the thin magnetic layer far exceeds the transverse component (to the
orientation of the magnetization of the thin layer) of the bare portion of
the incoming spin-polarized current (the part proportional to the electric
field). Due to the presence of long longitudinal spin-diffusion lengths,
the longitudinal and transverse components of the spin accumulations
become intertwined from one layer to the next, leading to a significant
amplification of the spin torque with respect to the treatments that
neglect spin accumulation about the interfaces. I estimate the scale of
the amplification as $\lambda_{sdl}/\lambda_J$, the ratio of the
spin-diffusion length to the characteristic length scale of the transverse
component of the spin accumulation. The large enhancment of the spin
current in the free ferromagnetic layer may lead to the reduction of the
critical current necessary to switch spintronics devices.


\appendix                           
\chapter{Appendices for interface resistance chapters}
\section{\label{app_algebra} Calculus related to the problem}
In this appendix, several auxiliary equations are obtained, which are used
in Chap.~\ref{chap_res_thr} to derive the equations for the chemical
potential profile in multilayers.
\subsection{\label{del_func} Dirac delta function in $\epsilon$ and 
and $v$ space}
The Fermi-Dirac equilibrium distribution function has the form
$$
f^0(\epsilon)=\frac{1}{\exp(\frac{\epsilon-\epsilon_F}{k_BT})+1},
$$ 
where $\epsilon$ is an electron energy, $\epsilon_F$ is Fermi energy, 
$k_B$ is Boltzmann constant, and $T$ is temperature. At $T\rightarrow 
0$, the derivative of the distribution function over the energy can be 
approximated by the delta function:
$$
\frac{\partial f(\epsilon)}{\partial\epsilon}
\sim\delta(\epsilon-\epsilon_F). 
$$
The electron's energy is a function of its velocity, so delta functions in 
$\epsilon$ space and $v$ space are related via
$$
\delta(\epsilon-\epsilon_F)=
\frac{\delta(v-v_F)}{\partial\epsilon/\partial v},
$$
and, in a free electron model,
\begin{equation}
\label{del_e_v}
\delta(\epsilon-\epsilon_F)=\frac{1}{mv}\delta(v-v_F),
\end{equation}
where $m$ is the electron's mass.
\subsection{\label{ang_av} Angular averaging}
For an arbitrary function $y(|\cos\theta|)$, $y(|\cos\theta|)\sin\theta$ 
is symmetric around $\theta=\pi/2$, and
\begin{equation}
\label{eq_ang_int}
\int_0^{\pi/2} y(|\cos\theta|)\sin\theta d\theta =
\int_{\pi/2}^\pi y(|\cos\theta|)\sin\theta d\theta.
\end{equation}
For example, since a correction to the equilibrium distribution function 
$g({\bf v},z)$ can be written as a function of $z$ and $|\cos\theta|$ 
only, where $\theta$ is an angle that an electron momentum makes with 
$z$ axis (Eq.~(\ref{eq_on_g})), the chemical potential $\mu(z)$ can be 
written as  
\begin{eqnarray*}
2\mu(z)&=&\int_0^\pi g(\theta,z)\sin\theta d\theta \\ \\
&=&\int_0^{\pi/2}
(g(|\cos\theta|,\cos\theta >0,z)+
g(|\cos\theta|,\cos\theta <0,z))\sin\theta d\theta 
\end{eqnarray*}
For an arbitrary function $y({\bf v})$, if it can be written as a 
function of $|\cos\theta|$ only, it follows from Eq.~(\ref{del_e_v}) and 
Eq.~(\ref{eq_ang_int}) that
\begin{eqnarray}
\label{veloc_int}
\int_{FS}d{\bf v}|v_z|\delta(\epsilon-\epsilon_F)y({\bf v})
&\sim&\int_0^\infty\int_0^\pi v^2 dv\,\, 
v|\cos\theta|\frac{1}{v}\delta(v-v_F)y(|cos\theta|)\sin\theta d\theta 
\nonumber \\
&=&v_F^2\int_0^{\pi/2}y(|cos\theta|)\cos\theta\sin\theta d\theta
\end{eqnarray}
\subsection{\label{duff_scat_term} Diffuse scattering term $F(g)$ in 
the boundary conditions}
By using the definition of the correction to the equilibrium distribution 
function $g({\bf v},z)$ (Eq.~(\ref{eq_on_g})), the term in the boundary 
conditions responsible for the diffuse scattering at an interface $F(g)$ 
(Eq.~(\ref{diff_scat_term_f})) can be written as 
$$
F(g)=\frac{1}{\Omega}\left(
\int d{\bf v}_2|v^<_{2z}|\delta(\epsilon-\epsilon_F)g(v_2^<,0)+
\int d{\bf v}_1|v^>_{1z}|\delta(\epsilon-\epsilon_F)g(v_1^>,0)
\right),
$$
where ${\bf v}_i$ is an electron velocity in $i$-th layer, and
$$
\Omega=\int d{\bf v}_2|v^<_{2z}|\delta(\epsilon-\epsilon_F)+
\int d{\bf v}_1|v^>_{1z}|\delta(\epsilon-\epsilon_F).
$$
Then, with the help of Eq.~(\ref{veloc_int}), 
\begin{eqnarray}
\label{func_F}
F(g)&=&\frac{2v^2_{F1}}{v^2_{F1}+v^2_{F2}}
\int_0^{\pi/2}g(v_1^>,0)\cos\theta\sin\theta d\theta \nonumber \\
&+&\frac{2v^2_{F2}}{v^2_{F1}+v^2_{F2}}
\int_0^{\pi/2}g(v_2^<,0)\cos\theta\sin\theta d\theta
\end{eqnarray}
is obtained.


\section{\label{app_curr_cons} Current conservation across an 
interface}
Electrical current $j(z)$ is defined in terms of the distribution 
function as 
$$
j_1(z)=\int_{FS}d{\bf v}_1|v_{1z}|(f(v^>_1,z)-f(v^<_1,z))
$$
in the left layer, and
$$
j_2(z)=\int_{FS}d{\bf v}_1|v_{2z}|(f(v^>_2,z)-f(v^<_2,z))
$$
in the right layer. At an interface, the electron distribution functions
$f(v^>_1,0)$, $f(v^<_1,0)$, $f(v^>_2,0)$, and $f(v^<_2,0)$ are related
via boundary conditions~(\ref{bound_cond_f})-(\ref{diff_scat_term_f}). 
Since the boundary condition is a linear combination of the specular
($S=1$) and diffuse ($S=0$) scattering terms, it is enough to consider
current conservation separately for $S=1$ and $S=0$.

If only specular scattering is present at an interface, current to the 
left of the interface, $j_1(z=0-, S=1)$, may be written as
$$
j_1(z=0-, S=1)=v^2_{F1}\int_0^{\pi/2}\cos\theta_1\sin\theta_1 d\theta_1
(g(v^>_1,0)(1-R_{12}(\theta_1))-T_{21}(\theta_2)g(v^<_2,0)),
$$
and current to the right of the interface, $j_2(z=0+)$, as
$$
j_2(z=0+, S=1)=v^2_{F2}\int_0^{\pi/2}\cos\theta_2\sin\theta_2 d\theta_2
(g(v^>_1,0)T_{12}(\theta_1)-(1-R_{21}(\theta_2))g(v^<_2,0)),
$$ 
where $\theta_1$ ($\theta_2$) is the angle of incidence (transmission) 
to the left (right) of the interface. In the process of specular 
scattering, transmission and reflection coefficients are defined so that 
$T_{ij}(\theta_i)=1-R_{ij}(\theta_i)$, and angles $\theta_1$ and 
$\theta_2$ are related (Appendix~\ref{app_refl_transm}), so that
$$
v^2_{F1}\cos\theta_1\sin\theta_1 d\theta_1=
v^2_{F2}\cos\theta_2\sin\theta_2 d\theta_2,
$$
and electrical current is conserved at the interface:
$$
j_1(z=0-, S=1)=j_2(z=0+, S=1).
$$

If only diffuse scattering is present at an interface, current to the
left of the interface, $j_1(z=0-, S=0)$, may be written as
\begin{eqnarray*}
j_1(z=0-, S=0)&=&v^2_{F1}\int_0^{\pi/2}\cos\theta_1\sin\theta_1 
d\theta_1
(g(v^>_1,0)-F(g)) \\
&=&v^2_{F1}\int_0^{\pi/2}\cos\theta_1\sin\theta_1 
d\theta_1g(v^>_1,0)-\frac{1}{2}v^2_{F1}F(g)
\end{eqnarray*}
where $F(g)$ is given by Eq.~(\ref{func_F}). Then, expression for 
$j_1(z=0-, S=0)$ takes the form
$$
j_1(z=0-, S=0)=\frac{v^2_{F1}v^2_{F2}}{v^2_{F1}+v^2_{F2}}
\int_0^{\pi/2}\cos\theta_1\sin\theta_1 d\theta_1
(g(v^>_1,0)-g(v^<_2,0)).
$$
Similarly, current to the right of the interface, $j_2(z=0+, S=0)$, is
$$
j_2(z=0+, S=0)=\frac{v^2_{F1}v^2_{F2}}{v^2_{F1}+v^2_{F2}}
\int_0^{\pi/2}\cos\theta_2\sin\theta_2 d\theta_2
(g(v^>_1,0)-g(v^<_2,0)).
$$
Since angles $\theta_1$ and $\theta_2$ are not related in the process of 
diffuse scattering, it follows that
$$
j_1(z=0-, S=0)=j_2(z=0+, S=0).
$$

If both specular and diffuse scattering are present at an interface, so 
that $S\neq 0$ and $S\neq 1$, the total electrical current 
$j_{1(2)}(z=0)$ can be written as a linear combination of 
$j_{1(2)}(z=0, S=1)$ and $j_{1(2)}(z=0, S=0)$:
$$
j_1(z=0-)=Sj_1(z=0-,S=1)+(1-S)j_1(z=0-,S=0)
$$
$$
j_2(z=0+)=Sj_2(z=0+,S=1)+(1-S)j_2(z=0+,S=0)
$$
Hence, total electrical current at the interface is conserved by the 
boundary conditions~(\ref{bound_cond_f})-(\ref{diff_scat_term_f}):
\begin{equation}
j_1(z=0-)=j_2(z=0+)
\end{equation}


\section{\label{app_refl_transm} Specular reflection and 
transmission at the potential step}

Consider an electron with the Fermi energy $\epsilon_F$ moving in a
one-dimensional step-like potential $V(z)=V_1$ if $z<0$, $V(z)=V_2$ if
$z>0$. (Fig.~\ref{spec_scat}a)) For example, this corresponds to a free
electron motion across an interface between two layers of different metals,
or layers of the same magnetic metals with antiparallel magnetization
directions.

If electron energy is referenced from $\epsilon_F$, so that
$\epsilon_F=0$, energy conservation law takes the following form:
\begin{equation}
\label{en_cons_tot}
V_1+\frac{mv^2_{F1}}{2}=V_2+\frac{mv^2_{F2}}{2}=\epsilon_F=0,
\end{equation}
where an effective electron mass $m$ is assumed to be the same in both
materials. Electron Fermi velocities $v_{F1}$ to the left of the step
($z<0$) and $v_{F2}$ to the right of the step ($z>0$) are related via
$$\frac{v_{F1}}{v_{F2}}=\sqrt{\frac{|V_1|}{|V_2|}}.$$ 
\begin{figure}
\includegraphics[width=\textwidth]{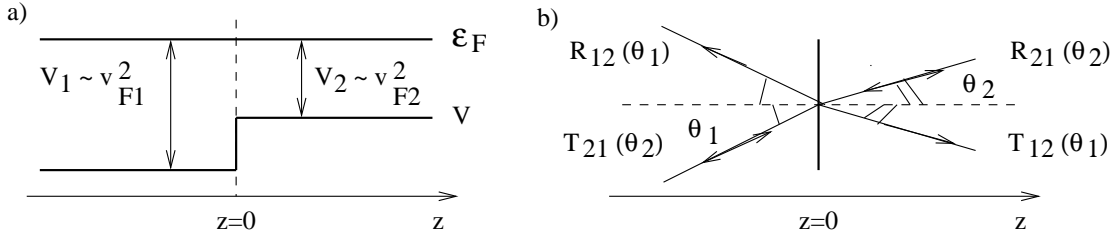}
\caption[Specular scattering at the potential barrier]{Specular 
scattering at the potential barrier: a) Motion in a step-like 
potential, b) reflection and transmission at an interface.}
\label{spec_scat}
\end{figure}

Electron velocity ${\bf v}$ and momentum ${\bf k}$ are related as $m{\bf
v}=\hbar{\bf k}$. From the one-dimensional geometry of the problem it
follows that the component of an electron momentum parallel to $z=0$
plane is conserved, so that the energy conservation law
(Eq.~(\ref{en_cons_tot})) can be written in terms of $z$-component of 
the momentum $k_z=k_F\cos\theta$ only:
\begin{equation}
\label{en_cons_kz}
-\frac{2m|V_1|}{\hbar^2}+k^2_{F1}\cos^2\theta_1=
-\frac{2m|V_2|}{\hbar^2}+k^2_{F2}\cos^2\theta_2,
\end{equation}
where $\theta_1$ ($\theta_2$) is the angle that an electron momentum 
makes with $z$ axis in the left (right) layer.

Condition~(\ref{en_cons_kz}) means that at an interface, the angle of 
incidence of an electron traveling from left to right $\theta_1$ and 
its angle of transmittance $\theta_2$, or, vice versa, the angle of 
incidence of an electron traveling from right to left $\theta_2$ and 
its angle of transmittance $\theta_1$ are related as follows
(Fig.~\ref{spec_scat}b)):
\begin{equation}
\label{theta2_theta1}
\left\{
\begin{array}{ccl}
\cos\theta_2&=&\cos\theta_1\frac{v_{F1}}{v_{F2}}
Re\sqrt{1-\frac{1}{\cos^2\theta_1}+
\frac{v^2_{F2}}{v^2_{F1}}\frac{1}{\cos^2\theta_1}} \\ \\
\sin\theta_2 d\theta_2&=&\sin\theta_1 d\theta_1\frac{v_{F1}}{v_{F2}}
\frac{1}{Re\sqrt{1-\frac{1}{\cos^2\theta_1}+
\frac{v^2_{F2}}{v^2_{F1}}\frac{1}{\cos^2\theta_1}}},
\end{array}
\right.
\end{equation}
or
\begin{equation}
\label{theta1_theta2}
\left\{
\begin{array}{ccl}
$$
\cos\theta_1&=&\cos\theta_2\frac{v_{F2}}{v_{F1}}
Re\sqrt{1-\frac{1}{\cos^2\theta_2}+
\frac{v^2_{F1}}{v^2_{F2}}\frac{1}{\cos^2\theta_2}}  \\ \\
\sin\theta_1 d\theta_1&=&\sin\theta_2 d\theta_2\frac{v_{F2}}{v_{F1}}
\frac{1}{Re\sqrt{1-\frac{1}{\cos^2\theta_2}+ 
\frac{v^2_{F1}}{v^2_{F2}}\frac{1}{\cos^2\theta_2}}}.
\end{array}
\right.
\end{equation}
  
Quantum mechanically, an electron is described by the wave function
$\psi$, which, in the case of the electron moving in a constant
potential, has the form of a plane wave, $\psi=e^{i{\bf k}{\bf r}}$.  At
the potential barrier along $z$ direction, an electron experiences
scattering, so that its wave function is only partially transmitted
across the barrier, and partially reflected. For an electron moving from
left to right, two plane waves, incident and reflected, are present in
the left layer, and transmitted plane wave is present in the right
layer:~(Fig.~\ref{spec_scat}b)) 
\begin{equation}
\label{psi_12}
\left\{
\begin{array}{ccl}
\psi_1&\sim&e^{ik_{1z}z}+re^{-ik_{1z}z} \\
\psi_2&\sim&te^{ik_{2z}z},
\end{array}
\right.
\end{equation}
where $r$ ($t$) is the reflection (transmission) coefficient for the
wave function $\psi$. At an interface between layers, both electrical
charge density, proportional to $|\psi|^2$, and quantum mechanical
flux density, proportional to
$\psi\psi^{\star\prime}-\psi^\star\psi^\prime$, have to be conserved,
which requires continuity of both $\psi$ and $\psi^\prime$ at $z=0$:
$$\psi_1(z=0)=\psi_2(z=0), \,\,\,\,
\psi^\prime_1(z=0)=\psi^\prime_2(z=0).$$
These continuity conditions, together with the expressions for the wave
function in the left and right layers~(\ref{psi_12}) lead to the
following expressions for the reflection and transmission coefficients
for the wave function:
\begin{equation}
\label{r_t}
\left\{
\begin{array}{ccl}
1+r&=&t \\
k_{1z}(1-r)&=&k_{2z}t
\end{array}
\right.
\Rightarrow
\left\{
\begin{array}{ccl}
r&=&\frac{k_{1z}-k_{2z}}{k_{1z}+k_{2z}}\\ \\
t&=&\frac{2k_{1z}}{k_{1z}+k_{2z}}
\end{array}
\right.
\end{equation}
In the boundary conditions~(\ref{bound_cond_f}) for the semiclassical 
Boltzmann equation, reflection and transmission coefficients for an 
electron {\it flux} are required. It follows from the quantum mechanical 
flux conservation condition that $r$ and $t$ are related via
$$1=|r|^2+\frac{k_{2z}}{k_{1z}}|t|^2,$$
so it is convenient to define the reflection (transmission) coefficient
for the flux in terms of the reflection (transmission) coefficient for
the wave function as
$$R=|r|^2, \,\,\,\,  T=\frac{k_{2z}}{k_{1z}}|t|^2,$$
so that the condition $T=1-R$ is satisfied. Reflection coefficient for 
the electron flux going from left to right takes the following form:
\begin{equation} 
\label{R_12}
R_{12}(\theta_1)=|r_{12}(\theta_1)|^2=\left|
\frac{1-\sqrt{1-\frac{1}{\cos^2\theta_1}+
\frac{v^2_{F2}}{v^2_{F1}}\frac{1}{\cos^2\theta_1}}}
{1+\sqrt{1-\frac{1}{\cos^2\theta_1}+
\frac{v^2_{F2}}{v^2_{F1}}\frac{1}{\cos^2\theta_1}}}\right|^2.
\end{equation}

In a similar fashion, reflection coefficient for the electron flux going 
from the right to the left layer can be obtained:
\begin{equation}
\label{R_21}
R_{21}(\theta_2)=|r_{21}(\theta_2)|^2=\left|
\frac{1-\sqrt{1-\frac{1}{\cos^2\theta_2}+
\frac{v^2_{F1}}{v^2_{F2}}\frac{1}{\cos^2\theta_2}}}
{1+\sqrt{1-\frac{1}{\cos^2\theta_2}+
\frac{v^2_{F1}}{v^2_{F2}}\frac{1}{\cos^2\theta_2}}}\right|^2.
\end{equation}

Expressions (\ref{R_12})-(\ref{R_21}) together with the relations on the 
angles $\theta_1$ and 
$\theta_2$ (\ref{theta1_theta2})-(\ref{theta2_theta1}) lead to the 
microscopic reversibility condition
\begin{equation}
R_{ij}(\theta_i)=R_{ji}(\theta_j).
\end{equation} 
Figure~\ref{refl_graph} shows schematically how the reflection 
coefficients $R_{12}$ and $R_{21}$ depend on the cosine of the angle of 
incidence $\theta_1$ or $\theta_2$ for two different geometries of the 
potential barrier, $V_2>V_1$ and  $V_2<V_1$.
\begin{figure}
\includegraphics[width=\textwidth]{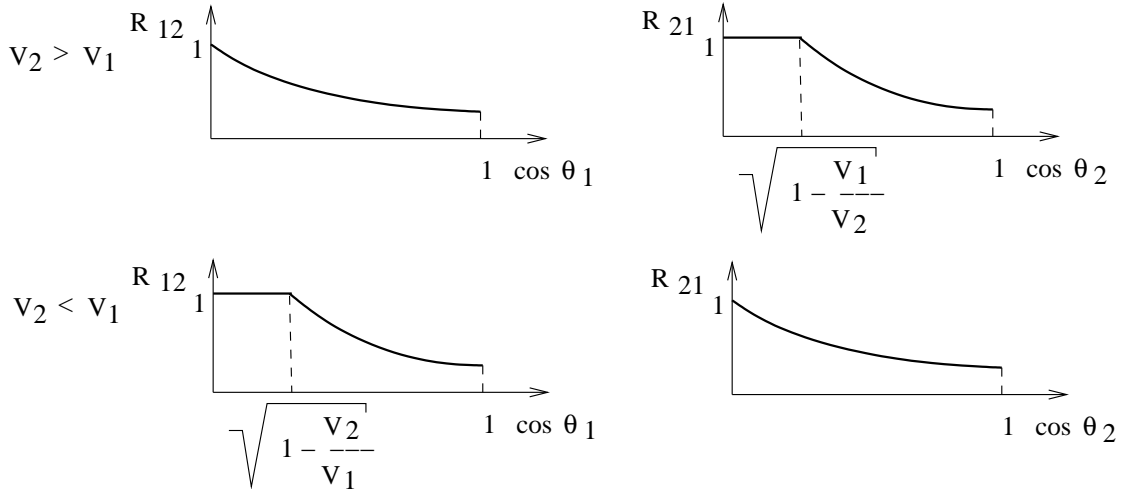}
\caption{Reflection coefficient at the potential barrier as a function of 
the angle of incidence}
\label{refl_graph}
\end{figure}


\section{\label{app_three_lrs} Equations for chemical potential for 
three-layered system}
In this appendix, equations for the chemical potential profile in the 
system consisting of three metallic layers (Fig.~\ref{three_layers}) are 
presented without derivation. The same electron relaxation time $\tau$ is 
assumed in all three metals, and a dimensionless variable $\xi=z/l_i$ is 
introduced, where $l_i=v_{Fi}\tau$ is an electron's mean free path in the 
$i$-th layer, so that $\mu_i=\mu_i(\xi=z/l_i)$, and $d_i=d/l_i$, where 
$2d$ is the thickness of the middle layer.
\begin{figure}
\centering
\includegraphics[width=4in]{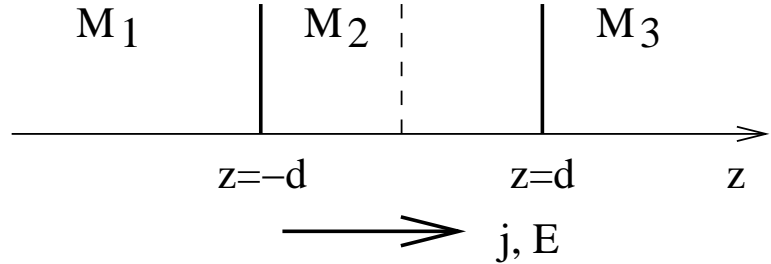}
\caption[System consisting of three layers]{System consisting of three 
layers.}
\label{three_layers}
\end{figure}

\begin{eqnarray}
\label{eq_mu3_1}
2\mu_1(\xi)&=&eE_1l_1\int_0^{\pi/2}
\left[(1+SR_{12}(\theta_1))\cos\theta_1
-\frac{v^2_{F1}}{v^2_{F2}}ST_{12}(\theta_1)\cos\theta_2(\theta_1)\right]
\nonumber \\
&\times&\exp{\left(\frac{\xi+d_1}{\cos\theta_1}\right)}\sin\theta_1d\theta_1
\\
&+&\int_{-\infty}^{-d_1}\mu_1(\xi^\prime)d\xi^\prime \nonumber \\
&\times&\int_0^{\pi/2}
\left(\exp{\left(-\frac{|\xi-\xi^\prime|}{\cos\theta_1}\right)}
+SR_{12}(\theta_1)\exp{\left(\frac{\xi+\xi^\prime+2d_1}{\cos\theta_1}\right)}
\right)\tan\theta_1d\theta_1 \nonumber \\
&+&\int_{-d_2}^{d_2}\mu_2(\xi^\prime)d\xi^\prime \nonumber \\
&\times&\int_0^{\pi/2}ST_{12}(\theta_1)
\exp{\left(\frac{-\xi^\prime-d_2}{\cos\theta_2(\theta_1)}\right)}
\exp{\left(\frac{-\xi+d_1}{\cos\theta_1}\right)}
\frac{\sin\theta_1}{\cos\theta_2(\theta_1)}d\theta_1 \nonumber \\
&+&\frac{2(1-S)v^2_{F1}}{v^2_{F1}+v^2_{F2}}
\int_{-\infty}^{-d_1}\mu_1(\xi^\prime)d\xi^\prime
\int_0^{\pi/2}\exp{\left(\frac{\xi^\prime+d_1}{\cos\theta}\right)}\sin\theta 
\nonumber \\
&\times&\int_0^{\pi/2}\exp{\left(\frac{\xi+d_1}{\cos\theta}\right)}\sin\theta
\nonumber \\
&+&\frac{2(1-S)v^2_{F2}}{v^2_{F1}+v^2_{F2}}
\int_{-d_2}^{d_2}\mu_2(\xi^\prime)d\xi^\prime
\int_0^{\pi/2}\exp{\left(\frac{-\xi^\prime-d_2}{\cos\theta}\right)}\sin\theta
\nonumber \\
&\times&\int_0^{\pi/2}\exp{\left(\frac{\xi+d_1}{\cos\theta}\right)}\sin\theta
\nonumber \\
&+&\int_0^{\pi/2}ST_{21}(\theta_2)A(\theta_2)
\exp{\left(\frac{-2d_2}{\cos\theta_2}\right)}
\exp{\left(\frac{\xi+d_1}{\cos\theta_1(\theta_2)}\right)}
\sin\theta_1(\theta_2)d\theta_1(\theta_2) \nonumber \\
&+&\frac{2(1-S)v^2_{F2}}{v^2_{F1}+v^2_{F2}}
\int_0^{\pi/2}A(\theta_2)\exp{\left(\frac{-2d_2}{\cos\theta}\right)}
\sin\theta d\theta
\int_0^{\pi/2}\exp{\left(\frac{\xi+d_1}{\cos\theta}\right)}
\sin\theta d\theta \nonumber
\end{eqnarray}
 
\begin{eqnarray}
\label{eq_mu3_2}
2\mu_2(\xi)&=&\int_{-d_2}^{d_2}\mu_2(\xi^\prime)d\xi^\prime
\int_0^{\pi/2}
\exp{\left(-\frac{|\xi-\xi^\prime|}{\cos\theta_2}\right)}\tan\theta_2 
d\theta_2 \nonumber \\
&+&\int_0^{\pi/2}A(\theta_2)
\exp{\left(\frac{\xi-d_2}{\cos\theta_2}\right)}\sin\theta_2 
d\theta_2 \\
&+&\int_0^{\pi/2}B(\theta_2)
\exp{\left(\frac{-\xi-d_2}{\cos\theta_2}\right)}\sin\theta_2 
d\theta_2 \nonumber
\end{eqnarray}

\begin{eqnarray}
\label{eq_mu3_3}
2\mu_3(\xi)&=&-eE_3l_3\int_0^{\pi/2}
\left[(1+SR_{32}(\theta_3))\cos\theta_3
-\frac{v^2_{F3}}{v^2_{F2}}ST_{32}(\theta_3)\cos\theta_2(\theta_3)\right]
\nonumber \\
&\times&\exp{\left(\frac{-\xi+d_3}{\cos\theta_3}\right)}\sin\theta_3d\theta_3
\\
&+&\int^{\infty}_{d_3}\mu_3(\xi^\prime)d\xi^\prime \nonumber \\
&\times&\int_0^{\pi/2}
\left(\exp{\left(-\frac{|\xi-\xi^\prime|}{\cos\theta_3}\right)}
+SR_{32}(\theta_3)\exp{\left(\frac{-\xi-\xi^\prime+2d_3}{\cos\theta_3}\right)}
\right)\tan\theta_3d\theta_3 \nonumber \\
&+&\int_{-d_2}^{d_2}\mu_2(\xi^\prime)d\xi^\prime \nonumber \\
&\times&\int_0^{\pi/2}ST_{32}(\theta_3)
\exp{\left(\frac{\xi^\prime-d_2}{\cos\theta_2(\theta_3)}\right)}
\exp{\left(\frac{-\xi+d_3}{\cos\theta_3}\right)}
\frac{\sin\theta_3}{\cos\theta_2(\theta_3)}d\theta_3 \nonumber \\
&+&\frac{2(1-S)v^2_{F3}}{v^2_{F3}+v^2_{F2}}
\int^{\infty}_{d_3}\mu_3(\xi^\prime)d\xi^\prime
\int_0^{\pi/2}\exp{\left(\frac{-\xi^\prime+d_3}{\cos\theta}\right)}\sin\theta 
\nonumber \\
&\times&\int_0^{\pi/2}\exp{\left(\frac{-\xi+d_3}{\cos\theta}\right)}\sin\theta
\nonumber \\
&+&\frac{2(1-S)v^2_{F2}}{v^2_{F3}+v^2_{F2}}
\int_{-d_2}^{d_2}\mu_2(\xi^\prime)d\xi^\prime
\int_0^{\pi/2}\exp{\left(\frac{\xi^\prime-d_2}{\cos\theta}\right)}\sin\theta
\nonumber \\
&\times&\int_0^{\pi/2}\exp{\left(\frac{-\xi+d_3}{\cos\theta}\right)}\sin\theta
\nonumber \\
&+&\int_0^{\pi/2}ST_{23}(\theta_2)B(\theta_2)
\exp{\left(\frac{-2d_2}{\cos\theta_2}\right)}
\exp{\left(\frac{-\xi+d_3}{\cos\theta_3(\theta_2)}\right)}
\sin\theta_3(\theta_2)d\theta_3(\theta_2) \nonumber \\
&+&\frac{2(1-S)v^2_{F2}}{v^2_{F3}+v^2_{F2}}
\int_0^{\pi/2}B(\theta_2)\exp{\left(\frac{-2d_2}{\cos\theta}\right)}
\sin\theta d\theta
\int_0^{\pi/2}\exp{\left(\frac{-\xi+d_3}{\cos\theta}\right)}
\sin\theta d\theta \nonumber
\end{eqnarray}
where
\begin{eqnarray}
\label{eq_A3}
A(\theta_2)&=&eE_2l_2|\cos\theta_2|(1+SR_{23}(\theta_2))
-ST_{23}(\theta_2)eE_3l_3|\cos\theta_3(\theta_2)| \nonumber \\
&+&SR_{23}(\theta_2)B(\theta_2)\exp{\left(\frac{-2d_2}{\cos\theta_2}\right)}
+\frac{2(1-S)v^2_{F2}}{v^2_{F3}+v^2_{F2}}
\int_0^{\pi/2}B(\theta)\exp{\left(\frac{-2d_2}{\cos\theta}\right)}
\sin\theta d\theta \nonumber \\
&+&SR_{23}(\theta_2)\int_{-d_2}^{d_2}
\exp{\left(\frac{\xi^\prime-d_2}{\cos\theta_2}\right)}
\frac{\mu_2(\xi^\prime)}{\cos\theta_2}d\xi^\prime \\
&+&ST_{23}(\theta_2)\int^{\infty}_{d_3}
\exp{\left(\frac{-\xi^\prime+d_3}{\cos\theta_3(\theta_2)}\right)}
\frac{\mu_3(\xi^\prime)}{\cos\theta_3(\theta_2)}d\xi^\prime \nonumber \\
&+&\frac{2(1-S)v^2_{F2}}{v^2_{F3}+v^2_{F2}}
\int_{-d_2}^{d_2}\mu_2(\xi^\prime)d\xi^\prime
\int_0^{\pi/2}\exp{\left(\frac{\xi^\prime-d_2}{\cos\theta}\right)}
\sin\theta d\theta \nonumber \\
&+&\frac{2(1-S)v^2_{F3}}{v^2_{F3}+v^2_{F2}}
\int^{\infty}_{d_3}\mu_3(\xi^\prime)d\xi^\prime
\int_0^{\pi/2}\exp{\left(\frac{-\xi^\prime+d_3}{\cos\theta}\right)}
\sin\theta d\theta \nonumber
\end{eqnarray}
  
\begin{eqnarray}
\label{eq_B3}
B(\theta_2)&=&-eE_2l_2|\cos\theta_2|(1+SR_{21}(\theta_2))
+ST_{21}(\theta_2)eE_1l_1|\cos\theta_1(\theta_2)| \nonumber \\
&+&SR_{21}(\theta_2)A(\theta_2)\exp{\left(\frac{-2d_2}{\cos\theta_2}\right)}
+\frac{2(1-S)v^2_{F2}}{v^2_{F1}+v^2_{F2}}
\int_0^{\pi/2}A(\theta)\exp{\left(\frac{-2d_2}{\cos\theta}\right)}
\sin\theta d\theta \nonumber \\
&+&SR_{21}(\theta_2)\int_{-d_2}^{d_2}
\exp{\left(\frac{-\xi^\prime-d_2}{\cos\theta_2}\right)}
\frac{\mu_2(\xi^\prime)}{\cos\theta_2}d\xi^\prime \\
&+&ST_{21}(\theta_2)\int_{-\infty}^{-d_1}
\exp{\left(\frac{\xi^\prime+d_1}{\cos\theta_1(\theta_2)}\right)}
\frac{\mu_1(\xi^\prime)}{\cos\theta_1(\theta_2)}d\xi^\prime \nonumber \\
&+&\frac{2(1-S)v^2_{F2}}{v^2_{F1}+v^2_{F2}}
\int_{-d_2}^{d_2}\mu_2(\xi^\prime)d\xi^\prime
\int_0^{\pi/2}\exp{\left(\frac{-\xi^\prime-d_2}{\cos\theta}\right)}
\sin\theta d\theta \nonumber \\
&+&\frac{2(1-S)v^2_{F1}}{v^2_{F1}+v^2_{F2}}
\int_{-\infty}^{-d_1}\mu_1(\xi^\prime)d\xi^\prime
\int_0^{\pi/2}\exp{\left(\frac{\xi^\prime+d_1}{\cos\theta}\right)}
\sin\theta d\theta \nonumber
\end{eqnarray}
  

\section{\label{app_five_lrs} Equations for chemical potential for 
five-layered system}
In this appendix, equations for the chemical potential profile in the
system consisting of five metallic layers (Fig.~\ref{five_layers}) are
presented without derivation. For simplicity, only symmetrical
structures $M_1$-$M_2$-$M_3$-$M_2$-$M_1$ are considered. The same electron
relaxation time $\tau$ is assumed in all three metals, and a dimensionless
variable $\xi=z/l_i$ is introduced, where $l_i=v_{Fi}\tau$ is an
electron mean free path in the $i$-th layer, so that 
$\mu_i=\mu_i(\xi=z/l_i)$, and $d_i=d/l_i$, $b_i=b/l_i$, where  
$2d$ is the thickness of the middle layer, $2b$ is the thickness of the 
second and fourth layers. 
\begin{figure}
\centering
\includegraphics[width=5in]{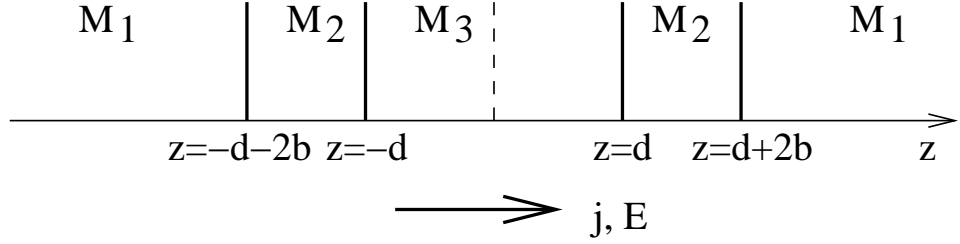}
\caption[System consisting of five layers]{System consisting of five 
layers.}
\label{five_layers}
\end{figure}

\begin{eqnarray}
\label{eq_mu5_1}
2\mu_1(\xi)&=&\int_{-\infty}^{-d_1-2b_1}\mu_1(\xi^\prime)d\xi^\prime
\int_0^{\pi/2}\exp{\left(-\frac{|\xi-\xi^\prime|}{\cos\theta_1}\right)}
\tan\theta_1d\theta_1 \\ 
&+&\int_0^{\pi/2}A(\theta_1)
\exp{\left(\frac{\xi+d_1+2b_1}{\cos\theta_1}\right)}
\sin\theta_1d\theta_1 \nonumber
\end{eqnarray}

\begin{eqnarray}
\label{eq_mu5_2}
2\mu_2(\xi)&=&\int_{-d_2-2b_2}^{-d_2}\mu_2(\xi^\prime)d\xi^\prime
\int_0^{\pi/2}\exp{\left(-\frac{|\xi-\xi^\prime|}{\cos\theta_2}\right)}
\tan\theta_2d\theta_2 \\
&+&\int_0^{\pi/2}B_1(\theta_2)
\exp{\left(-\frac{\xi+d_2+2b_2}{\cos\theta_2}\right)}
\sin\theta_2d\theta_2  \nonumber \\
&+&\int_0^{\pi/2}B_2(\theta_2)
\exp{\left(\frac{\xi+d_2}{\cos\theta_2}\right)}
\sin\theta_2d\theta_2  \nonumber  
\end{eqnarray}

\begin{eqnarray}
\label{eq_mu5_3}
2\mu_3(\xi)&=&\int_{-d_3}^{d_3}\mu_3(\xi^\prime)d\xi^\prime
\int_0^{\pi/2}\exp{\left(-\frac{|\xi-\xi^\prime|}{\cos\theta_3}\right)}
\tan\theta_3d\theta_3 \\
&+&\int_0^{\pi/2}C(\theta_3)
\left(\exp{\left(\frac{-\xi-d_3}{\cos\theta_3}\right)}
-\exp{\left(\frac{\xi-d_3}{\cos\theta_3}\right)}\right)
\sin\theta_3d\theta_3 \nonumber
\end{eqnarray}

\begin{eqnarray}
\label{eq_mu5_4}
\mu_4(\xi)=-\mu_2(-\xi)
\end{eqnarray}

\begin{eqnarray}
\label{eq_mu5_5}
\mu_5(\xi)=-\mu_1(-\xi)
\end{eqnarray}
where
\begin{eqnarray}
\label{eq_A5}
A(\theta_1)&=&eE_1l_1(1+SR_{12}(\theta_1))\cos\theta_1
-eE_2l_2ST_{12}(\theta_1)\cos\theta_2(\theta_1) \nonumber \\
&+&SR_{12}(\theta_1)\int_{-\infty}^{-d_1-2b_1}
\exp{\left(\frac{\xi^\prime+d_1+2b_1}{\cos\theta_1}\right)}
\frac{\mu_1(\xi^\prime)}{\cos\theta_1}d\xi^\prime \\
&+&\frac{2(1-S)v^2_{F1}}{v^2_{F1}+v^2_{F2}}
\int_{-\infty}^{-d_1-2b_1}\mu_1(\xi^\prime)d\xi^\prime
\int_0^{\pi/2}\exp{\left(\frac{\xi^\prime+d_1+2b_1}{\cos\theta}\right)}
\sin\theta d\theta \nonumber \\
&+&ST_{12}(\theta_1)\int_{-d_2-2b_2}^{-d_2}
\exp{\left(\frac{-\xi^\prime-d_2-2b_2}{\cos\theta_2(\theta_1)}\right)}
\frac{\mu_2(\xi^\prime)}{\cos\theta_2(\theta_1)}d\xi^\prime \nonumber \\
&+&\frac{2(1-S)v^2_{F2}}{v^2_{F1}+v^2_{F2}}
\int_{-d_2-2b_2}^{-d_2}\mu_2(\xi^\prime)d\xi^\prime
\int_0^{\pi/2}\exp{\left(\frac{-\xi^\prime-d_2-2b_2}{\cos\theta}\right)}
\sin\theta d\theta \nonumber \\
&+&ST_{12}(\theta_1)B_2(\theta_2(\theta_1))
\exp{\left(\frac{-2b_2}{\cos\theta_2(\theta_1)}\right)} \nonumber \\
&+&\frac{2(1-S)v^2_{F2}}{v^2_{F1}+v^2_{F2}}
\int_0^{\pi/2}B_2(\theta)
\exp{\left(\frac{-2b_2}{\cos\theta}\right)}\cos\theta\sin\theta d\theta 
\nonumber
\end{eqnarray}

\begin{eqnarray}
\label{eq_B5_1}
B_1(\theta_1)&=&-eE_2l_2(1+SR_{21}(\theta_2))\cos\theta_2
+eE_1l_1ST_{21}(\theta_2)\cos\theta_1(\theta_2) \nonumber \\
&+&ST_{21}(\theta_2)\int_{-\infty}^{-d_1-2b_1}
\exp{\left(\frac{\xi^\prime+d_1+2b_1}{\cos\theta_1(\theta_2)}\right)}
\frac{\mu_1(\xi^\prime)}{\cos\theta_1(\theta_2)}d\xi^\prime \\
&+&\frac{2(1-S)v^2_{F1}}{v^2_{F1}+v^2_{F2}}
\int_{-\infty}^{-d_1-2b_1}\mu_1(\xi^\prime)d\xi^\prime
\int_0^{\pi/2}\exp{\left(\frac{\xi^\prime+d_1+2b_1}{\cos\theta}\right)}
\sin\theta d\theta \nonumber \\
&+&SR_{21}(\theta_2)\int_{-d_2-2b_2}^{-d_2}
\exp{\left(\frac{-\xi^\prime-d_2-2b_2}{\cos\theta_2}\right)}
\frac{\mu_2(\xi^\prime)}{\cos\theta_2}d\xi^\prime \nonumber \\
&+&\frac{2(1-S)v^2_{F2}}{v^2_{F1}+v^2_{F2}}
\int_{-d_2-2b_2}^{-d_2}\mu_2(\xi^\prime)d\xi^\prime
\int_0^{\pi/2}\exp{\left(\frac{-\xi^\prime-d_2-2b_2}{\cos\theta}\right)}
\sin\theta d\theta \nonumber \\
&+&SR_{21}(\theta_2)B_2(\theta_2)
\exp{\left(\frac{-2b_2}{\cos\theta_2}\right)} \nonumber \\
&+&\frac{2(1-S)v^2_{F2}}{v^2_{F1}+v^2_{F2}}
\int_0^{\pi/2}B_2(\theta)
\exp{\left(\frac{-2b_2}{\cos\theta}\right)}\cos\theta\sin\theta d\theta
\nonumber
\end{eqnarray}

\begin{eqnarray}
\label{eq_B5_2}
B_2(\theta_2)&=&eE_2l_2(1+SR_{23}(\theta_2))\cos\theta_2
-eE_3l_3ST_{23}(\theta_2)\cos\theta_3(\theta_2) \nonumber \\
&+&SR_{23}(\theta_2)\int_{-d_2-2b_2}^{-d_2}
\exp{\left(\frac{\xi^\prime+d_2}{\cos\theta_2}\right)}
\frac{\mu_2(\xi^\prime)}{\cos\theta_2}d\xi^\prime \\
&+&\frac{2(1-S)v^2_{F2}}{v^2_{F3}+v^2_{F2}}
\int_{-d_2-2b_2}^{-d_2}\mu_2(\xi^\prime)d\xi^\prime
\int_0^{\pi/2}\exp{\left(\frac{\xi^\prime+d_2}{\cos\theta}\right)}
\sin\theta d\theta \nonumber \\
&+&ST_{23}(\theta_2)\int_{-d_3}^{d_3}
\exp{\left(\frac{-\xi^\prime-d_3}{\cos\theta_3(\theta_2)}\right)}
\frac{\mu_3(\xi^\prime)}{\cos\theta_3(\theta_2)}d\xi^\prime 
\nonumber \\
&+&\frac{2(1-S)v^2_{F3}}{v^2_{F3}+v^2_{F2}}
\int_{-d_3}^{d_3}\mu_3(\xi^\prime)d\xi^\prime
\int_0^{\pi/2}\exp{\left(\frac{-\xi^\prime-d_3}{\cos\theta}\right)}
\sin\theta d\theta \nonumber \\
&+&SR_{23}(\theta_2)B_1(\theta_2)
\exp{\left(\frac{-2b_2}{\cos\theta_2}\right)} \nonumber \\
&+&\frac{2(1-S)v^2_{F2}}{v^2_{F3}+v^2_{F2}}
\int_0^{\pi/2}B_1(\theta)
\exp{\left(\frac{-2b_2}{\cos\theta}\right)}\cos\theta\sin\theta d\theta
\nonumber \\
&-&ST_{23}(\theta_2)C(\theta_2)
\exp{\left(\frac{-2d_3}{\cos\theta_3(\theta_2)}\right)} \nonumber \\
&-&\frac{2(1-S)v^2_{F3}}{v^2_{F3}+v^2_{F2}}
\int_0^{\pi/2}C(\theta)
\exp{\left(\frac{-2d_3}{\cos\theta}\right)}\cos\theta\sin\theta d\theta
\nonumber
\end{eqnarray}

\begin{eqnarray}
\label{eq_C5}
C(\theta_3(1&+&SR_{32}(\theta_3))\exp{\left(\frac{2d_3}{\cos\theta_3}\right)})
\nonumber \\
&=&-eE_3l_3(1+SR_{32}(\theta_3))\cos\theta_3
+eE_2l_2ST_{32}(\theta_3)\cos\theta_2(\theta_3) \nonumber \\
&+&ST_{32}(\theta_3)\int_{-d_2-2b_2}^{-d_2}
\exp{\left(\frac{\xi^\prime+d_2}{\cos\theta_2(\theta_3)}\right)}
\frac{\mu_2(\xi^\prime)}{\cos\theta_2(\theta_3)}d\xi^\prime \\
&+&\frac{2(1-S)v^2_{F2}}{v^2_{F3}+v^2_{F2}}
\int_{-d_2-2b_2}^{-d_2}\mu_2(\xi^\prime)d\xi^\prime
\int_0^{\pi/2}\exp{\left(\frac{\xi^\prime+d_2}{\cos\theta}\right)}
\sin\theta d\theta \nonumber \\
&+&SR_{32}(\theta_3)\int_{-d_3}^{d_3}
\exp{\left(\frac{-\xi^\prime-d_3}{\cos\theta_3}\right)}
\frac{\mu_3(\xi^\prime)}{\cos\theta_3}d\xi^\prime
\nonumber \\
&+&\frac{2(1-S)v^2_{F3}}{v^2_{F3}+v^2_{F2}}
\int_{-d_3}^{d_3}\mu_3(\xi^\prime)d\xi^\prime
\int_0^{\pi/2}\exp{\left(\frac{-\xi^\prime-d_3}{\cos\theta}\right)}
\sin\theta d\theta \nonumber \\
&+&ST_{32}(\theta_2)C(\theta_2(\theta_3))
\exp{\left(\frac{-2b_2}{\cos\theta_2(\theta_3)}\right)} \nonumber \\
&+&\frac{2(1-S)v^2_{F2}}{v^2_{F3}+v^2_{F2}}
\int_0^{\pi/2}B_1(\theta)
\exp{\left(\frac{-2b_2}{\cos\theta}\right)}\cos\theta\sin\theta d\theta
\nonumber \\
&-&\frac{2(1-S)v^2_{F3}}{v^2_{F3}+v^2_{F2}}
\int_0^{\pi/2}C(\theta)
\exp{\left(\frac{-2d_3}{\cos\theta}\right)}\cos\theta\sin\theta d\theta
\nonumber
\end{eqnarray}

\section{\label{app_numerics} Numerical procedure of solving the 
Fredholm equation of the second kind}
In this Appendix, the numerical procedure of solving the Fredholm equation
of the second kind, Eq.~(\ref{sol_mu}), will be briefly discussed. The
general form of the Fredholm equation of the second kind is the following:
\begin{equation}
\label{fred_eq}
f(x)=\int_a^b K(x,y)f(y)dy +g(x),
\end{equation}
where $f(x)$ is an unknown function defined on the interval $[a,b]$ of 
the real axis, $K(x,y)$ is the kernel of the integral, and $g(x)$ is the 
source. 

The Nystrom routine is used in to solve the
equation~(\ref{fred_eq}).~\cite{num_rec} It transforms the integral
equation into the system of $N$ linear equations for the unknown vector
$f(x_i)$, where $x_i$ are the points of the mesh at the interval $[a,b]$,
given by the $N$-point Gauss-Legendre rule. This standard
rule~\cite{num_rec} gives the points of the mesh with weights:
$$
gauleg(a,b,{\bf x},{\bf w},N),
$$
where $a$ and $b$ are the boundaries of the interval of interest, ${\bf
x}$ is the vector of the points of the mesh, ${\bf w}$ is the vector of
the weights, and $N$ is the number of points in the mesh. The equation  
for the vector $f(x_i)$ takes the following form:
\begin{equation}
\label{lin_syst}
f(x_i)=\sum^N_{j=1}K(x_i,y_j)f(y_j)w_j + g(x_i),
\end{equation}
or
\begin{equation}
\label{lin_syst_2}
f_i=\sum^N_{j=1}{\bar K}_{ij}f_j + g_i,
\end{equation}
where ${\bar K}_{ij}=w_jK(x_i,y_j)$, or, in the matrix form,
\begin{equation}
\label{lin_syst_3}
(\hat{{\bf 1}}-\hat{{\bar K}}){\bf f}={\bf g}. 
\end{equation}

In the case of a singularity in the kernel, for instance if 
$K(x,y)\rightarrow\infty$ when $x=y$, this singularity is removed from the 
integral by writing 
$$
\int_a^b K(x,y)f(y)dy=\int_a^b K(x,y)(f(y)-f(x))dy + r(x)f(x),
$$ 
where $r(x)=\int_a^b K(x,y)dy$. Instead of the Eq.~(\ref{lin_syst_2}), the 
following system of equations is obtained:
\begin{equation}
\label{lin_syst_4}
f_i(1-r_i+\sum^N_{j=1, j\neq i}{\bar K}_{ij})=\sum^N_{j=1, j\neq i}{\bar 
K}_{ij}f_j + g_i.
\end{equation}

The systems of linear equations~(\ref{lin_syst_2})  or~(\ref{lin_syst_4})
are solved using the {\it ludcmp} and {\it lubksb}
routines.~\cite{num_rec} In order to work with the vectors and matrices,
the standard NR.h package is used, which contains also the standard 
numerical routines $gauleg$, $ludcmp$, and $lubksb$. 

The numerical procedure for solving the system of the Fredholm equations
of the second kind, Eq.~(\ref{sol_mu_many}), is identical to the one for
solving a single equation, and will not be covered here.
 
\section{\label{app_real_struct} Parameters entering the equations for the
chemical potentials for Co-Cu and Fe-Cr systems}
\subsection{\label{cocu_ratio} Electrons Fermi velocities ratios for Co-Cu 
system}
In this appendix, the ratio of the electrons Fermi velocities in the Co-Cu
structure $v_{FCu}/v_{FCo}$ for majority and minority electrons is
obtained using the results presented in Ref.~\cite{stiles_jap} for the
transmission probabilities averaged over the Fermi surface. The
probability for an electron to be transmitted through an interface between
metals $i$ and $j$ as a function of the angle of incidence takes the
following form (see Appendix~\ref{app_refl_transm}):
\begin{equation}
\label{transm_coeff}
T_{ij}(\cos\theta)=\frac{4\sqrt{1+(v^2_{Fj}/v^2_{Fi}-1)/\cos^2\theta}}
{1+\sqrt{1+(v^2_{Fj}/v^2_{Fi}-1)/\cos^2\theta}},
\end{equation}
where $v^2_{Fi}$ and $v^2_{Fj}$ are the electron Fermi velocities in the
metals $i$ and $j$. After averaging over the Fermi surface, the expression
for the transmission probability takes the following form:
\begin{equation}   
\label{transm_avg}
T(A)=\frac{1}{2}\int_0^\pi T_{ij}(\cos\theta, A)\sin\theta d\theta=
4A\int_0^{1/A}\frac{\sqrt{1+1/x^2}}{1+\sqrt{1+1/x^2}},
\end{equation}
where $A=v^2_{Fj}/v^2_{Fi}-1$. The equation $T(A)=T^\star$ is solved
numerically, with $T^\star$ equal to 0.44 for the minority electrons, and
0.66 for the majority electrons.~\cite{stiles_jap} In this illustration, a
(110) interface has been chosen, and the electrons were assumed to go
from Cu into Co. The ratio $v_{FCu}/v_{FCo}$ is found to be 1.05 for
majority electrons, and 1.18 for minority electrons. If the motion from Co
into Cu is considered, the same ratio of $v_{FCu}/v_{FCo}$ is obtained for
majority electrons, but different ratio is obtained for minority
electrons; this suggests that the free electron approximation used to
obtain the transmission probability expression~(\ref{transm_coeff}) is
reasonable for majority states, but not for minority states.
Within the free electron model, the averaging over the Fermi surface is 
performed, and the information about the true electron momenta is lost, 
which leads to different values of $v_{FCu}/v_{FCo}$ for different 
directions of motion.

\subsection{\label{table} The summary of the Fermi velocities ratios 
entering the equations for the chemical potentials}
The following table summarizes the correspondence between the equations
that have to be solved in order to find the resistances of up and down
electrons $R\uparrow$ and $R\downarrow$ in different geometries and
magnetization configurations (Fig.~\ref{real_struct}) and the electron 
Fermi velocities ratios entering these equations:
\begin{center}
\begin{tabular}{|c|c|c|c|c|}\cline{3-5}
  \multicolumn{2}{c|}{}  & Two layers & \multicolumn{2}{c|}{Three or 
five layers} \\ 
\cline{3-5}
  \multicolumn{2}{c|}{}  & Eqs.(\ref{eq_mu_1})-(\ref{eq_mu_2}) & 
\multicolumn{2}{c|}{Eqs.(\ref{eq_mu3_1})-(\ref{eq_B3}) or 
(\ref{eq_mu5_1})-(\ref{eq_C5})} 
\\ \cline{3-5}
  \multicolumn{2}{c|}{} & & parallel & anti-parallel \\ \hline 
CoCu& $R\uparrow$ & $\frac{v_{F2}}{v_{F1}}=1.05$ & 
                     $\frac{v_{F2}}{v_{F1}}=\frac{v_{F2}}{v_{F3}}=1.05$ & 
                     $\frac{v_{F2}}{v_{F1}}=1.05$; 
                     $\frac{v_{F2}}{v_{F3}}=1.18$ 
                     \\ \cline{2-5}
     & $R\downarrow$ & $\frac{v_{F2}}{v_{F1}}=1.18$ &
                       $\frac{v_{F2}}{v_{F1}}=\frac{v_{F2}}{v_{F3}}=1.18$ &
                       $\frac{v_{F2}}{v_{F1}}=1.18$; 
                       $\frac{v_{F2}}{v_{F3}}=1.05$ 
\\ \hline
FeCr& $R\uparrow$ & $\frac{v_{F2}}{v_{F1}}=0.837$ & 
                     $\frac{v_{F2}}{v_{F1}}=\frac{v_{F2}}{v_{F3}}=0.837$ & 
                     $\frac{v_{F2}}{v_{F1}}=0.837$; 
                     $\frac{v_{F2}}{v_{F3}}=1.003$ 
                     \\ \cline{2-5}
     & $R\downarrow$ & $\frac{v_{F2}}{v_{F1}}=1.003$ &
                       $\frac{v_{F2}}{v_{F1}}=\frac{v_{F2}}{v_{F3}}=1.003$ &
                       $\frac{v_{F2}}{v_{F1}}=1.003$; 
                       $\frac{v_{F2}}{v_{F3}}=0.837$ 
\\ \hline 
\end{tabular}
\end{center}

\chapter{Appendices for torques chapters}

\section{\label{app_bound_cond}
Derivation of the boundary conditions for spin-ac\-cumulation and 
spin-current between the layers}

In this appendix, the boundary conditions at the interfaces between the
layers in Fig.~\ref{pic_multi} are derived. To achieve this goal, I
consider a sub-system shown in Fig.~\ref{pic_bound_cond} which consists
of a semi-infinite FM layer with $x<0$ with the local magnetization
${\bf M}=\cos \theta {\bf e}_{z}-\sin \theta {\bf e}_{y}$, a diffuse
interfacial layer I between $0<x<d^I$ with the same local magnetization
as in the FM layer, and a semi-infinite NM layer for $x>d^I$. When $d^I$
is infinitesimally small, this sub-system represents the three FM-NM
interfaces in Fig.~\ref{pic_multi}, i.e., between the thick FM and
spacer layers, between the spacer and thin FM layers (when spatially
inverted), and between the thin FM and back NM layers. Both spin
accumulation and current are assumed to be continuous at the FM-I and
I-NM interfaces, and one can derive the relation between spin
accumulation and current at $x=0$ with the same quantities at $x=d^I$ as
the thickness of the interfacial layer $d^I$ goes to zero. In this limit
the parameters of the interfacial layer, such as $\lambda_{mfp}^{I}$,
$\tau_{sf}^{I}$, $J^{I}$, $\lambda _{J}^{I}$, and, most important, its
resistance $AR_{I}$ remain constant; the latter condition implies that
the diffusion constant of the interfacial layer $D_{0}^{I}\sim d^I$ as
$d^I\rightarrow 0$.
\begin{figure} 
\centering
\includegraphics[width=4.5in]{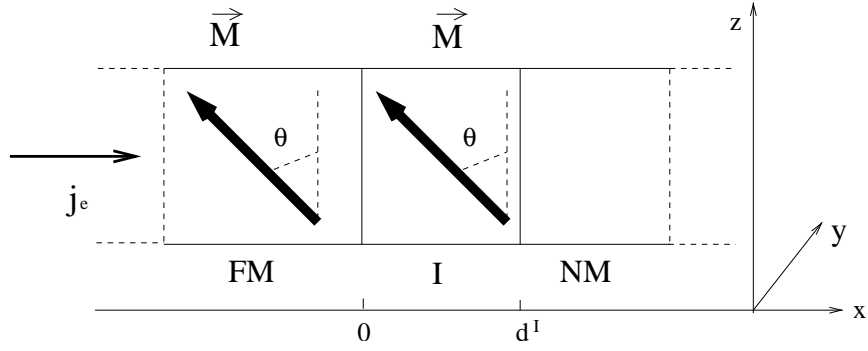}
\caption[Structure of the interface between the ferromagnetic 
and non-magnetic layers]{Structure of the interface between the layers 
at Fig.\ref{pic_multi}. FM is a semi-infinite ferromagnetic layer 
with the local magnetization 
${\bf M}_d^{(2)}=\cos\theta{\bf e}_z-\sin\theta{\bf e}_y$, I is a 
diffuse interfacial layer with the same local magnetization as in FM 
layer, and NM is a semi-infinite nonmagnetic layer.} 
\label{pic_bound_cond} 
\end{figure}

Equations~(\ref{eq_m_long}),~(\ref{eq_m_trans}) for the spin 
accumulation are solved, and
the Eq.~(\ref{jm_fin}) is used to find spin current in the ferromagnetic 
and
interfacial layers. By adopting a set of local coordinates
$(\bar{x},\bar{y},\bar{z})$ such that the {\it local} magnetization is
${\bf M}_{\bar{z}}={\bf e}_{\bar{z}}$ the spin accumulation and current
in the FM layer take the form:
\begin{equation}
\left\{
\begin{array}{l}
m^F_{\bar{x}}=2Re\left( G_{2}\exp (\frac{x}{l_{+}^{F}})\right) \\
\\
m^F_{\bar{y}}=2Im\left( G_{2}\exp (\frac{x}{l_{+}^{F}})\right) \\
\\
m^F_{\bar{z}}=G_{1}\exp (\frac{x}{\lambda _{sdl}^{F}}),
\end{array}
\right.
\end{equation}
and
\begin{equation}
\left\{
\begin{array}{l}
j_{m,\bar{x}}^F=-4D_{0}^{F}Re\left( \frac{G_{2}}{l_{+}^{F}}\exp 
(\frac{x}{l_{+}^{F}})\right) \\
\\
j_{m,\bar{y}}^F=-4D_{0}^{F}Im\left( \frac{G_{2}}{l_{+}^{F}}\exp 
(\frac{x}{l_{+}^{F}})\right) \\
\\
j_{m,\bar{z}}^F=\beta j_{e}-\frac{2D_{0}^{F}(1-\beta \beta ^{\prime 
})}{\lambda _{sdl}^{F}}G_{1}\exp (\frac{x}{\lambda _{sdl}^{F}}).
\end{array}
\right.
\end{equation}
In the interfacial layer, spin accumulation and spin current take the 
form:
\begin{equation}
\left\{
\begin{array}{l}
m_{\bar{x}}^I=2Re\left( G_{5}\exp (\frac{x}{l_{+}^{I}})\right) 
+2Re\left(
G_{6}\exp (-\frac{x}{l_{+}^{I}})\right) \\
\\
m_{\bar{y}}^I=2Im\left( G_{5}\exp (\frac{x}{l_{+}^{I}})\right)+2Im\left(
G_{6}\exp (-\frac{x}{l_{+}^{I}})\right) \\
\\
m_{\bar{z}}^I=G_{3}\exp (\frac{x}{\lambda _{sdl}^{I}})+G_{4}\exp 
(-\frac{x}{\lambda_{sdl}^{I}}),
\end{array}
\right.
\end{equation}
and
\begin{equation}
\left\{
\begin{array}{l}
j_{m,\bar{x}}^I=-4D_{0}^{I}\left[ Re\left( \frac{G_{5}}{l_{+}^{I}}\exp 
(\frac{x
}{l_{+}^{I}})\right) -Re\left( \frac{G_{6}}{l_{+}^{I}}\exp (-\frac{x}{
l_{+}^{I}})\right) \right] \\
\\
j_{m,\bar{y}}^I=-4D_{0}^{I}\left[ Im\left( \frac{G_{5}}{l_{+}^{I}}\exp 
(\frac{x
}{l_{+}^{I}})\right) -Im\left( \frac{G_{6}}{l_{+}^{I}}\exp (-\frac{x}{
l_{+}^{I}})\right) \right] \\
\\
j_{m,\bar{z}}^I=\gamma j_{e}-\frac{2D_{0}^{I}(1-\gamma \gamma ^{\prime 
})}{
\lambda _{sdl}^{I}}\left[ G_{3}\exp (\frac{x}{\lambda 
_{sdl}^{I}})-G_{4}\exp
(-\frac{x}{\lambda _{sdl}^{I}})\right] .
\end{array}
\right.
\end{equation}
Here $\beta $, $\beta ^{\prime}$ are spin-polarization parameters for
the conductivity and diffusion constant in the bulk of the 
ferromagnetic layer, defined in Chap.~\ref{chap_swt_thr}; $\gamma$,
$\gamma^{\prime}$ are similar parameters for the conductivity 
and diffusion constant in  the interfacial layer, 
defined as 
\begin{eqnarray}
\label{gam_gam_prime}
{\bf C}_{I}&=&\gamma C_{0}^{I}{\bf M}_{d}, \nonumber \\
{\bf D}_{I}&=&\gamma^{\prime}D_{0}^{I}{\bf M}_{d}.
\end{eqnarray}
and 
\begin{equation}
(l_{+}^{F,I})^{-1}=\sqrt{\frac{1}{(\lambda _{sf}^{F,I})^{2}}-\frac{i}{
(\lambda _{J}^{F,I})^{2}}}\approx \frac{1-i}{\sqrt{2}\lambda
_{J}^{F,I}},
\end{equation} 
when $\lambda _{J}^{F,I}\ll \lambda _{sf}^{F,I}$.

The boundary conditions for the continuity of the spin accumulation and
current at the interface between ferromagnetic and interfacial layer $x=0$
take the form:
\begin{equation}
\left\{
\begin{array}{l}
\label{m0}
2ReG_{2}=2ReG_{5}+2ReG_{6} \\
\\
2ImG_{2}=2ImG_{5}+2ImG_{6} \\
\\
G_{1}=G_{3}+G_{4},
\end{array}
\right.
\end{equation}
and
\begin{equation}
\left\{
\begin{array}{l}
\label{j0}
-4D_{0}^{F}Re(\frac{G_{2}}{l_{+}^{F}}
)=-4D_{0}^{I}Re(\frac{G_{5}-G_{6}}{l_{+}^{I}}) \\
\\
-4D_{0}^{F}Im(\frac{G_{2}}{l_{+}^{F}})=
-4D_{0}^{I}Im(\frac{
G_{5}-G_{6}}{l_{+}^{I}}) \\
\\
\beta j_{e}-\frac{2D_{0}^{F}(1-\beta \beta ^{\prime 
})}{
\lambda _{sdl}^{F}}G_{1}=\gamma j_{e}-\frac{2D_{0}^{I}(1-\gamma \gamma
^{\prime })}{\lambda _{sdl}^{I}}(G_{3}-G_{4}).
\end{array}
\right.
\end{equation}

To relate ${\bf m}^{F}(0)$ to ${\bf m}^{I}(d^I)$, and ${\bf j}_m^F(0)$
to ${\bf j}_m^I(d^I)$, one can use the assumption that as the thickness
of the interfacial layer goes to zero, other parameters of the
interfacial layer, such as $\lambda _{sdl}^{I}$, $J^{I}$, and $\lambda
_{J}^{I}$ remain constant, but the diffusion constant $D_{0}^{I}$ goes
to zero with the same rate as $d^I$, so that $d^I/D_{0}^{I}=const$.
Then, for example, for small $d^I\ll \lambda_{J}^{I}$ the
$\bar{x}$-component of the spin-accumulation at $x=d^I$ may be written
as
\begin{equation}
m_{\bar{x}}^I(d^I\rightarrow 0)\approx 2Re(G_{5}+G_{6})+2Re\left(
(G_{5}-G_{6})\frac{d^I}{l_{+}^{I}}\right) .  \label{bb}
\end{equation}

By comparing this expression with Eqs.~(\ref{m0}) and~(\ref{j0}), the
following relation between the $\bar{x}$-components of spin accumulation
and current at $x=0$ and $x=d^I$ is obtained:
\begin{equation}
m_{\bar{x}}^I(d^I\rightarrow 
0)=m_{\bar{x}}^F(0)-j_{m,\bar{x}}^F(0)\frac{d^I}{
2D_{0}^{I}},  \label{mx}
\end{equation}
and similarly,
\begin{equation}
m_{\bar{y}}^I(d^I\rightarrow 
0)=m_{\bar{y}}^F(0)-j_{m,\bar{y}}^F(0)\frac{d^I}{
2D_{0}^{I}},
\end{equation}
\begin{equation}
m_{\bar{z}}^I(d^I\rightarrow 0)=m_{\bar{z}}^F(0)+j_{e}\frac{\gamma }{
2(1-\gamma \gamma ^{\prime 
})}\frac{d^I}{D_{0}^{I}}-j_{m,\bar{z}}^F(0)\frac{1}{
2(1-\gamma \gamma ^{\prime })}\frac{d^I}{D_{0}^{I}}.
\end{equation}
In a manner similar to Eq.~(\ref{bb}) the $\bar{x}$-component of spin 
current
at $x=d^I$ may be written as
\[
j_{m,\bar{x}}^I(d^I\rightarrow 0)\approx -4D_{0}^{I}Re\left( 
\frac{G_{5}-G_{6}
}{l_{+}^{I}}\right) -2Re\left( i(G_{5}+G_{6})\right) 
\frac{d^{I}J^{I}}{\hbar }.
\]
By comparing this expression with Eqs.~(\ref{m0}) and~(\ref{j0}), one
finds the continuity condition for the $\bar{x}$-component of spin
current:
\begin{equation}
j_{m,\bar{x}}^I(d^I\rightarrow 
0)=j_{m,\bar{x}}^F(0)-m_{\bar{y}}^F(0)\frac{
d^{I}J^{I}}{\hbar },
\end{equation}
and, similarly,
\begin{equation}
j_{m,\bar{y}}^I(d^I\rightarrow 
0)=j_{m,\bar{y}}^F(0)+m_{\bar{x}}^F(0)\frac{
d^{I}J^{I}}{\hbar },
\end{equation}
and
\begin{equation}
j_{m,\bar{z}}^I(d^I\rightarrow 
0)=j_{m,\bar{z}}^F(0)-m_{\bar{z}}^F(0)\frac{d^I}{
\tau_{sf}^{I}}.  \label{jz}
\end{equation}
With these relations, the boundary conditions at the three interfaces in
the multilayered structure depicted in Fig.~\ref{pic_multi} can now
be obtained.

By using the conditions Eqs.~(\ref{mx})-(\ref{jz}), the boundary
conditions at the interface between the thin (first) ferromagnetic and
non-magnetic (N) layers of the structure shown in
Fig.~\ref{pic_multi} at $x=t_{F}$ may be written immediately, since
in the thin FM layer the local coordinate system $(\bar{x},\bar{y},\bar{z})$ 
coincides with the global axes $(x,y,z)$; one finds
\begin{equation}
\left\{
\begin{array}{l}
\label{bctFm} 
m^N_x(t_{F})-m^{(1)}_x(t_{F})=-rj^{(1)}_{m,x}(t_{F}) \\
\\
m^N_y(t_{F})-m^{(1)}_y(t_{F})=-rj^{(1)}_{m,y}(t_{F}) \\
\\
m^N_z(t_{F})-m^{(1)}_z(t_{F})=rj_{e}\frac{\gamma }{1-\gamma \gamma 
^{\prime }}
-rj^{(1)}_{m,z}(t_{F})\frac{1}{1-\gamma\gamma^{\prime }},
\end{array}
\right.
\end{equation}
and
\begin{equation}
\left\{
\begin{array}{l}
\label{bctFj} 
j^N_{m,x}(t_F)-j^{(1)}_{m,x}(t_{F})=
-m^{(1)}_{y}(t_F)\frac{d^{I}J^I}{\hbar}\\
\\
j^N_{m,y}(t_{F})-j^{(1)}_{m,y}(t_{F})=
m^{(1)}_{x}(t_{F})\frac{d^{I}J^I}{\hbar} 
\\
\\
j^N_{m,z}(t_{F})-j^{(1)}_{m,z}(t_{F})=
-m^{(1)}_{z}(t_{F})\frac{d^{I}}{\tau^I_{sf}},
\end{array}
\right.
\end{equation}
where $r=d^{I}/2D_{0}^{I}$. Similarly, the boundary conditions at the
interface between the non-magnetic spacer (S) and the thin (first) FM
layer at $x=0$ take the form:
\begin{equation}
\left\{
\begin{array}{l}
\label{SpFMm} 
m^S_{x}(0)-m^{(1)}_{x}(0)=rj^{(1)}_{m,x}(0) \\
\\
m^S_{y}(0)-m^{(1)}_{y}(0)=rj^{(1)}_{m,y}(0) \\
\\
m^S_{z}(0)-m^{(1)}_{z}(0)=-rj_{e}\frac{\gamma }{1-\gamma \gamma 
^{\prime}}
+rj^{(1)}_{m,z}(0)\frac{1}{1-\gamma \gamma ^{\prime }},
\end{array}
\right.
\end{equation}
and
\begin{equation}
\left\{
\begin{array}{l}
\label{SpFMj} 
j^S_{m,x}(0)-j^{(1)}_{m,x}(0)=m^{(1)}_{y}(0)\frac{d^{I}J^I}{\hbar} \\
\\
j^S_{m,y}(0)-j^{(1)}_{m,y}(0)=-m^{(1)}_{x}(0)\frac{d^{I}J^I}{\hbar} \\
\\
j^S_{m,z}(0)-j^{(1)}_{m,z}(0)=m^{(1)}_{z}(0)\frac{d^{I}}{\tau^I_{sf}},
\end{array}
\right.
\end{equation}

Note that spin-current conservation condition at the interfaces, which
means that there are no torques acting at the interfaces, is due to the
infinitely small thickness of the interfacial layers $d^{I}\rightarrow
0$. To write the boundary conditions at the interface between
the thick FM and NM spacer layers at $x=0$, one have to change from the
local coordinate system $(\bar{x} ,\bar{y},\bar{z})$, related to the
magnetization direction in the thick FM layer, to the global $(x,y,z)$
system. Any vector ${\bf a}$ will be transformed according to the
following rule:
\begin{equation}
\label{rot_matr}
\left\{
\begin{array}{l}
a_{\bar{x}}=a_{x} \\
\\
a_{\bar{y}}=a_{y}\cos \theta +a_{z}\sin \theta \\
\\
a_{\bar{z}}=-a_{y}\sin \theta +a_{z}\cos \theta .
\end{array}
\right.
\end{equation}
By applying this transformation to the conditions,
Eqs.~(\ref{mx})-(\ref{jz}), one obtain the following boundary conditions
at the interface between the thick (second) FM and non-magnetic spacer
(S) layers:
\begin{equation}
\left\{
\begin{array}{c}
\label{FMSpm}
m^S_{x}(0)-m^{(2)}_{x}(0)=-rj^{(2)}_{m,x}(0) \\
\\
m^S_{y}(0)-m^{(2)}_{y}(0)=
-rj_{e}\frac{\gamma }{1-\gamma \gamma ^{\prime}}\sin\theta 
-rj^{(2)}_{m,y}(0)\frac{1-\gamma \gamma ^{\prime }\cos ^{2}\theta }{
1-\gamma \gamma ^{\prime }} \\
\\
+rj^{(2)}_{m,z}(0)\sin \theta \cos \theta \frac{\gamma \gamma ^{\prime 
}}{1-\gamma\gamma ^{\prime }} \\
\\
m^S_{z}(0)-m^{(2)}_{z}(0)=rj_{e}\frac{\gamma }{1-\gamma \gamma ^{\prime 
}}\cos\theta 
+rj^{(2)}_{m,y}(0)\sin \theta \cos \theta \frac{\gamma \gamma ^{\prime 
}}{
1-\gamma \gamma ^{\prime }} \\
\\
+rj^{(2)}_{m,z}(0)\frac{1-\gamma \gamma ^{\prime }\sin ^{2}\theta 
}{1-\gamma\gamma ^{\prime }},
\end{array}
\right.
\end{equation}
and
\begin{equation}
\left\{
\begin{array}{c}
\label{FMSpj}
j^S_{m,x}(0)-j^{(2)}_{m,x}(0)=-m^{(2)}_{y}(0)\cos \theta 
\frac{d^{I}J^{I}}{\hbar
}-m^{(2)}_{z}(0)\sin \theta \frac{d^{I}J^{I}}{\hbar } \\
\\
j^S_{m,y}(0)-j^S_{m,y}(0)=m^{(2)}_{x}(0)\cos \theta 
\frac{d^{I}J^{I}}{\hbar }
-m^{(2)}_{y}(0)\sin ^{2}\theta \frac{d^{I}}{\tau _{sf}^{I}} \\
\\
+m^{(2)}_{z}(0)\sin \theta \cos \theta \frac{d^{I}}{\tau _{sf}^{I}} \\
\\
j^S_{m,z}(0)-j^{(2)}_{m,z}(0)=m^{(2)}_{x}(0)\sin \theta 
\frac{d^{I}J^{I}}{\hbar }
+m_{2y}(0)\sin \theta \cos \theta \frac{d^{I}}{\tau _{sf}^{I}} \\
\\
-m^{(2)}_{z}(0)\cos ^{2}\theta \frac{d^{I}}{\tau _{sf}^{I}},
\end{array}
\right.
\end{equation}

Note that as the thickness of the interfacial layer $d^{I}$ goes to zero,
for diffuse scattering considered here, large discontinuities in the
spin-accumulation (Eqs.~(\ref{bctFm}), (\ref{SpFMm}), (\ref{FMSpm})) are
produced, proportional to finite $r=d^{I}/2D_{0}^{I}$, but small
discontinuities in the spin-currents (Eqs.~(\ref{bctFj}), (\ref{SpFMj}),
(\ref{FMSpj})) proportional to $d^{I}$, because $J_{I}$ does not increase
and $\tau _{sf}^{I}$ does not decrease as $d^{I}\rightarrow 0$. Since the
torque acting at the interface is proportional to the discontinuity of
the spin-current at the interface (see Eq.~(\ref{tot_torque})), the
finite thickness of the interfacial layer is essential for torque
production at the interface. As the infinitely small interfacial
thicknesses $d^{I}\rightarrow 0$ are considered, one obtains spin-current
conservation conditions at each interface:
\begin{equation}
{\bf j}^N_{m}(t_{F})={\bf j}^{(1)}_{m}(t_{F}),  \label{bctFjc}
\end{equation}
\begin{equation}
{\bf j}^S_{m}(0)={\bf j}^{(1)}_{m}(0),  \label{SpFMjc}
\end{equation}
and
\begin{equation}
{\bf j}^{(2)}_{m}(0)={\bf j}^S_{m}(0).  \label{FMSpjc}
\end{equation}

By eliminating $m^{S}(0)$ and $j^S_{m}(0)$ from Eqs.~(\ref{SpFMm}),
(\ref {FMSpm}), (\ref{SpFMjc}), and (\ref{FMSpjc}), one obtains the
boundary conditions at the interface between thick (second) and thin
(first) FM layers at $x=0$ (of course there is the NM spacer in-between,
however its thickness $t_{N}$ is irrelevant for these boundary
conditions as long as $ t_{N}\ll \lambda _{sdl}^{N}$):
\begin{equation}
\left\{
\begin{array}{c}
\label{bc0m}
m^{(1)}_{x}(0)-m^{(2)}_{x}(0)=-2rj^{(1)}_{m,x}(0) \\
\\
m^{(1)}_{y}(0)-m^{(2)}_{y}(0)=
-rj_{e}\frac{\gamma }{1-\gamma \gamma ^{\prime }}\sin\theta 
-rj^{(1)}_{m,y}(0)\frac{2-\gamma \gamma ^{\prime }(1+\cos ^{2}\theta 
)}{
1-\gamma \gamma ^{\prime }} \\
\\
+rj^{(1)}_{m,z}(0)\sin \theta \cos \theta 
\frac{\gamma \gamma ^{\prime }}{1-\gamma\gamma ^{\prime }} \\
\\
m^{(1)}_{z}(0)-m^{(2)}_{z}(0)=
rj_{e}\frac{\gamma }{1-\gamma \gamma ^{\prime }}(1+\cos\theta )+
rj^{(1)}_{m,y}(0)\sin \theta \cos \theta \frac{\gamma \gamma ^{\prime 
}}{
1-\gamma \gamma ^{\prime }} \\
\\
-rj^{(1)}_{m,z}(0)\frac{2-\gamma \gamma ^{\prime }\sin ^{2}\theta 
}{1-\gamma
\gamma ^{\prime }},
\end{array}
\right.
\end{equation}
and
\begin{equation}
{\bf j}^{(1)}_{m}(0)={\bf j}^{(2)}_{m}(0)  \label{bc0j}
\end{equation}

Finally, I show that parameter $r=d^{I}/2D_{0}^{I}$ is proportional to
the interface resistance $AR_{I}$ found from CPP transport measurements
~\cite{BP_jmmm_99}. By considering the expression (\ref{je}) for the
electrical current in the interfacial layer, and the assumptions that $
D_{0}^{I}\sim d^{I}$ and $\lambda _{sdl}^{I}$ remains constant as the
thickness of the interfacial layer $d^{I}\rightarrow 0$, one finds
$AR_{I}=d^{I}/2C_{0}^{I}$, or
$r=\frac{d^{I}}{2D_{0}^{I}}=AR_{I}\frac{C_0^I}{D_{0}^{I}}$. 
Parameters $C_{0}^{I}$ and $D_{0}^{I}$ may be related via Einstein's
relation
\begin{equation}
\label{einst_rel}
\hat{C}_{I}=e^{2}\hat{N}_{I}(\epsilon _{F})\hat{D}_{I}, 
\end{equation}
where
\begin{equation}
\left\{
\begin{array}{l}
\label{cdn_matrix}
\hat{C}=C_0{\hat I}+\mbox{\boldmath$\sigma$}\cdot{\bf C}\nonumber \\
\hat{D}=D_0{\hat I}+\mbox{\boldmath$\sigma$}\cdot{\bf D} \\
\hat{N}=N_0{\hat I}+\mbox{\boldmath$\sigma$}\cdot{\bf N},\nonumber 
\end{array}
\right.
\end{equation} 
$\mbox{\boldmath$\sigma$}$ is the Pauli matrix. In 
Eqs.~(\ref{cdn_matrix}), as well as in the equations below, the 
subscript $I$ is omitted for simplisity, as only the parameters of the 
interfacial layer are considered. 

Using the definitions of $\gamma$ and $\gamma^\prime$,
Eqs.~(\ref{gam_gam_prime}), and introducing a spin-polarization
parameter for the density of states $\gamma^{\prime\prime}$, such that
\begin{equation}
{\bf N}=\gamma^{\prime\prime}N_0{\bf M}_d
\end{equation}
equations~(\ref{cdn_matrix}) may be written as
\begin{equation}
\left\{
\begin{array}{l}
\label{cdn_sigmaz}
\hat{C}=C_0({\hat I}+\gamma{\hat\sigma}_z)\nonumber \\
\hat{D}=D_0({\hat I}+\gamma^\prime{\hat\sigma}_z) \\
\hat{N}=N_0({\hat I}+\gamma^{\prime\prime}{\hat\sigma}_z). \nonumber
\end{array}   
\right.
\end{equation}
The product of $\hat N$ and $\hat D$ then takes the form
$$
{\hat N}{\hat D}=N_0D_0((1+\gamma^\prime\gamma^{\prime\prime}){\hat I}+
(\gamma^\prime+\gamma^{\prime\prime}){\hat\sigma}_z),
$$
and from the Einstein's relation~(\ref{einst_rel}) it follows that
\begin{equation}
\left\{
\begin{array}{l} 
\label{c0_d0}
C_0=e^2N_0D_0(1+\gamma^\prime\gamma^{\prime\prime}) \nonumber \\
\gamma C_0=e^2N_0D_0(\gamma^\prime+\gamma^{\prime\prime}).
\end{array}   
\right.
\end{equation}
The experimental value of $\gamma$ may be found in the 
Ref.~\cite{BP_jmmm_99}, the value of $\gamma^{\prime\prime}$ may be 
found from the density of states of up and down electrons, $N_\uparrow$ 
and $N_\downarrow$ correspondingly, since 
$\hat N$ may be written as
$$
\hat N=
\bordermatrix{&   &      \cr
              & N_\uparrow  & 0   \cr
              & 0  & N_\downarrow  \cr},
$$
and comparing this matrix with the expression for $\hat N$ 
in~(\ref{cdn_sigmaz}), one obtains
\begin{equation}  
\left\{
\begin{array}{l}
\gamma^{\prime\prime}=\frac{N_\uparrow-N_\downarrow}{N_\uparrow+N_\downarrow}
\nonumber \\ \\
N_0=\frac{N_\uparrow+N_\downarrow}{2}.
\end{array}
\right.
\end{equation}
From the equation~(\ref{c0_d0}) the spin-polarization parameter 
$\gamma^{\prime}$ can be expressed in terms of $\gamma$ and 
$\gamma^{\prime\prime}$ as 
\begin{equation}
\label{gam_prime}
\gamma^{\prime}=
\frac{\gamma-\gamma^{\prime\prime}}{1-\gamma\gamma^{\prime\prime}},
\end{equation}
so that the parameter $r$ takes the form: 
\begin{equation}
r=AR_{I}e^{2}N_{0}^{I}(\epsilon_{F})\frac{1-\gamma^{\prime\prime
2}}{1-\gamma\gamma^{\prime}}, 
\label{res} 
\end{equation} 
where $e$ is the electron charge , $N_{0}^{I}(\epsilon _{F})$ is the
density of states at the interface at Fermi energy.

\section{\label{app_sol_three}
Solution of the diffusion equation for spin-accumu\-lation}
In this appendix, I present the expressions for the spin-accumulation 
and spin-current in the multilayered system shown on the 
Fig.~\ref{pic_three_lay}.
\begin{figure}
\centering
\includegraphics[width=4.5in]{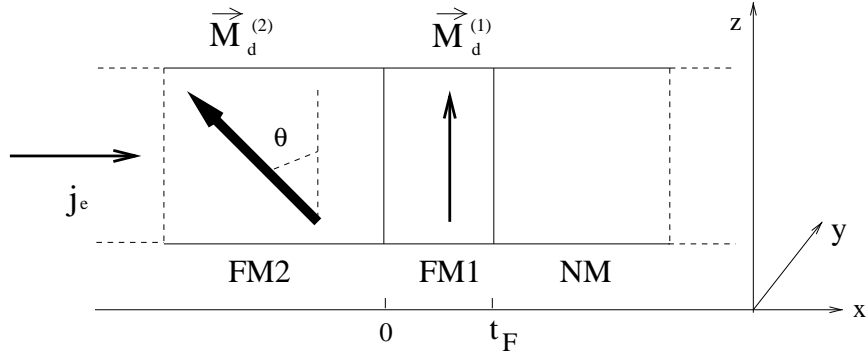}
\caption[Three-layered structure used for current induced reversal of a 
magnetic layer]
{Three-layered structure used for current induced reversal of a magnetic
layer. FM2 is a thick ferromagnetic layer with the thickness exceeding
$\lambda^F_{sdl}$ and local magnetization ${\bf
M}_d^{(2)}=\cos\theta{\bf e}_z-\sin\theta{\bf e}_y$, FM1 is a thin 
ferromagnetic layer with the thickness
$t_F$ and local magnetization ${\bf M}_d^{(1)}={\bf e}_z$, and NM is a
nonmagnetic back layer.}
\label{pic_three_lay}
\end{figure}

Equations.~(\ref{eq_m_long}),~(\ref{eq_m_trans}) for the spin 
accumulation in each of
three layers are solved, and the spin currents are found using
Eq.~(\ref{jm_fin}).  In the thick ferromagnetic layer, spin accumulation 
and
current take the form
\begin{equation}
\left\{
\begin{array}{l}
m^{(2)}_{x}=2Re\left( G_{2}\exp 
(\frac{x}{l_{+}})\right) \\
\\
m^{(2)}_{y}=2Im\left( G_{2}\exp 
(\frac{x}{l_{+}})\right) 
\cos \theta -G_{1}\exp (
\frac{x}{\lambda^F_{sdl}})\sin \theta \\
\\
m^{(2)}_{z}=2Im\left( G_{2}\exp 
(\frac{x}{l_{+}})\right) 
\sin \theta +G_{1}\exp (
\frac{x}{\lambda^F_{sdl}})\cos \theta ,
\end{array}
\right.
\end{equation}
and
\begin{equation}
\left\{
\begin{array}{c}
j^{(2)}_{m,x}=-4D_{0}Re\left( \frac{G_{2}}{l_{+}}
\exp (\frac{x}{l_{+}})\right) 
\\
\\
j^{(2)}_{m,y}=-\beta j_{e}\sin \theta -4D_{0}Im\left( 
\frac{G_{2}}{l_{+}}\exp (
\frac{x}{l_{+}})\right) \cos \theta \\
\\
+\frac{2D_{0}(1-\beta \beta ^{\prime })}{\lambda 
_{sdl}}G_{1}\exp (\frac{x}{
\lambda^F_{sdl}})\sin \theta \\
\\
j^{(2)}_{m,z}=\beta j_{e}\cos \theta -4D_{0}Im\left( 
\frac{G_{2}}{l_{+}}\exp (
\frac{x}{l_{+}})\right) \sin \theta 
-\frac{2D_{0}(1-\beta \beta ^{\prime })}{
\lambda^F_{sdl}}G_{1}\exp 
(\frac{x}{\lambda^F_{sdl}})\cos \theta .
\end{array}
\right.
\end{equation}
In the thin ferromagnetic layer
\begin{equation}
\left\{
\begin{array}{l}
m^{(1)}_{x}=2Re\left( G_{5}\exp 
(-\frac{x}{l_{+}})\right) 
+2Re\left( G_{6}\exp (
\frac{x-t_{F}}{l_{+}})\right) \\
\\
m^{(1)}_{y}=2Im\left( G_{5}\exp 
(-\frac{x}{l_{+}})\right) 
+2Im\left( G_{6}\exp (
\frac{x-t_{F}}{l_{+}})\right) \\
\\
m^{(1)}_{z}=G_{3}\exp (-\frac{x}{\lambda^F_{sdl}})+
G_{4}\exp (\frac{x-t_{F}}{\lambda^F_{sdl}}),
\end{array}
\right.
\end{equation}
and
\begin{equation}
\left\{
\begin{array}{l}
j^{(1)}_{m,x}=4D_{0}\left[ Re\left( 
\frac{G_{5}}{l_{+}}\exp 
(-\frac{x}{l_{+}}
)\right) -Re\left( \frac{G_{6}}{l_{+}}\exp 
(\frac{x-t_{F}}{l_{+}})\right)
\right] \\
\\
j^{(1)}_{m,y}=4D_{0}\left[ Im\left( 
\frac{G_{5}}{l_{+}}\exp 
(-\frac{x}{l_{+}}
)\right) -Im\left( \frac{G_{6}}{l_{+}}\exp 
(\frac{x-t_{F}}{l_{+}})\right)
\right] \\
\\
j^{(1)}_{m,z}=\beta j_{e}+\frac{2D_{0}(1-\beta \beta 
^{\prime })}
{\lambda^F_{sdl}}
\left[ G_{3}\exp (-\frac{x}{\lambda^F_{sdl}})-G_{4}\exp 
(\frac{x-t_{F}}{
\lambda^F_{sdl}})\right] .
\end{array}
\right.
\end{equation}
where $l_{+}^{-1}=\sqrt{\frac{1}{\lambda 
_{sf}^{2}}-\frac{i}{\lambda _{J}^{2}
}}\approx \frac{1-i}{\sqrt{2}\lambda _{J}}$, and 
$\lambda^F_{sdl}$ is
spin-diffusion length in FM layer. In the non-magnetic 
layer,
\begin{equation}
{\bf m}^{N}={\bf A}\exp (-\frac{x-t_{F}}{\lambda 
_{sdl}^{N}}),
\end{equation}
and
\begin{equation}
{\bf j}^N_{m}=\frac{2D_{0}^{N}}{\lambda _{sdl}^{N}}{\bf A}\exp 
(-\frac{x-t_{F}
}{\lambda _{sdl}^{N}}).
\end{equation}

To obtain the 12 unknown constants $A_{x}$, $A_{y}$, $A_{z}$, $G_{1}$, 
$ReG_{2}$, $ImG_{2}$, $G_{3}$, $G_{4}$, $ReG_{5}$, $ImG_{5}$, $ReG_{6}$, 
$ImG_{6}$, one can use the boundary conditions (see 
Appendix~\ref{app_bound_cond}, Eqs.~(\ref{bctFm}), (\ref{bc0m}), 
(\ref{bctFjc}), and (\ref{bc0j})):
\begin{equation}
\left\{
\begin{array}{l}
m^N_{x}(t_{F})-m^{(1)}_{x}(t_{F})=-rj^{(1)}_{m,x}(t_{F}) 
\\
\\
m^N_{y}(t_{F})-m^{(1)}_{y}(t_{F})=-rj^{(1)}_{m,y}(t_{F}) 
\\
\\
m^N_{z}(t_{F})-m^{(1)}_{z}(t_{F})=
rj_{e}\frac{\gamma }{1-\gamma \gamma ^{\prime }}
-rj^{(1)}_{m,z}(t_{F})\frac{1}{1-\gamma \gamma ^{\prime 
}},
\end{array}
\right.
\end{equation}
and
\begin{equation}
\left\{
\begin{array}{c}
m^{(1)}_{x}(0)-m^{(2)}_{x}(0)=-2rj^{(1)}_{m,x}(0) \\
\\
m^{(1)}_{y}(0)-m^{(2)}_{y}(0)=-rj_{e}\frac{\gamma}
{1-\gamma\gamma^\prime}\sin\theta-rj^{(1)}_{m,y}(0)
\frac{2-\gamma \gamma ^{\prime }(1+\cos^{2}\theta)}{
1-\gamma \gamma ^{\prime }} \\
\\
+rj^{(1)}_{m,z}(0)\sin\theta\cos\theta 
\frac{\gamma \gamma^\prime}{1-\gamma
\gamma^\prime} \\
\\
m^{(1)}_{z}(0)-m^{(2)}_{z}(0)=
rj_{e}\frac{\gamma }{1-\gamma\gamma^\prime}(1+\cos\theta)
+rj^{(1)}_{m,y}(0)\sin\theta\cos\theta 
\frac{\gamma\gamma^\prime}{1-\gamma\gamma^\prime} \\
\\
-rj^{(1)}_{m,z}(0)\frac{2-\gamma \gamma^\prime
\sin ^{2}\theta}{1-\gamma\gamma^\prime},
\end{array}
\right.   
\label{w}
\end{equation}
where the parameter $r$ is proportional to the interface resistance
$AR_{I}$ , $r=AR_{I}e^{2}N_{0}(1-\gamma^{\prime\prime 2})/(1-\gamma
\gamma^\prime)$, $e$ is the electron charge, $N_{0}$ is the density
of states at the interface, $\gamma$, $\gamma^\prime$, $\gamma
^{\prime\prime}$ are the spin polarization parameters for the
conductivity, diffusion constant, and density of states at the
interfaces (see Appendix~\ref{app_bound_cond}). The other six boundary
conditions come from the conservation of spin current at the interfaces:
\begin{equation}
{\bf j}^N_{m}(t_{F})={\bf j}^{(1)}_{m}(t_{F}),
\end{equation}
\begin{equation}
{\bf j}^{(1)}_{m}(0)={\bf j}^{(2)}_{m}(0).
\end{equation}

In order to find the torque $a$ and effective field $b$ acting on the 
thin ferromagnetic layer, one can use the definition of $a$ and $b$ 
(Eq.~(\ref{def_a_b})):
\begin{equation}
\label{def_a_b_3}
J{\bf m}^{(1)}_\perp=a{\bf M}_d^{(2)}\times{\bf M}_d^{(1)}
+b{\bf M}_d^{(1)}\times({\bf M}_d^{(2)}\times{\bf 
M}_d^{(1)})=-a\sin\theta{\bf e}_x-b\sin\theta{\bf e}_y.
\end{equation}
From the other hand, 
\begin{equation}
\label{m_perp_a_b}
J{\bf m}^{(1)}_\perp=Jm^{(1)}_x{\bf e}_x+Jm^{(1)}_y{\bf e}_y.
\end{equation}
Comparing these expressions, one can see that the torque $a$ and
effective field $b$ per unit length can be obtained by averaging the $x$-
and $y$- components of the spin accumulation over the thickness $t_F$ of
the thin layer.


\bibliography{BIB/thesprop}

\bibliographystyle{prsty}

\end{thesisbody}
\end{document}